\documentclass[aps,amsmath,amssymb,superscriptaddress,twocolumn,prb]{revtex4}
\usepackage{graphicx}
\usepackage{epsfig}
\usepackage{ulem}
\usepackage{amsmath}
\usepackage{amssymb}
\usepackage[usenames]{color}
\definecolor{orange}{rgb}{1,0.5,0}
\definecolor{violet}{rgb}{0.5,0,0.5}
\definecolor{mycolor}{rgb}{0.122, 0.435, 0.698}
\usepackage{mathtools}
\newcommand{\defeq}{\vcentcolon=}

\begin{document}

\title{Unbiased charge oscillations in DNA monomer-polymers and dimer-polymers}

\date{\today}

\author{K. Lambropoulos}
\affiliation{National and Kapodistrian University of Athens, Faculty of Physics,
Panepistimiopolis, 15784 Zografos, Athens, Greece}

\author{M. Chatzieleftheriou}
\affiliation{National and Kapodistrian University of Athens, Faculty of Physics,
Panepistimiopolis, 15784 Zografos, Athens, Greece}

\author{A. Morphis}
\affiliation{National and Kapodistrian University of Athens, Faculty of Physics,
Panepistimiopolis, 15784 Zografos, Athens, Greece}

\author{K. Kaklamanis}
\affiliation{National and Kapodistrian University of Athens, Faculty of Physics,
Panepistimiopolis, 15784 Zografos, Athens, Greece}

\author{M. Theodorakou}
\affiliation{National and Kapodistrian University of Athens, Faculty of Physics,
Panepistimiopolis, 15784 Zografos, Athens, Greece}

\author{C. Simserides}
\email{csimseri@phys.uoa.gr}
\homepage{http://users.uoa.gr/~csimseri/physics_of_nanostructures_and_biomaterials.html}
\affiliation{National and Kapodistrian University of Athens, Faculty of Physics,
Panepistimiopolis, 15784 Zografos, Athens, Greece}






\begin{abstract}
We call {\it monomer} a B-DNA base-pair and examine, analytically and numerically, electron or hole oscillations in monomer- and dimer-polymers, i.e.,
periodic sequences with repetition unit made of one or two monomers.
We employ a tight-binding (TB) approach at the base-pair level
to readily determine the spatiotemporal evolution of a single extra carrier along a $N$ base-pair polymer.
We study HOMO and LUMO eigenspectra as well as the mean over time probabilities to find the carrier at a particular monomer.
We use the pure mean transfer rate $k$ to evaluate the easiness of charge transfer. The inverse decay length $\beta$ for exponential fits $k(d)$, where $d$ is the charge transfer distance, and the exponent $\eta$ for power law fits $k(N)$ are computed; generally power law fits are better. We illustrate that increasing the number of different parameters involved in the TB description, the fall of $k(d)$ or $k(N)$ becomes steeper and show the range covered by $\beta$ and $\eta$.
Finally, both for the time-independent and the time-dependent problem, we analyze the {\it palindromicity} and the {\it degree of eigenspectrum dependence}
of the probabilities to find the carrier at a particular monomer.
\end{abstract}

\maketitle

\section{\label{sec:introduction} Introduction} 
Recently, we studied B-DNA dimers, trimers and polymers (we call monomer a B-DNA base-pair), with a tight-binding (TB) approach at the base-pair level, using the relevant on-site energies of the base-pairs and the hopping parameters between successive base-pairs~\cite{Simserides:2014,LKGS:2014}. Our method allows us to readily determine the spatiotemporal evolution of holes or electrons along a $N$ base-pair DNA segment through the solution of $N$ coupled differential equations~\cite{Simserides:2014,LKGS:2014}. We showed that for all dimers and for trimers made of identical monomers the carrier movement is periodic with frequencies in the mid- and far-infrared i.e. approximately in the THz domain~\cite{Simserides:2014,LKGS:2014}, a region of intense research~\cite{Yin:2012}. Increasing the number of monomers above three, periodicity is generally lost~\cite{Simserides:2014,LKGS:2014}. Even for the simplest tetramer, the carrier movement is not periodic\cite{Lambropoulos:2014}.
For periodic cases, we defined~\cite{Simserides:2014,LKGS:2014} the maximum transfer percentage $p$ (e.g. the maximum probability to find the carrier at the last monomer having placed it initially at the first monomer) and the pure maximum transfer rate $\frac{p}{T}=pf$ ($T$ being the period, $f$ the frequency).
For all cases, either periodic or not, the pure mean transfer rate $k$ (cf.~Eq.~\ref{meantransferrateN}) and the speed $u = kd$, where $d = (N-1) \times $ 3.4 {\AA} is the charge transfer distance, can be used to characterize the system. Our analytical calculations and numerical results show that for dimers $k=2\frac{p}{T}$ and for trimers made of identical monomers $k \approx 1.3108 p$.
Using $k$ to evaluate the easiness of charge transfer, one can calculate the inverse decay length $\beta$ for exponential fits $k(d)$ and the exponent $\eta$ for power law fits $k(N)$. Studying a few polymers and segments taken from experiments~\cite{Simserides:2014}, we found that $\beta \approx$ 0.2 - 2 {\AA}$^{-1}$ and
$\eta  \approx$ 1.7 - 17.

Carrier oscillations within ``molecular'' systems have been occasionally studied in the literature.
Real-Time Time-Dependent Density Functional Theory (RT-TDDFT)~\cite{RT-TDDFT} simulations predicted oscillations ($\approx$ 0.1-10 PHz)
within p-nitroaniline and FTC chromophore~\cite{Takimoto:2007},
zinc porphyrin, green fluorescent protein chromophores and adenine-thymine base-pair~\cite{LopataGovind:2011}.
In a simplified single-stranded helix of 101 bases, subjecting the system to a collinear uniform electric field, THz Bloch oscillations could be induced~\cite{Malyshev:2009}.
Single and multiple charge transfer within a typical DNA dimer in connection to a bosonic bath,
where each base-pair is approximated by a single site, as in our TB approach, has been studied in Ref.~\cite{Tornow:2010}.
In the subspace of single charge transfer between base-pairs and having initially placed the charge at the donor site,
the authors obtain a period slightly greater than 10 fs, having used a ``typical hopping matrix element'' 0.2 eV.
Using our equation~\cite{Simserides:2014,LKGS:2014} $f= \frac{1}{T} = \frac{\sqrt{(2t)^2 + \Delta^2}}{h}$, putting $ t = $ 0.2 eV for the ``typical hopping matrix element'' and identical dimers i.e. difference of the on-site energies $\Delta = 0$, we obtain a period $T \approx$ 10.34 fs in accordance with the dotted line in Fig.~4 of Ref.~\cite{Tornow:2010}.

In the present article, we focus on periodic DNA polymers with a repetition unit made of one or two monomers and analyze the parameters which favor charge transfer. A synopsis of the theory behind the calculations is given in Section~\ref{sec:theory} both for the time-dependent and the time-independent problem. In Section~\ref{sec:resdis} we distinguish three special types of polymers, we discuss analytical solutions, and we present and discuss our numerical and analytical results. Both for the time-independent and the time-dependent problem, we introduce two important properties of the probabilities to find the carrier at a particular monomer: {\it palindromicity} and {\it degree of eigenspectrum dependence}.
Moreover, we use $k$ to characterize the easiness of charge transfer and compare exponential fits of $k(d)$ with power law fits of $k(N)$. We also compare fits including any number of monomers with fits including only odd or only even numbers. In Section~\ref{sec:conclusion} we summarize our conclusions.
Finally, we just mention that although for $N \ge 4$ the oscillations are not periodic, Fast Fourier Transform analysis shows that the frequency content remains strong in the THz domain.

\section{\label{sec:theory} Theory} 
By YX we denote two successive base-pairs, according to the convention
\begin{eqnarray}
   &\vdots& \nonumber \\
5' &      & 3' \nonumber \\
\textrm{Y}  &   -  & \textrm{Y}_{\textrm{compl}} \nonumber \\
\textrm{X}  &   -  & \textrm{X}_{\textrm{compl}} \nonumber  \\
3' &      & 5' \nonumber \\
   &\vdots&
\label{bpdimer}
\end{eqnarray}
for the DNA strands orientation.
We denote by X, X$_{\textrm{compl}}$, Y, Y$_{\textrm{compl}}$ DNA bases, where
X$_{\textrm{compl}}$ (Y$_{\textrm{compl}}$) is the complementary base of X (Y).
In other words, the notation YX means that the bases Y and X of
two successive base-pairs are located at the same strand in the direction $5'-3'$.
X-X$_{\textrm{compl}}$ is the one base-pair and
Y-Y$_{\textrm{compl}}$ is the other base-pair,
separated and twisted by 3.4 {\AA} and $36^{\circ}$, respectively,
relatively to the first base-pair.
For example, the notation GT denotes that one strand contains G and T in the direction $5'-3'$ and the complementary
strand contains C and A in the direction $3'-5'$.

For each base-pair or {\it monomer},
the highest occupied molecular orbital (HOMO) and the lowest unoccupied molecular orbital (LUMO) play a key role,
since we suppose that an extra hole or electron inserted in a DNA segment travels through HOMOs or LUMOs.


\subsection{\label{subsubsec:base-pair-level-time-dependent} Time-dependent problem} 
For a TB description at the base-pair level, the time-dependent single carrier (hole/electron) wave function of the DNA polymer,
$\Psi^{DNA}_{H/L}({\bf r},t)$,
is written as a linear combination of base-pair wave functions with time-dependent coefficients, i.e.,
\begin{equation}\label{psi-total}
\Psi^{DNA}_{H/L}({\bf r},t) = \sum_{\mu=1}^{N} A_{\mu}(t) \; \Psi^{bp(\mu)}_{H/L}({\bf r}).
\end{equation}
$\Psi^{bp(\mu)}_{H/L}({\bf r})$ is the $\mu^{th}$ base-pair's HOMO or LUMO wave function ($H/L$).
The sum is over all base-pairs of the DNA polymer.
$|A_{\mu}(t)|^2$ gives the probability to find the carrier at the base-pair $\mu$, at the time $t$.

Using the time-dependent Schr\"{o}dinger equation
\begin{equation}\label{tdse}
i \hbar \frac{\partial \Psi^{DNA}_{H/L}({\bf r},t)}{\partial t} = \hat{H}^{DNA} \Psi^{DNA}_{H/L}({\bf r},t)
\end{equation}
and Eq.~\ref{psi-total} as well as the details described by Hawke {\it et al.}~\cite{HKS:2010-2011},
we find that the time evolution of the coefficients $A_\mu(t)$
obeys the TB system of differential equations
\begin{equation}\label{TBbp}
i \hbar \frac{ dA_{\mu}} {dt} = E^{bp (\mu)}_{H/L} A_{\mu}
+ t^{bp (\mu;\mu-1)}_{H/L} A_{\mu-1}
+ t^{bp (\mu;\mu+1)}_{H/L} A_{\mu+1}.
\end{equation}
$E^{bp (\mu)}_{H/L}$ is the HOMO/LUMO on-site energy of base-pair $\mu$, and
$t^{bp (\mu; \mu ')}_{H/L}$ is the hopping parameter between base-pair $\mu$ and base-pair $\mu '$.
The values of $E^{bp (\mu)}_{H/L}$ and $t^{bp (\mu; \mu ')}_{H/L}$ used in the present work are the same employed in Refs.~\cite{Simserides:2014,LKGS:2014}.

To solve Eq.~\ref{TBbp} we define the vector matrix
\begin{equation}\label{x}
\vec{x}(t) = \left[
\begin{array}{c}
A_1(t) \\
A_2(t) \\
\vdots \\
A_N(t)  \end{array} \right].
\end{equation}
Then, Eq.~\ref{TBbp} reads
\begin{equation}\label{xdotmathcalAx}
\dot{\vec{x}}(t) = \widetilde{\mathcal{A}} \vec{x}(t),
\end{equation}
where
$\widetilde{\mathcal{A}} = - \frac{i}{\hbar} \textrm{A}$ and
the matrix $\textrm{A}$ is a symmetric tridiagonal matrix shown in the Appendix~\ref{apA}. We solve Eq.~\ref{xdotmathcalAx} using the {\it eigenvalue method}, i.e. looking for solutions of the form $\vec{x}(t) = \vec{v} e^{\tilde{\lambda} t} \Rightarrow \dot{\vec{x}}(t) = \tilde{\lambda} \vec{v} e^{\tilde{\lambda} t}$. Hence, Eq.~\ref{xdotmathcalAx} reads
\begin{equation}\label{mathcalA}
\widetilde{\mathcal{A}} \vec{v} = \tilde{\lambda} \vec{v}
\end{equation}
or, with $ \tilde{\lambda} = - \frac{i}{\hbar} \lambda$,
\begin{equation}\label{textrmA}
\textrm{A} \vec{v} = \lambda \vec{v},
\end{equation}
i.e. we have to solve an eigenvalue problem. Provided that the normalized eigenvectors $\vec{v_k}$ corresponding to the eigenvalues $\lambda_k$ of Eq.~\ref{textrmA} are linearly independent (which holds in all cases studied), the solution to our problem is
\begin{equation}
\vec{x}(t) = \sum_{k=1}^{N} c_k \vec{v_k} e^{-\frac{i}{\hbar} \lambda_k t}.
\end{equation}

If we initially place the carrier at base-pair 1 and we want to see how the carrier will evolve, time passing, then
the initial condition would be
\begin{equation}\label{x0}
\vec{x}(0) = \left[
\begin{array}{c}
A_1(0) \\
A_2(0) \\
\vdots \\
A_N(0)  \end{array} \right]
=
\left[
\begin{array}{c}
1 \\
0 \\
\vdots \\
0  \end{array} \right].
\end{equation}
However, one could initially place the carrier at another base-pair (hence 1 would be at the corresponding place at the right-hand side) or even imagine to initially distribute the carrier probability equally among monomers (hence all right-hand side components would be $1/\sqrt{N}$).
From the initial conditions we determine $c_i$.

An estimation of the transfer rate can be obtained~\cite{Simserides:2014} as follows:
Supposing that initially i.e. for $t = 0$ we place the carrier at the first monomer (Eq.~\ref{x0}), then $|A_{1}(0)|^2 = 1$, while all other $|A_{j}(0)|^2 = 0$, $j=2, \dots, N$. Hence, for a polymer consisting of $N$ monomers, a \textit{pure} mean transfer rate can be defined as
\begin{equation}\label{meantransferrateN}
k = \frac{\langle |A_{N}(t)|^2 \rangle}{{t_{N}}_{mean}},
\end{equation}
where ${t_{N}}_{mean}$ is the first time $|A_{N}(t)|^2$ becomes equal to $\langle |A_{N}(t)|^2 \rangle$ i.e.
``the mean transfer time''.
Finally, the speed of charge transfer could be defined  as $ u = k d$; $d = (N-1) \times $ 3.4 {\AA} is the charge transfer distance.

\subsection{\label{subsubsec:base-pair-level-time-independent} Time-independent problem} 
The time-independent Schr\"{o}dinger equation
\begin{equation}\label{tidse}
\hat{H}^{DNA} \Psi^{DNA}_{H/L}({\bf r}) = \lambda \Psi^{DNA}_{H/L}({\bf r})
\end{equation}
can be solved expanding the time-independent single carrier (hole/electron) wave function of the DNA polymer,
$\Psi^{DNA}_{H/L}({\bf r})$ as a linear combination of base-pair wave functions with time-independent coefficients, i.e.,
\begin{equation}\label{psi-tid-total}
\Psi^{DNA}_{H/L}({\bf r}) = \sum_{\mu=1}^{N} \Gamma_{\mu} \; \Psi^{bp(\mu)}_{H/L}({\bf r}).
\end{equation}
$|\Gamma_{\mu}|^2$ gives the probability to find the carrier at the base-pair $\mu$.
The problem specified in Eqs.~(\ref{tidse})-(\ref{psi-tid-total}) is equivalent with $\textrm{A}  \vec{v} = \lambda \vec{v}$ (Eq.~\ref{textrmA}), with
\begin{equation}\label{vk}
\vec{v} = \left[
\begin{array}{c}
\Gamma_1 \\
\Gamma_2 \\
\vdots   \\
\Gamma_N  \end{array} \right].
\end{equation}
In other words, the eigenvalues and eigenvectors of Eq.~\ref{tidse} are $\lambda_k$ and $\vec{v}_k$, respectively; $v_{\mu k} = \Gamma_{\mu k}$.




\section{\label{sec:resdis} Periodic polymers with repetition unit made of one or two monomers} 
Let us focus on periodic DNA polymers with repetition unit made of either one monomer or two monomers (a dimer).
We distinguish three types of DNA polymers: \\
\textbf{(type $\alpha'$)} poly(dG)-poly(dC) and poly(dA)-poly(dT), \\
\textbf{(type $\beta'$)}  GCGC..., CGCG..., ATAT..., TATA..., and \\
\textbf{(type $\gamma'$)} TCTC... $\equiv$ GAGA..., CTCT... $\equiv$ AGAG..., ACAC... $\equiv$ GTGT..., CACA... $\equiv$ TGTG... .

Let us define $\Delta \defeq |E^{bp(o)} - E^{bp(e)}|$, where $E^{bp(o)}$ is the on-site energy of the carrier at {\it odd} monomers ($\mu = $ 1, 3, 5, ...) and $E^{bp(e)}$ is the on-site energy of the carrier at {\it even} monomers  ($\mu = $ 2, 4, 6, ...).
Let us by the way define $\Sigma \defeq E^{bp(o)} + E^{bp(e)}$.
Further, counting from the start, let us call $t^{bp}$ the hopping parameter {\it from odd to even} monomers (between $\mu = $ 1 and $\mu = $ 2 ...) and $t^{bp'}$ the hopping parameter {\it from even to odd} monomers (between $\mu = $ 2 and $\mu = $ 3 ...) .
For simplicity, we have dropped the indices ${H/L}$.

Then, we realize that the intricacy of the energy structure -- i.e. the number of different parameters involved in the TB description -- increases from type $\alpha'$ to type $\beta'$ and further to type $\gamma'$: In type $\alpha'$, $\Delta = 0$ and $t^{bp'}=t^{bp}$, so, we only have one non-zero TB parameter. In type $\beta'$, still  $\Delta = 0$ but $t^{bp'} \neq t^{bp}$, so, we have two non-zero TB parameters. Finally, in type $\gamma'$, $\Delta \neq 0$ and $t^{bp'} \neq t^{bp}$, so, we have  three non-zero TB parameters.

The eigenproblems we have to solve refer to
a tridiagonal   Toeplitz matrix of order $N$ for type $\alpha'$ polymers (cf. Eq.~\ref{Atypea}, the analytical solution is rather simple) and
a tridiagonal 2-Toeplitz matrix of order $N$ for type $\beta'$  (cf. Eq.~\ref{Atypeb}) and $\gamma'$ (cf. Eq.~\ref{Atypec}) polymers.
These eigenproblems have been studied in Ref.~\cite{Gover:1994} where the characteristic polynomial of a tridiagonal 2-Toeplitz matrix is shown to be
closely connected to polynomials satisfying the three point Chebyshev recurrence formula -- an extension of the well-known result for a tridiagonal Toeplitz matrix. Two theorems (2.3 and 2.4) describe the eigenvalues for odd and even $N$~\cite{Gover:1994}.
When $N$ is odd the eigenvalues can be expressed explicitly in terms of Chebyshev zeros~\cite{Gover:1994}.
Although for even $N$ there is no explicit formula, a recipe to produce the eigenvalues is given~\cite{Gover:1994}.
Specifically, these theorems refer to the tridiagonal 2-Toeplitz matrix of order $n$, given by Eq.~2.8 of Ref.~\cite{Gover:1994} (for us $ n = N $.):
\begin{equation}\label{Bn}
B_n = \left[
\begin{array}{ccccc}
\alpha_1   & \beta_1  &          &          & \textbf{0} \\
\gamma_1   & \alpha_2 & \beta_2  &          &            \\
           & \gamma_2 & \alpha_1 & \beta_1  &            \\
           &          & \gamma_1 & \alpha_2 & \ddots     \\
\textbf{0} &          &          & \ddots   & \ddots
\end{array}
\right]
\end{equation}
\noindent {\it Theorem 2.3 of Ref.~\cite{Gover:1994}}:
The eigenvalues of the tridiagonal 2-Toeplitz matrix of order $2m+1$ given in Eq.~\ref{Bn} (Eq.~2.8 of Ref.~\cite{Gover:1994}) are $\alpha_1$ and the solutions of the quadratic equations
\begin{equation}\label{theorem2.3}
(\alpha_1-\lambda)(\alpha_2-\lambda)-\left[ \beta_1 \gamma_1 + \sqrt{\beta_1 \beta_2 \gamma_1 \gamma_2} P_r + \beta_2 \gamma_2 \right] = 0,
\end{equation}
where $P_r=2 \cos{\frac{r\pi}{m+1}}$, $r=1,2,\dots,m$, are the zeros of $p_m'(\mu)$ defined by Equations \ref{equation2.30} and \ref{equation2.32p}. \\
\noindent {\it Theorem 2.4 of Ref.~\cite{Gover:1994}}:
The eigenvalues of the tridiagonal 2-Toeplitz matrix of order $2m$ given in Eq.~\ref{Bn} (Eq.~2.8 of Ref.~\cite{Gover:1994}) are the solutions of the quadratic equations
\begin{equation}\label{theorem2.4}
(\alpha_1-\lambda)(\alpha_2-\lambda)-\left[ \beta_1 \gamma_1 + \sqrt{\beta_1 \beta_2 \gamma_1 \gamma_2} Q_r + \beta_2 \gamma_2 \right] = 0,
\end{equation}
where $Q_r$, $r=1,2,\dots,m$, are the zeros of $q_m'(\mu)$ defined by Equations \ref{equation2.31} and \ref{equation2.32q}.

Equations~\ref{equation2.30}-\ref{equation2.31} are the Chebyshev three point recurrence formula.
\begin{equation}\label{equation2.30}
p_{m+1}'(\mu) = \mu p_m'(\mu)-p_{m-1}'(\mu),
\end{equation}
\begin{equation}\label{equation2.31}
q_{m+1}'(\mu) = \mu q_m'(\mu)-q_{m-1}'(\mu).
\end{equation}
The initial polynomials are
\begin{equation}\label{equation2.32p}
p_0'(\mu) = 1 \quad p_1'(\mu) = \mu,
\end{equation}
\begin{equation}\label{equation2.32q}
q_0'(\mu) = 1 \quad q_1'(\mu) = \mu +\beta.
\end{equation}
Finally,
\begin{equation}
\beta^2 = \frac{\beta_2 \gamma_2}{\beta_1 \gamma_1},
\end{equation}
\begin{equation}
\mu = \frac{\nu-(1+\beta^2)}{\beta},
\end{equation}
\begin{equation}
\nu = \frac{(\alpha_1-\lambda)(\alpha_2-\lambda)}{\beta_1 \gamma_1}.
\end{equation}
The eigenvectors are found in terms of polynomials satisfying the three point recurrence relationship~\cite{Gover:1994}.
For the tridiagonal 2-Toeplitz matrices of our interest, up to our knowledge, explicit eigenvalues have been found only for odd $N$~\cite{Kouachi:2006,Alvarez:2005} which agree with the results of Ref.\cite{Gover:1994}.
Throughout this work we solve the eigenproblems numerically and additionally we compare to analytical results.

\subsection{\label{subsec:eigen} Stationary states (time-independent problem): Eigenvalues and Eigenvectors } 
In this work we calculate the eigenvalues and eigenvectors numerically. However, in some cases, we compare with analytical solutions.
To simplify the notation, below we write $E^{bp} = E$, $t^{bp} = t$ and $t^{bp'} = t'$.

\subsubsection{\label{eigen:typea} type $\alpha'$ (and type $\alpha'$ {\it cyclic})} 
\begin{figure*}[h!]
\includegraphics[width=6cm]{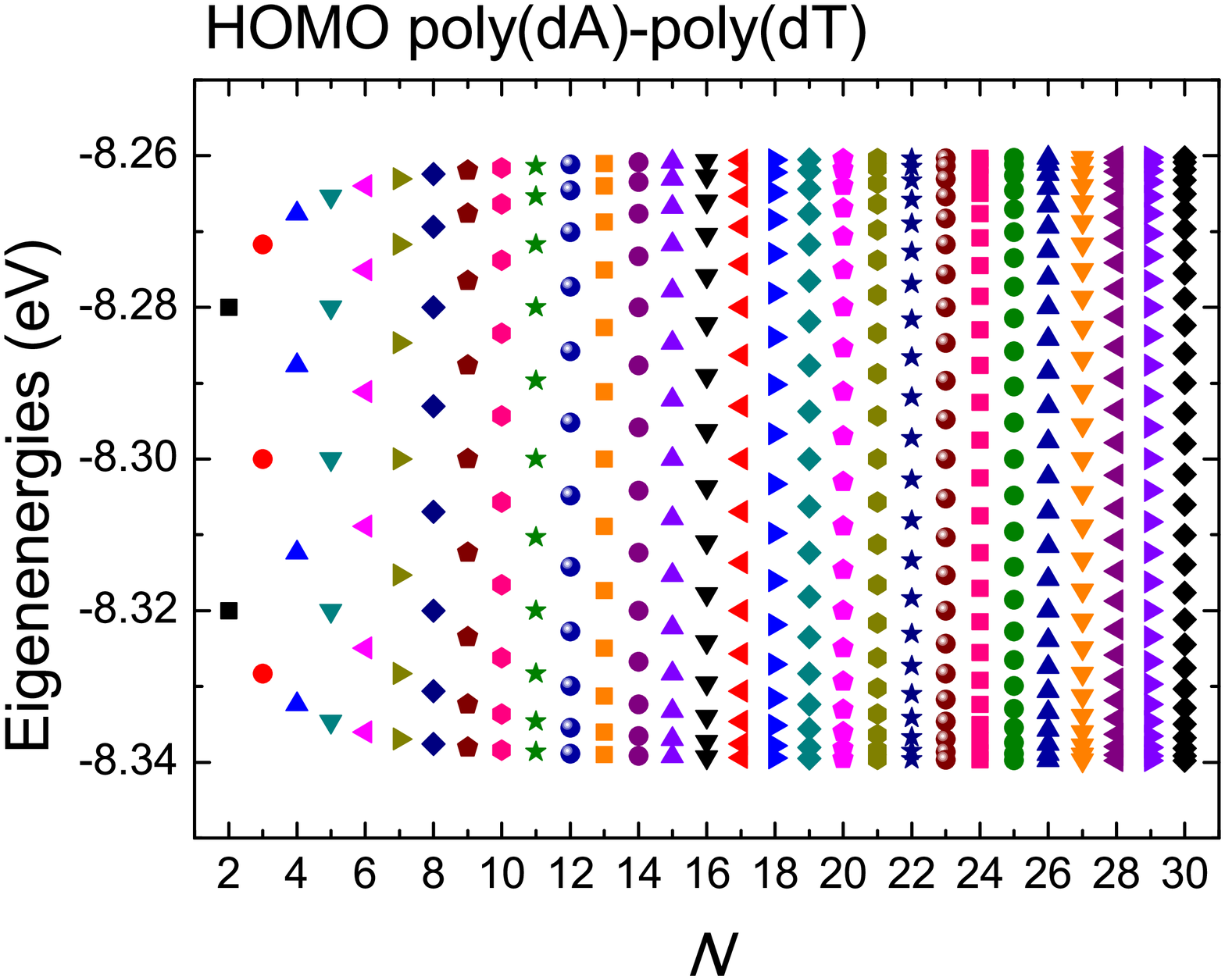}
\includegraphics[width=6cm]{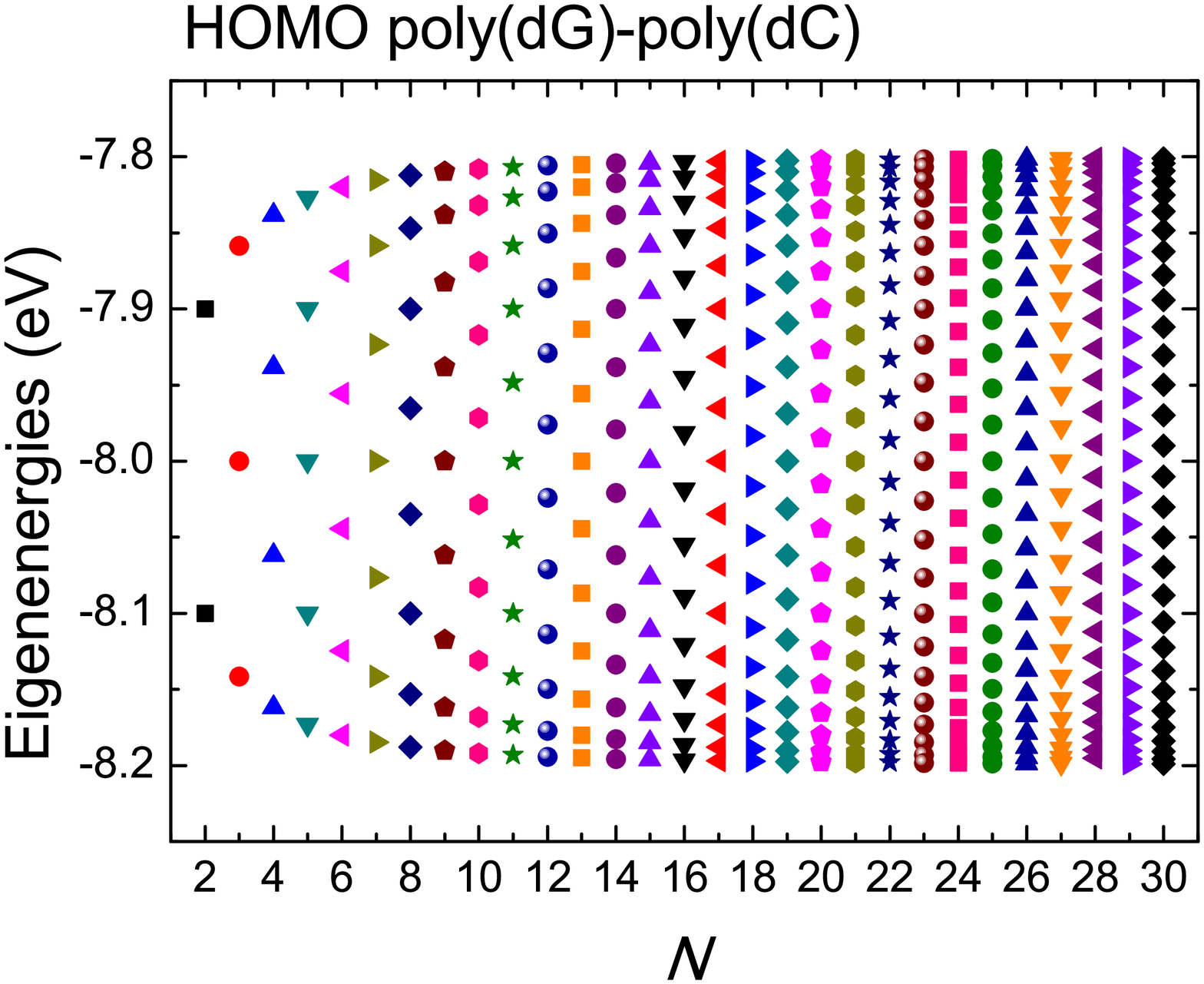}
\includegraphics[width=6cm]{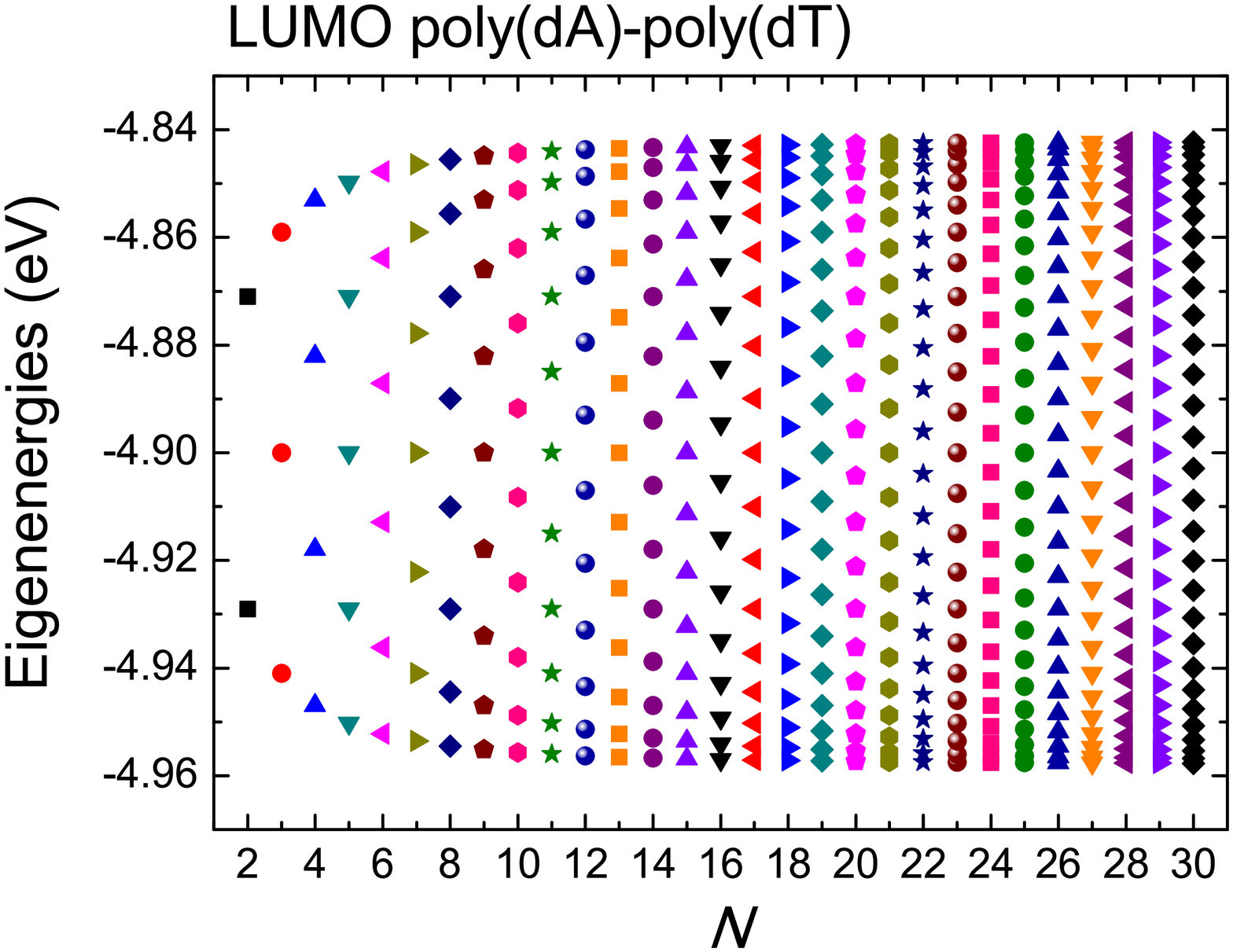}
\includegraphics[width=6cm]{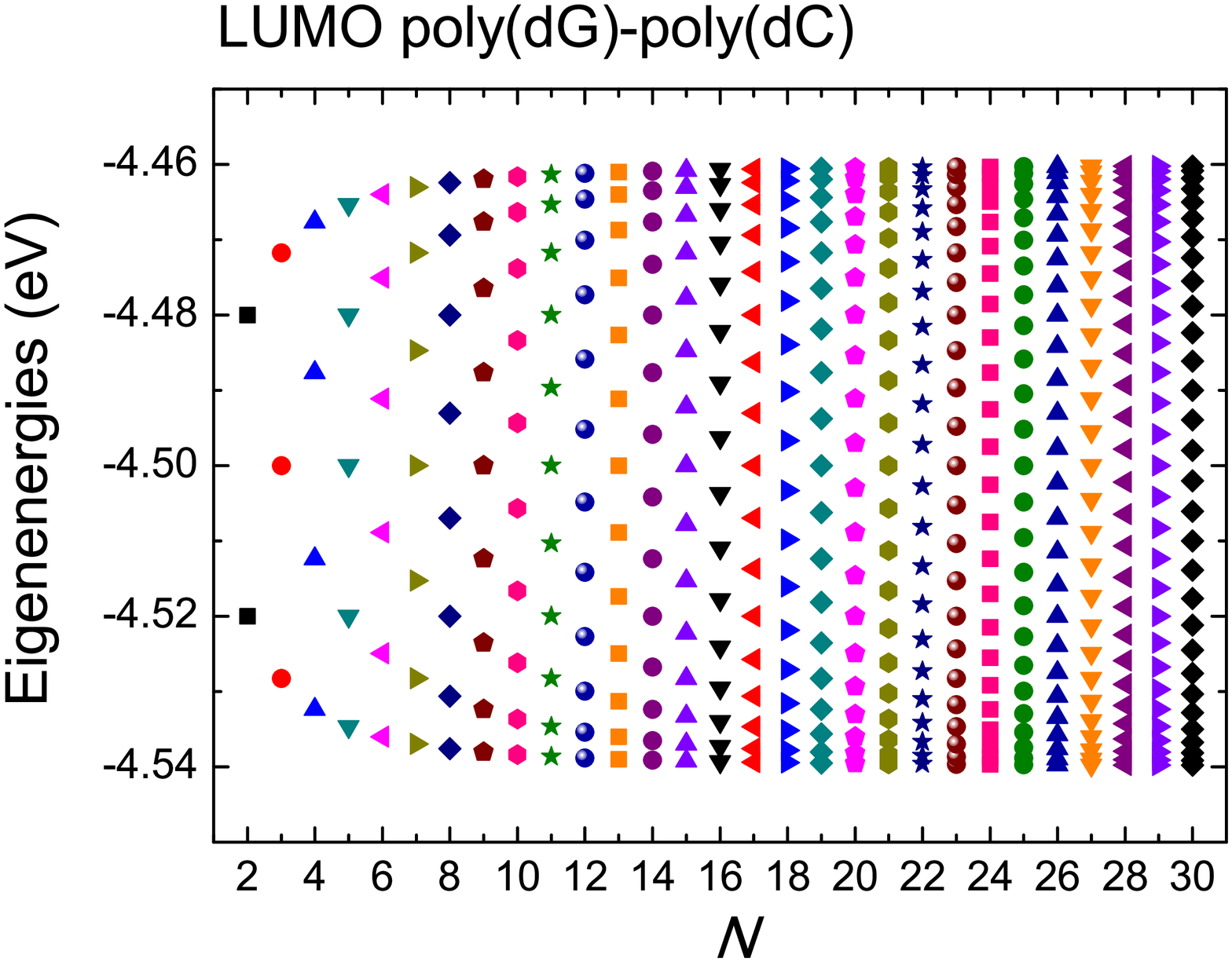}
\includegraphics[width=6cm]{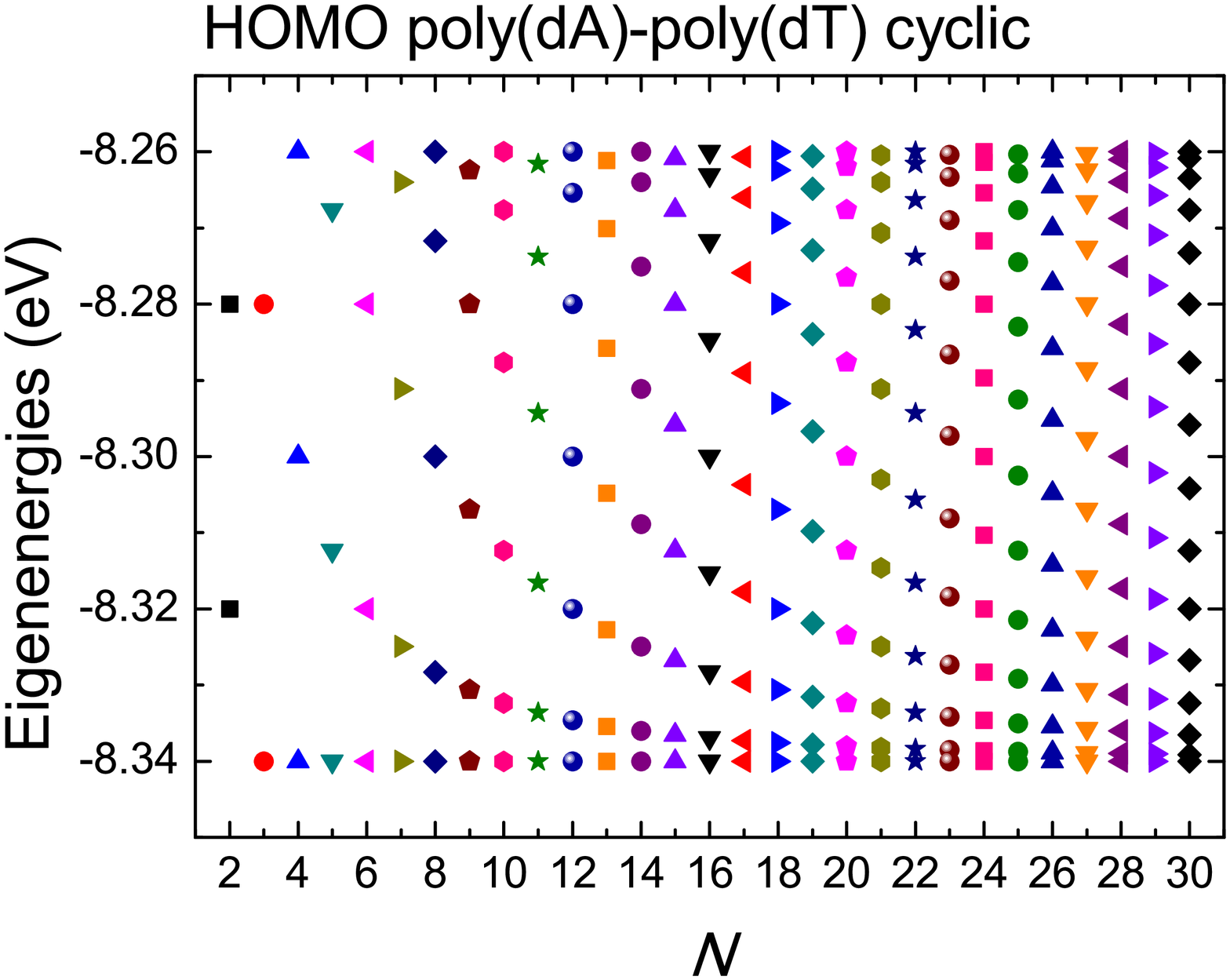}
\includegraphics[width=6cm]{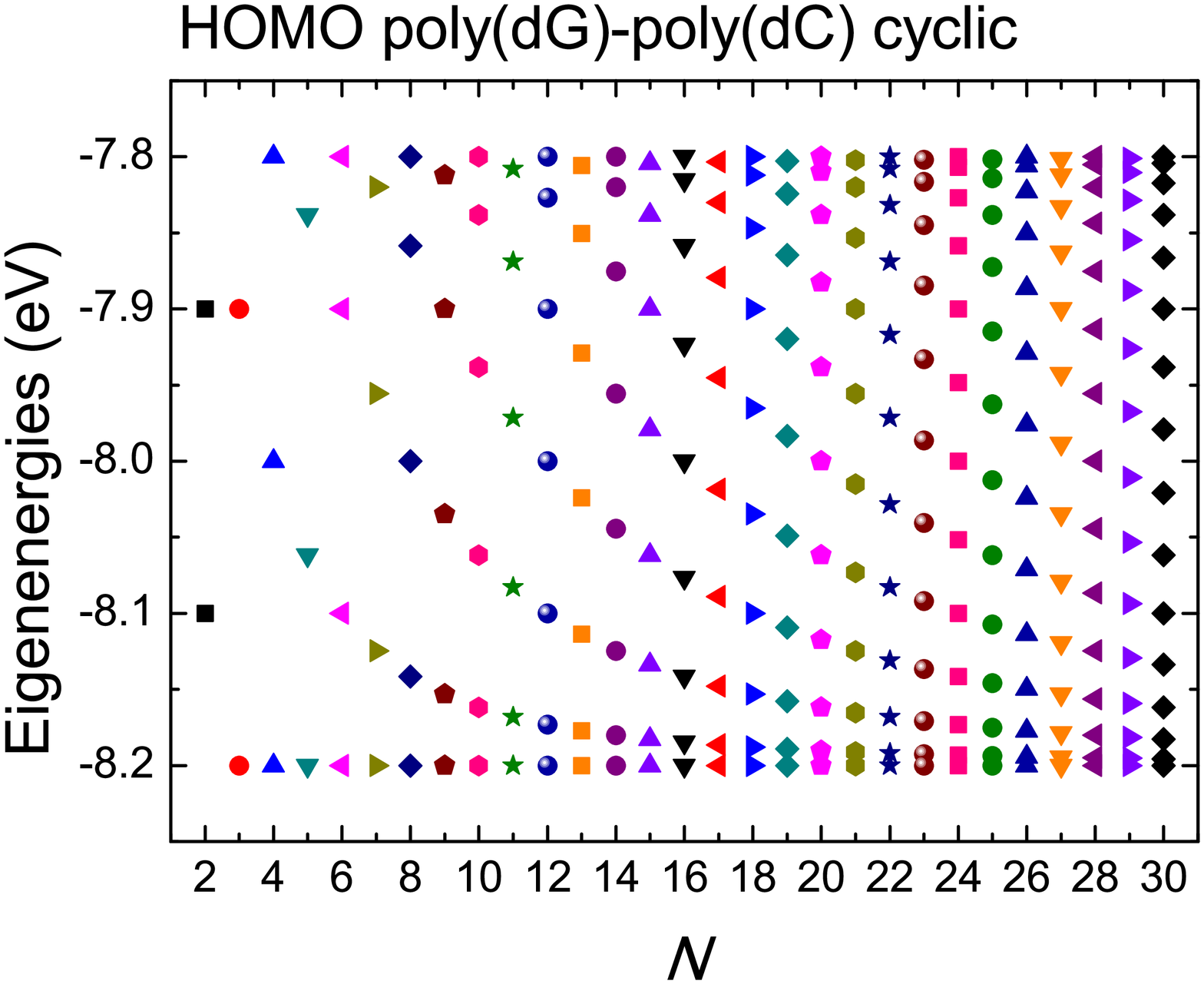}
\includegraphics[width=6cm]{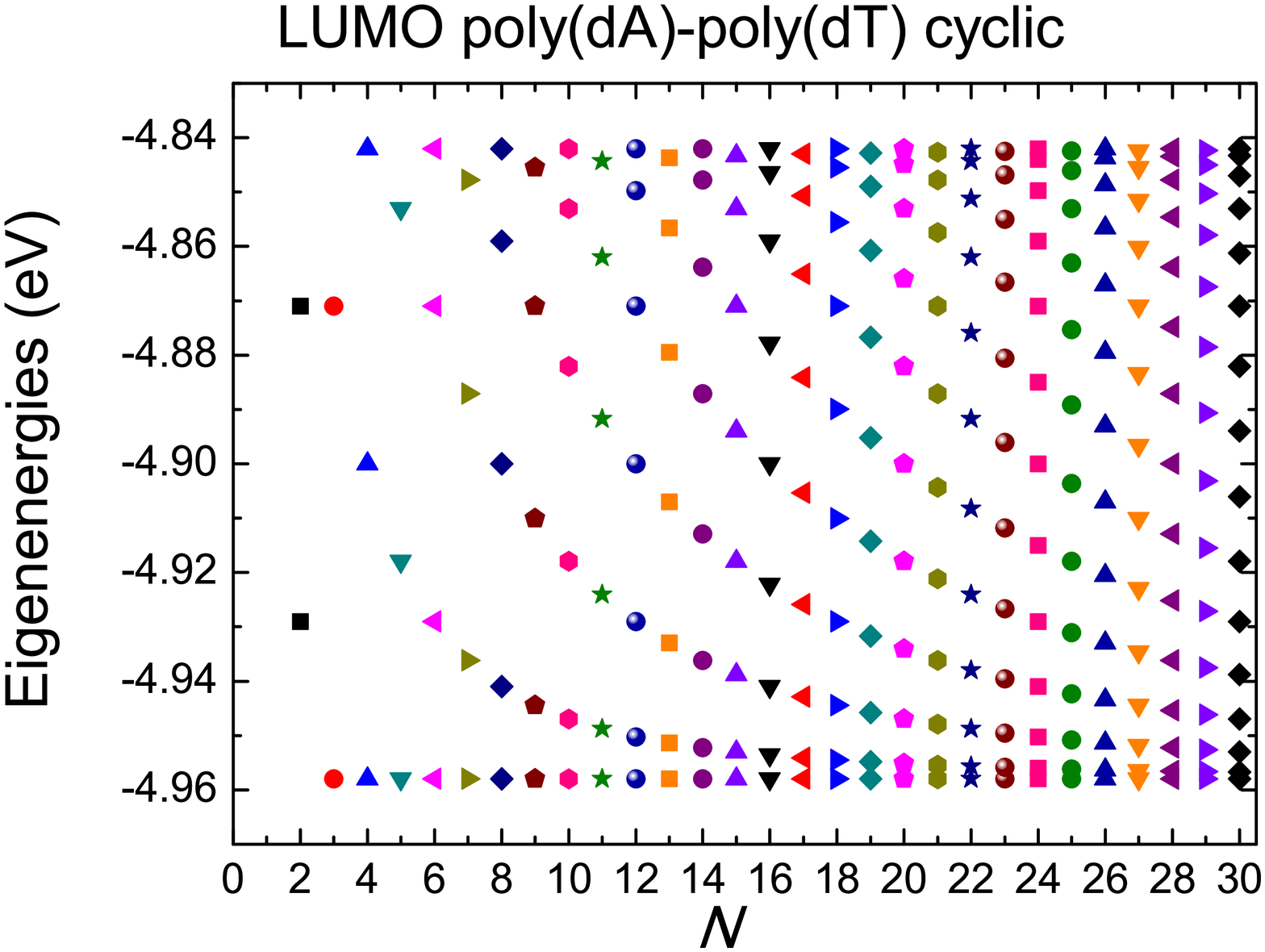}
\includegraphics[width=6cm]{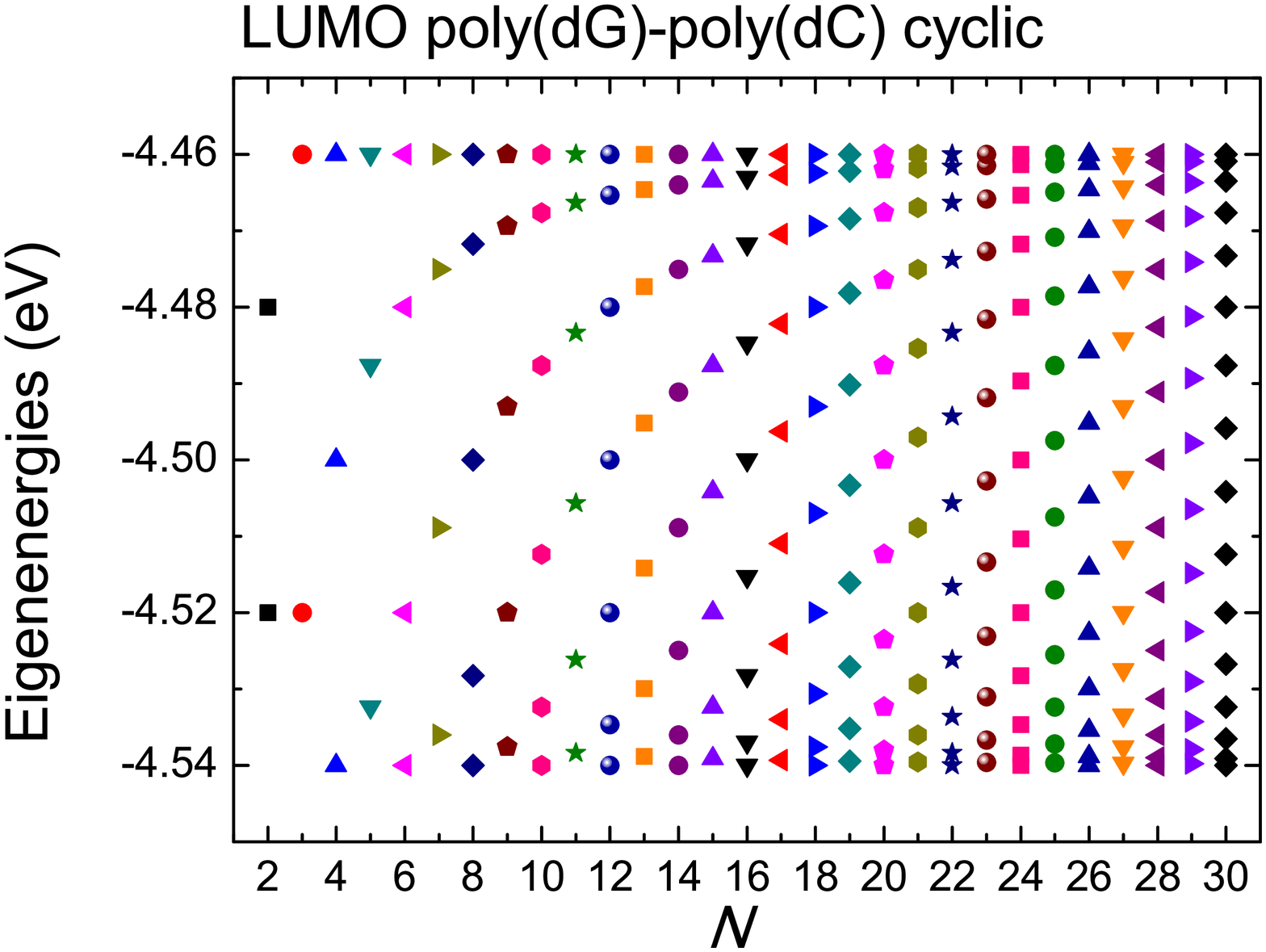}
\caption{\label{fig:eigenspectrum-ta} Eigenspectra of type $\alpha'$ polymers and type $\alpha'$ \textit{cyclic} polymers.}
\end{figure*}

For type $\alpha'$ [poly(dG)-poly(dC) and poly(dA)-poly(dT)], the matrix $\textrm{A}$ is a symmetric tridiagonal uniform matrix
\begin{equation}\label{Atypea}
\textrm{A} = \left[
\begin{array}{ccccccc}
E     & t    & 0    &\cdots& 0    & 0    & 0 \\
t     & E    & t    &\cdots& 0    & 0    & 0 \\
\vdots&\vdots&\vdots&\vdots&\vdots&\vdots&\vdots\\
0     & 0    & 0    &\cdots& t    & E    & t \\
0     & 0    & 0    &\cdots& 0    & t    & E \end{array} \right]
\end{equation}
with eigenvalues
\begin{equation}\label{eigenvaluestypea}
\lambda_k = E + 2 t \cos \left( \frac{k\pi}{N+1} \right),
\end{equation}
where $k=1,2,\dots,N$. All eigenvalues are real and distinct (non degenerate) since the matrix is symmetric ($\textrm{A} = \textrm{A}^\textrm{T}$),
all eigenvalues are symmetric around $E$,
for odd $N$ the trivial eigenvalue ($ = E$) exists, and
all eigenvalues lie in the interval $(E-2t,E+2t)$.
The eigenspectrum of type $\alpha'$ polymers is shown in Fig.~\ref{fig:eigenspectrum-ta}.
The $\mu$ component of the $k$ eigenvector is given by
\begin{equation}\label{eigenvectorstypea}
v_{\mu k} = \sqrt{\frac{2}{N+1}} \sin \left( \frac{\mu k \pi}{N+1} \right),
\end{equation}
where $k = 1,2,\dots,N$ and $\mu=1,2,\dots,N$.
Since $v_{\mu k}$ do not depend on $E$ or $t$, then, for any $k$,
the probability to find the carrier at a particular monomer $\mu$, $|v_{\mu k}|^2$
also does not depend on $E$ or $t$.
This property (let's call it \textbf{eigenspectrum independence} of the probabilities) is conserved in the time-dependent case (cf. subsubsection~\ref{subsubsec:meanprob-alpha}).
Since $\sin(\frac{(N-\mu+1) k \pi}{N+1}) = \pm \sin(\frac{\mu k \pi}{N+1})$ it follows that for each eigenstate $k$, $|v_{\mu k}|^2$
are {\it palindromes} i.e. the occupation probability for the $\mu$-th monomer is equal to the occupation probability of the $(N-\mu+1)$-th monomer.
This property (let's call it \textbf{palindromicity}) is conserved in the time-dependent case (cf. subsubsection~\ref{subsubsec:meanprob-alpha}).
The eigenspectra of type $\alpha'$ polymers are shown in the first two rows of Fig.~\ref{fig:eigenspectrum-ta}.

Furthermore, one could imagine the {\it cyclic} polymers with
$ \textrm{A}(1,N) = t^{bp(1;N)}_{H/L} = \textrm{A}(N,1) = t^{bp(N;1)}_{H/L} \neq 0 $.
For {\it cyclic} type $\alpha'$ polymers,
the matrix $\textrm{A}$ is a symmetric tridiagonal uniform matrix with two ``perturbed corners''
\begin{equation}\label{Atypeac}
\textrm{A} = \left[
\begin{array}{ccccccc}
E & t & 0 &\cdots& 0 & 0 & t \\
t & E & t &\cdots& 0 & 0 & 0 \\
\vdots&\vdots&\vdots&\vdots&\vdots&\vdots&\vdots\\
0 & 0 & 0 &\cdots& t & E & t \\
t & 0 & 0 &\cdots& 0 & t & E \end{array} \right]
\end{equation}
whose eigenvalues are
\begin{equation}\label{eigenvaluestypeac}
\lambda_k = E + 2t \cos \left( \frac{2 k \pi}{N} \right),
\end{equation}
where $k = 1, 2, \dots, N$.
Generally, all eigenvalues are not distinct but degeneracies exist.
The number of discrete eigenvalues $M=\frac{N+1}{2}$ for $N$ odd and $M=\frac{N+2}{2}$ for $N$ even.
Eigenvalues of this and other tridiagonal Toeplitz matrices with four perturbed corners can also be found in Ref.~\cite{YuehCheng:2008}.
The $\mu$ component of the $k$ eigenvector is given by
\begin{equation}\label{eigenvectorstypeac}
v_{\mu k} = \frac{1}{\sqrt{N}} \exp \left( \frac{i \mu 2 k \pi}{N} \right),
\end{equation}
where $\mu = 1, 2, \dots, N$.
Since $|v_{\mu k}|^2=\frac{1}{N}$, for any eigenstate $k$, the occupation probability is equal for all monomers.
The eigenspectra of type $\alpha'$ \textit{cyclic} polymers are shown in the last two rows of Fig.~\ref{fig:eigenspectrum-ta}.

\subsubsection{\label{eigen:typeb} type $\beta'$} 
For type $\beta'$ polymers, the matrix $\textrm{A}$ is
\begin{equation}\label{Atypeb}
\textrm{A} = \left[
\begin{array}{ccccc}
E    & t  & 0  & 0  & \cdots  \\
t    & E  & t' & 0  & \cdots  \\
0    & t' & E  & t  & \cdots  \\
 \vdots   & \vdots  & \vdots  & \vdots & \vdots  \end{array} \right]
\end{equation}

\begin{figure*}[h!]
\includegraphics[width=6cm]{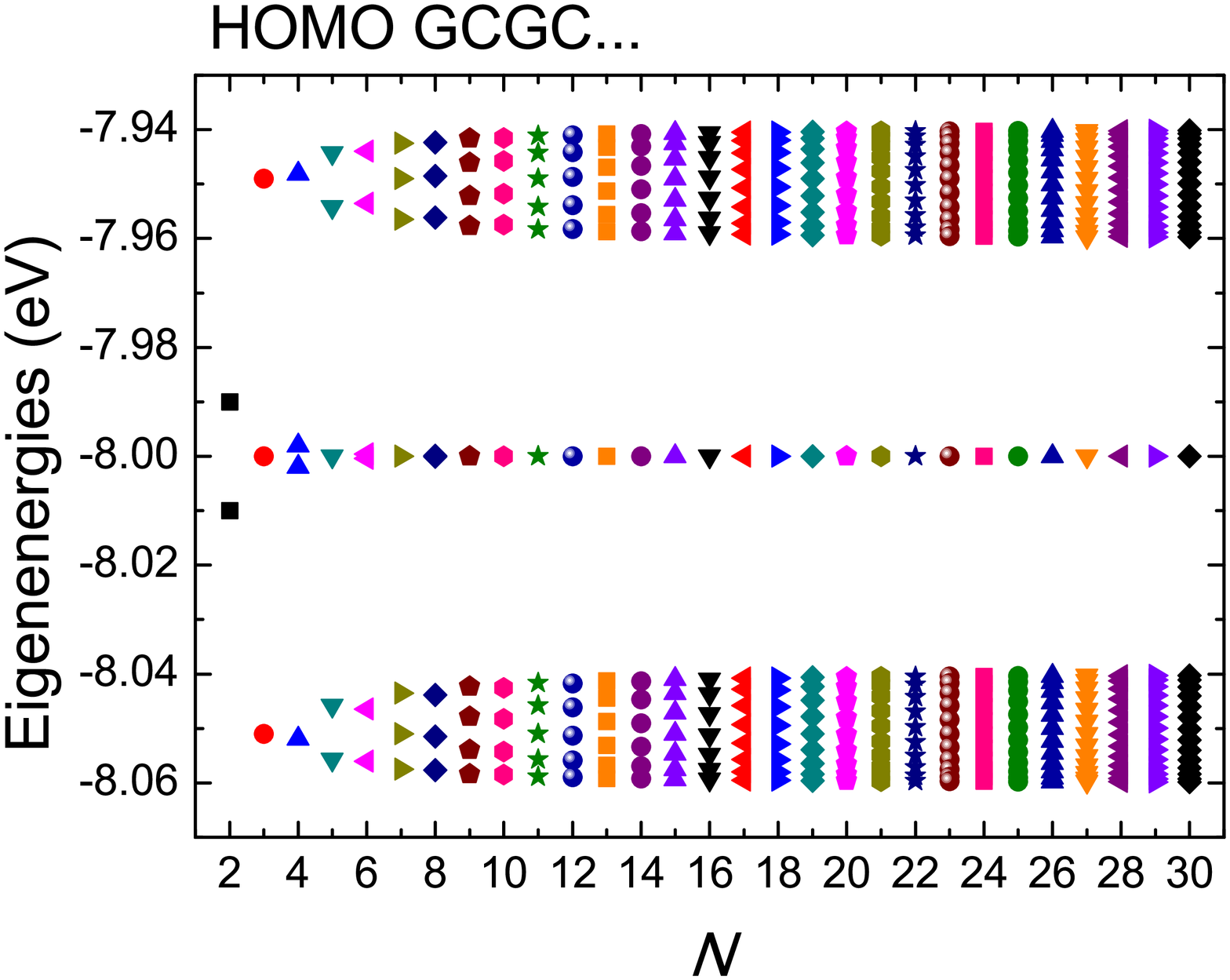}
\includegraphics[width=6cm]{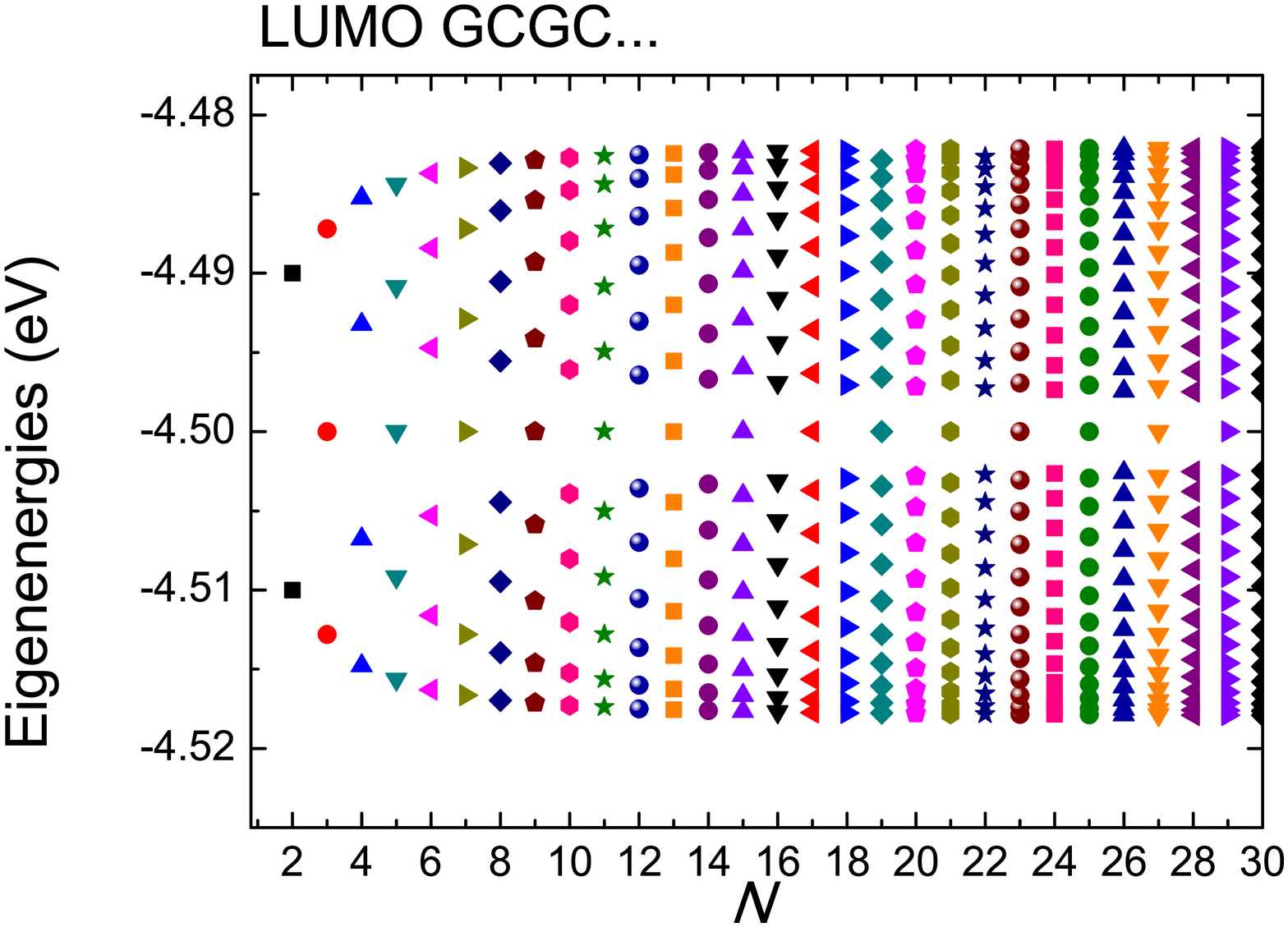}
\includegraphics[width=6cm]{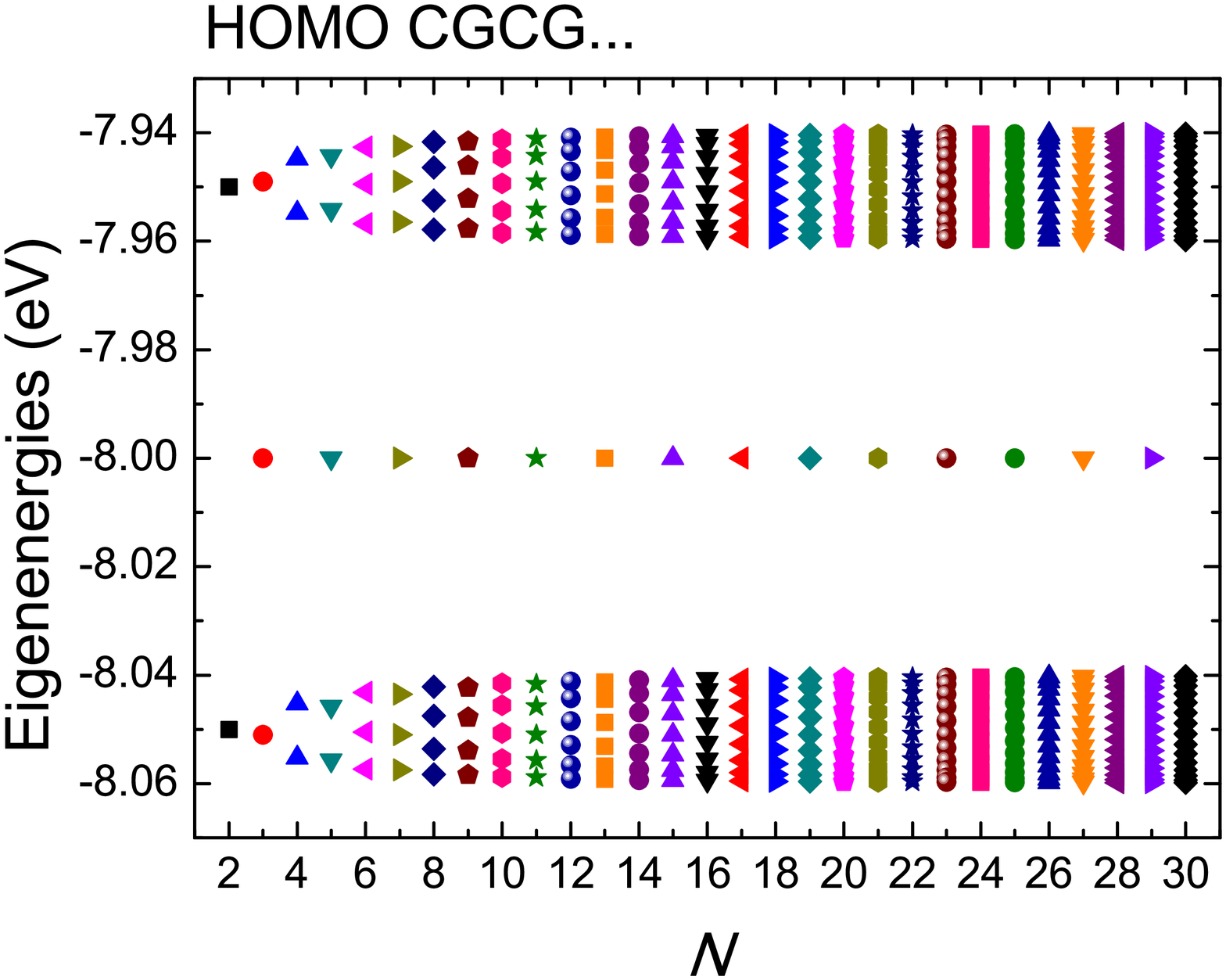}
\includegraphics[width=6cm]{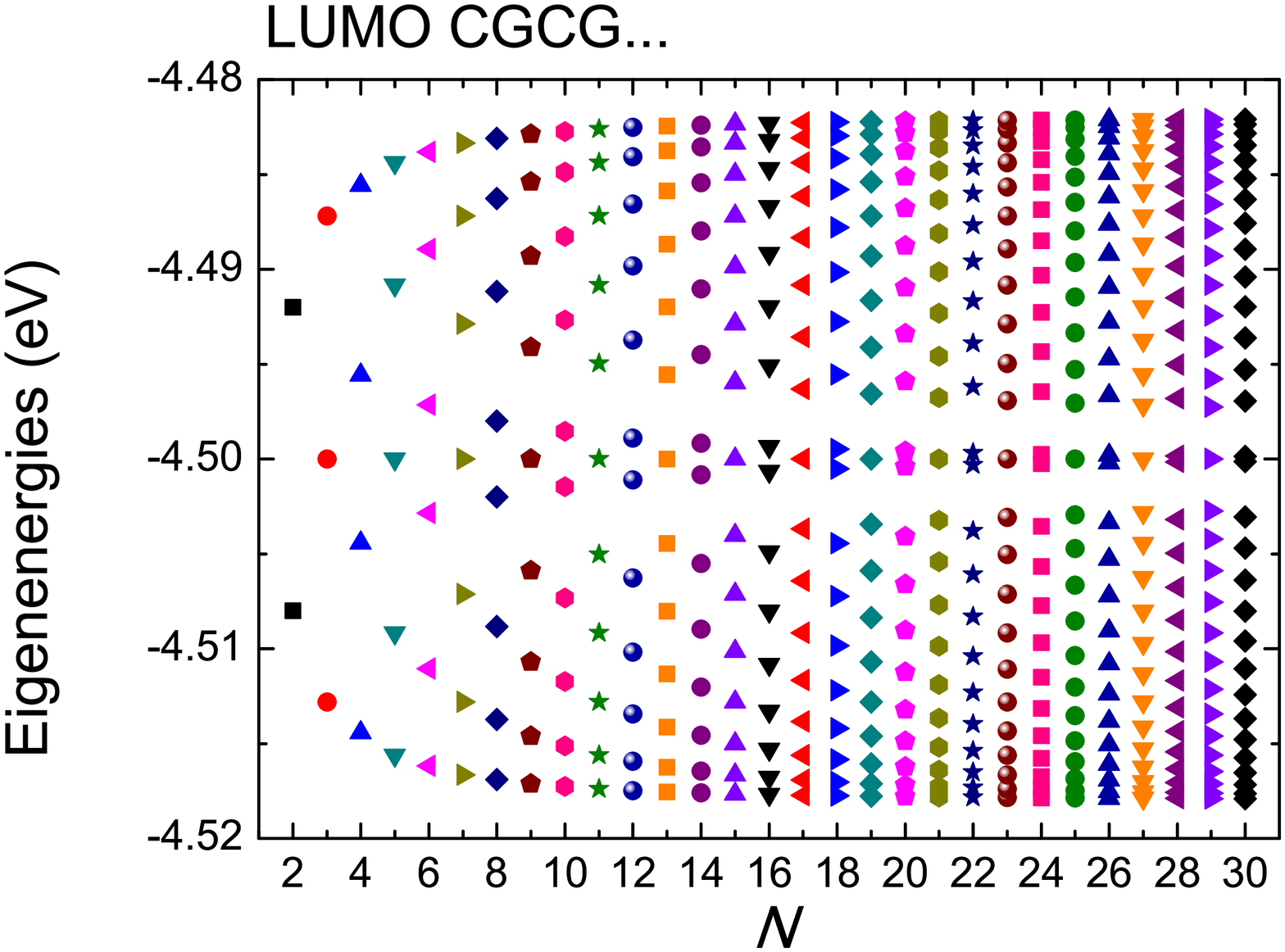}
\includegraphics[width=6cm]{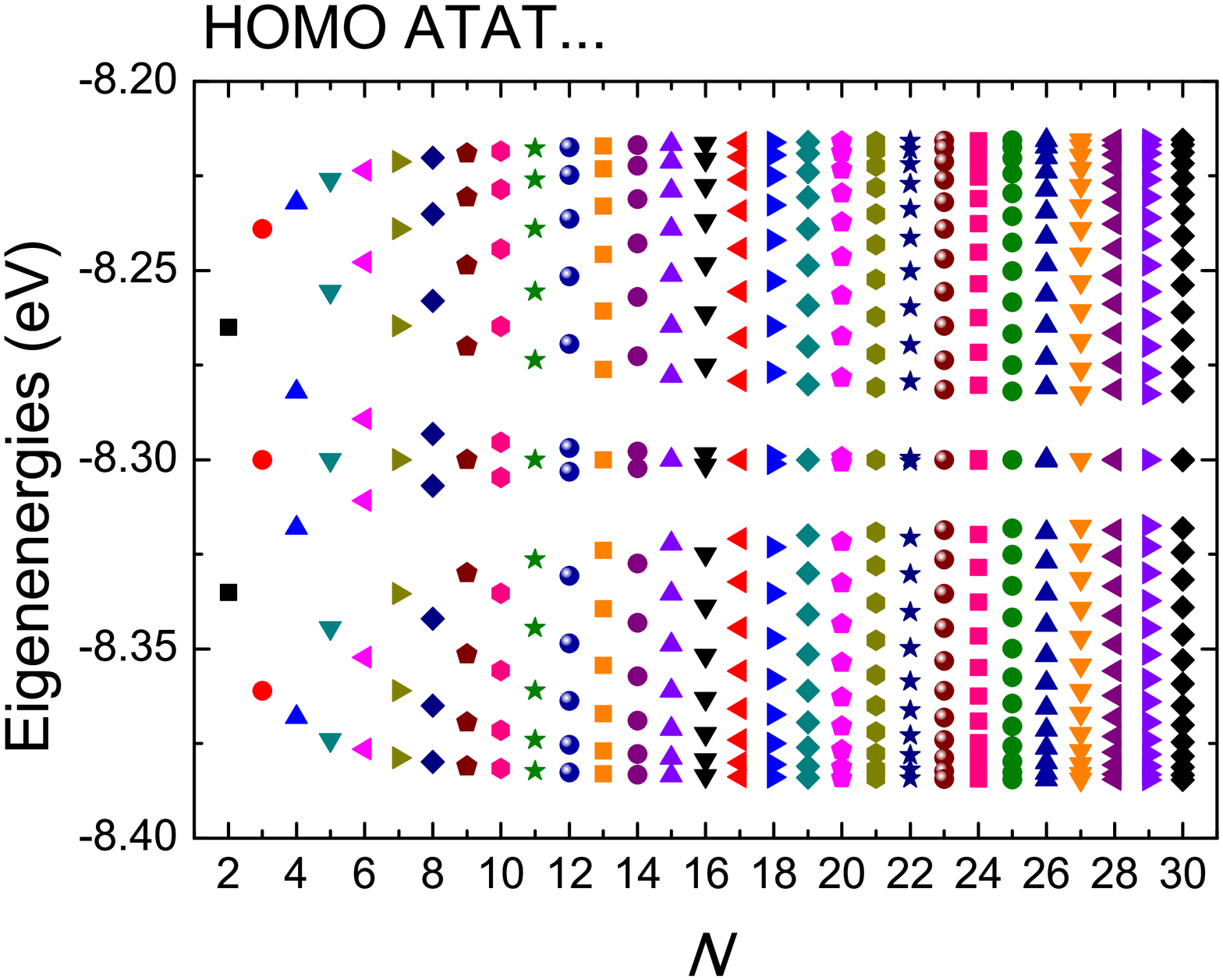}
\includegraphics[width=6cm]{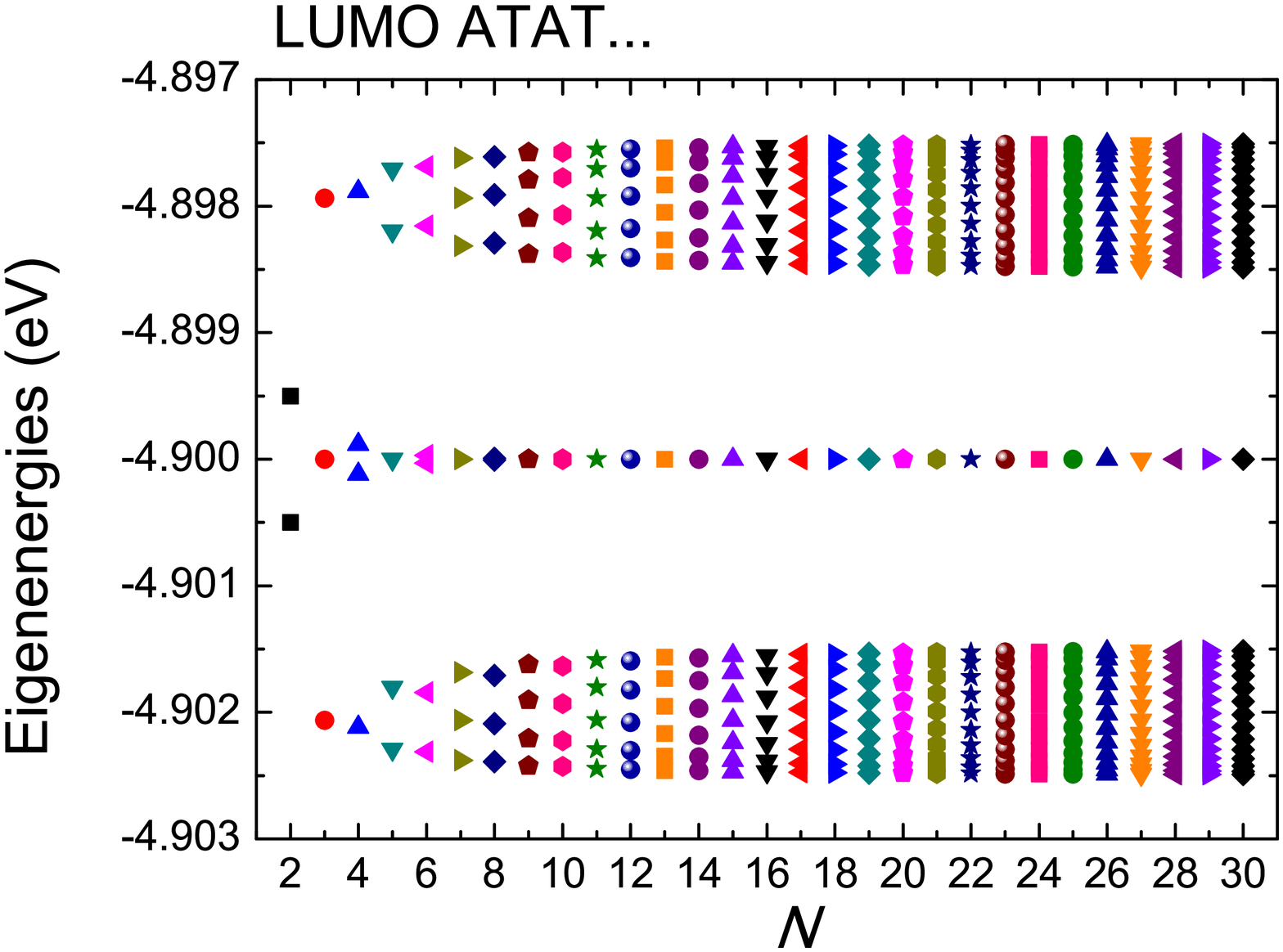}
\includegraphics[width=6cm]{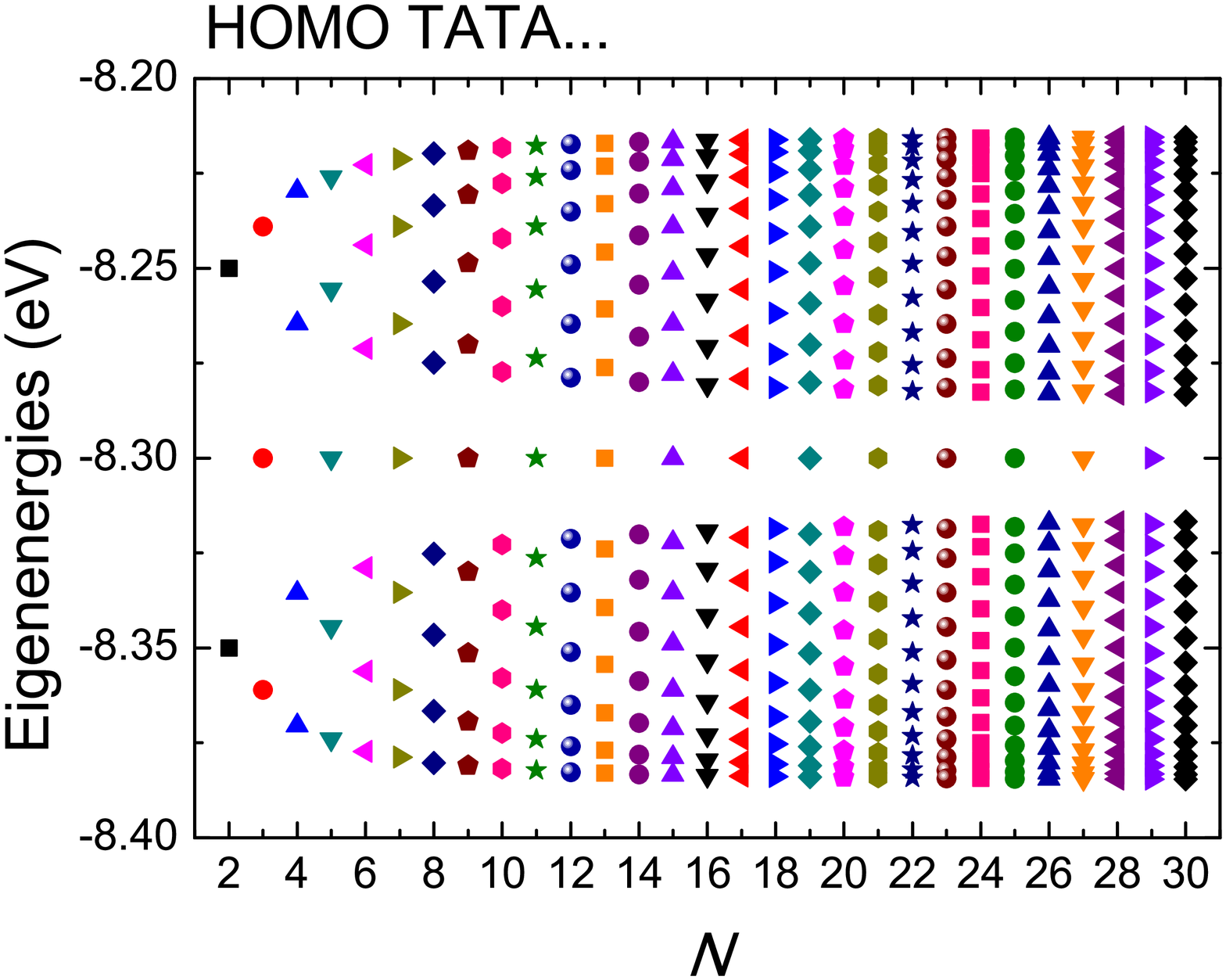}
\includegraphics[width=6cm]{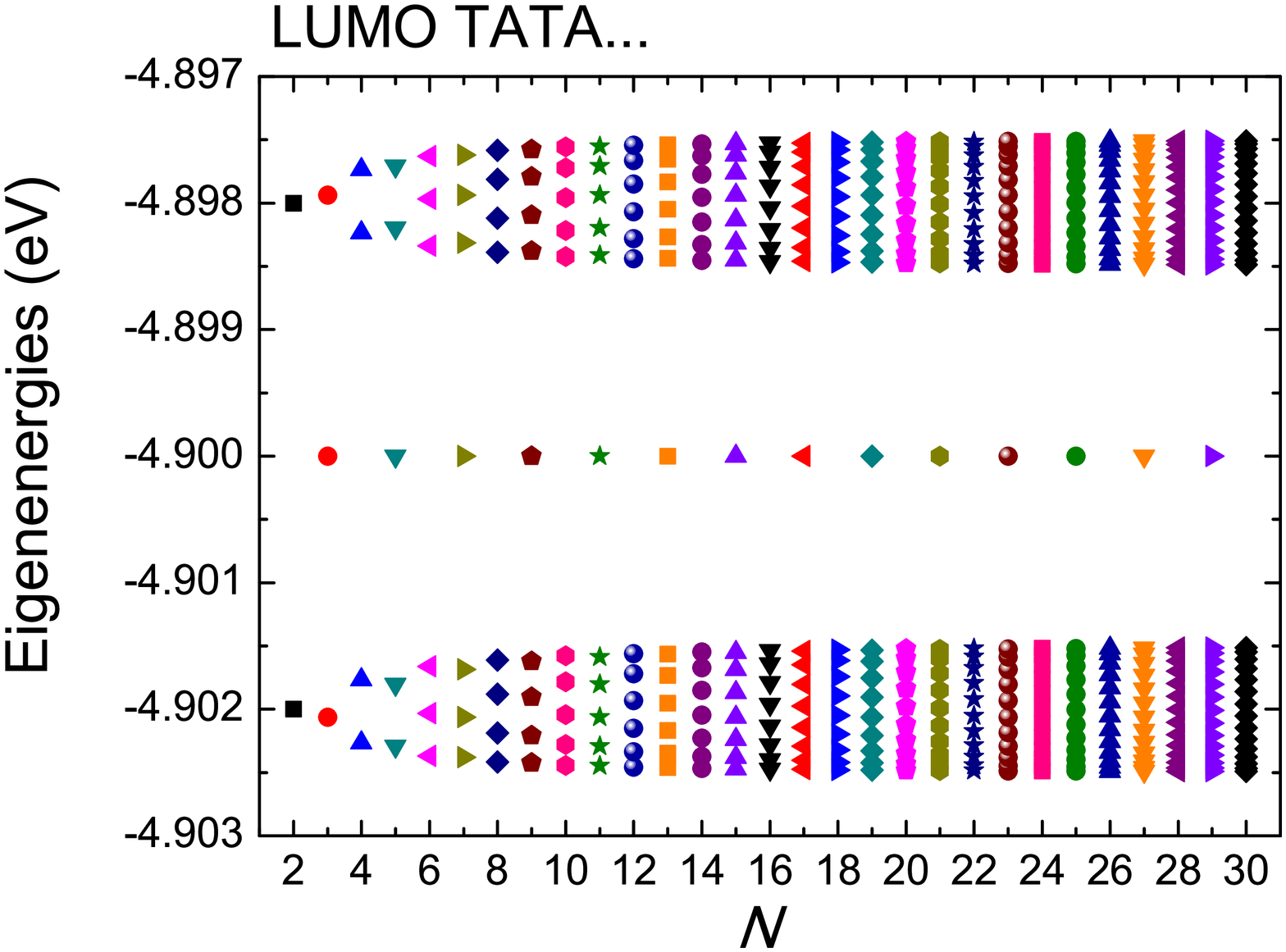}
\caption{\label{fig:eigenspectrum-tb} Eigenspectra of type $\beta'$ polymers.}
\end{figure*}

For odd $N$, \textrm{A} has the same number of $t$ and $t'$. Hence, for odd $N$,
its eigenvalues and eigenvectors have some noteworthy properties:
For odd $N$, for the same set of parameters $\{ E,t,t'\}$,
the set of eigenvalues $\{\lambda_k\}$ remains the same if we interchange the sequence of base-pairs,
i.e. $\{\lambda_k\}(\textrm{XY...})$=$\{\lambda_k\}(\textrm{YX...})$;
e.g. $\{\lambda_k\}$ is the same for HOMO GCGCGCG and HOMO CGCGCGC.
Moreover,  for odd $N$, for the same set of parameters $\{ E,t,t' \}$, the eigenvectors have the properties
$|v_{\mu k}(\textrm{XY...})| = |v_{(N-\mu+1) k}(\textrm{YX...})|$ and
$|v_{\mu k}(\textrm{XY...})| = |v_{\mu (N-k+1)}(\textrm{XY...})|$.
For odd $N$, the eigenvalues can be written~\cite{Kouachi:2006} as
\begin{equation}\label{eigenvaluestypebKouachi}
\lambda_k = \left\{
\begin{array}{ll}
E + \sqrt{t^2 + {t'}^2 + 2  t t' \cos \theta_k }, & k = 1,   \dots, m \\
E - \sqrt{t^2 + {t'}^2 + 2  t t' \cos \theta_k }, & k = m+1, \dots, 2m\\
E, &k = N \end{array} \right.
\end{equation}
\begin{equation}
\theta_k = \left\{
\begin{array}{ll}
\frac{2k\pi}{N+1}     & k = 1,   \dots, m \\
\frac{2(k-m)\pi}{N+1} & k = m+1, \dots, 2m
\end{array} \right.,
\end{equation}
which are equivalent with the resulting eigenvalues in Ref.~\cite{Gover:1994}
\begin{equation}\label{eigenvaluestypeb}
\{ \lambda_k \} = \left\{
\begin{array}{l}
E, \quad \textrm{and} \\
E  \pm \sqrt{t^2 + {t'}^2 + 2 t t' \cos \left( \frac{r\pi}{m+1} \right)}
\end{array} \right.
\end{equation}
where $m=\frac{N-1}{2}$ and $r = 1, 2, \dots m$.
Analytical expressions for the eigenvectors, for odd $N$, can be found in Ref.~\cite{Kouachi:2006}.
For odd $N$, it is worth noting that the eigenvectors $v_{\mu k}$ depend on $t$ and $t'$, hence, for any $k$,
the probability to find the carrier at a particular monomer $\mu$, $|v_{\mu k}|^2$
also depends on $t$ and $t'$. Hence, in contrast to type $\alpha'$ polymers,
now we have \textbf{partial eigenspectrum dependence} of the probabilities i.e. dependence on the hopping parameters but not on the on-site energy.
For odd $N$, $|v_{\mu k}|^2$ are palindromes only for even $\mu$; this property is conserved in the time-dependent case (cf. subsubsection~\ref{subsubsec:meanprob-beta}).
For even $N$, the situation is more complicated~\cite{Gover:1994}.
We have not encountered an analytical solution, in the literature, yet. For even $N$, \textrm{A} does not have the same number of $t$ and $t'$.
For even $N$, $|v_{\mu k}|^2$ are palindromes for all $\mu$; this property is conserved in the time-dependent case (cf. subsubsection~\ref{subsubsec:meanprob-beta}).
\textbf{Hence, we have palindromicity for $N$ even, but for $N$ odd only partial palindromicity.}
The eigenspectra of type $\beta'$ polymers for odd and even $N$ are shown in Fig.~\ref{fig:eigenspectrum-tb}.

\subsubsection{\label{eigen:typec} type $\gamma'$} 
\begin{figure*}[h!]
\includegraphics[width=6cm]{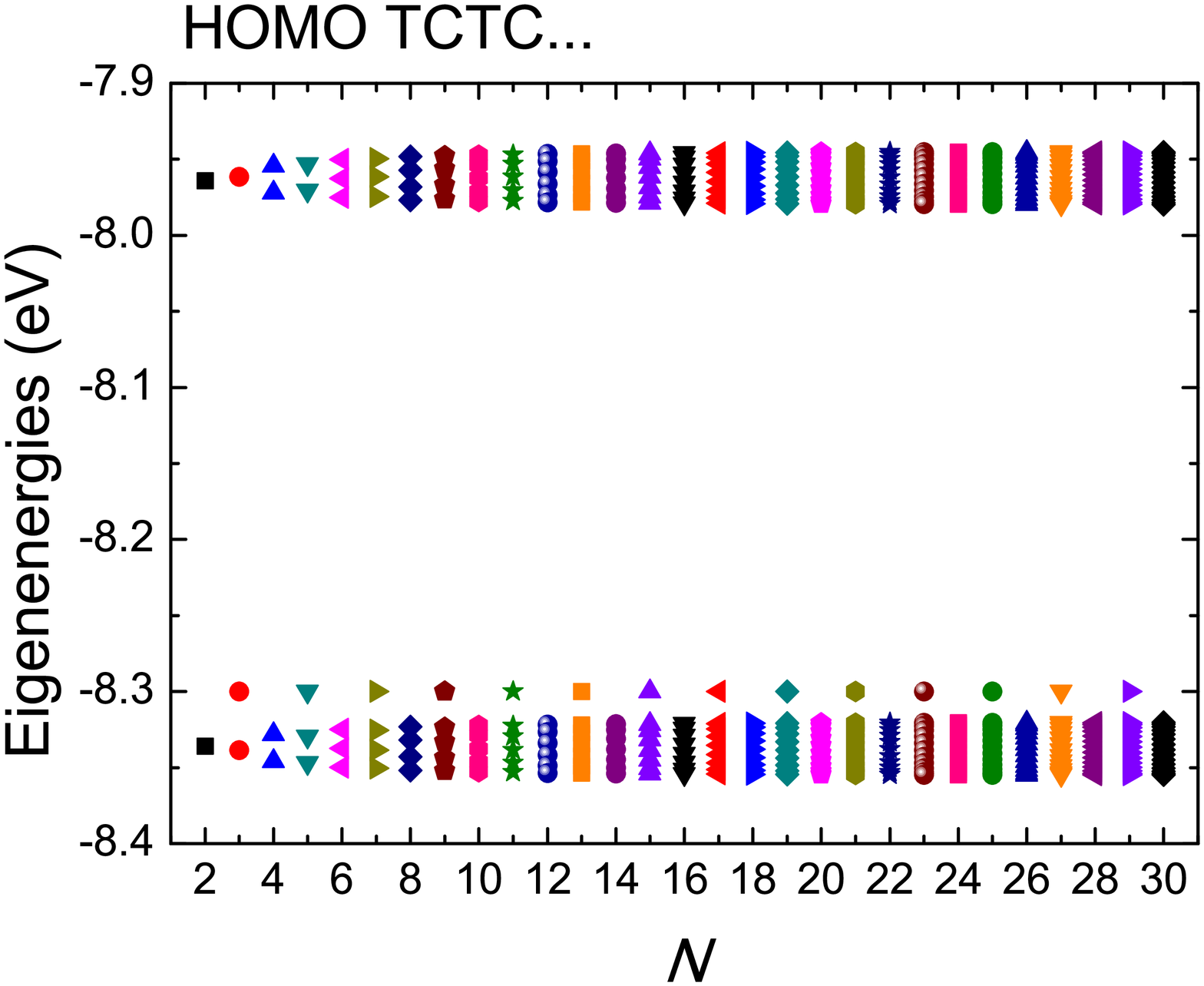}
\includegraphics[width=6cm]{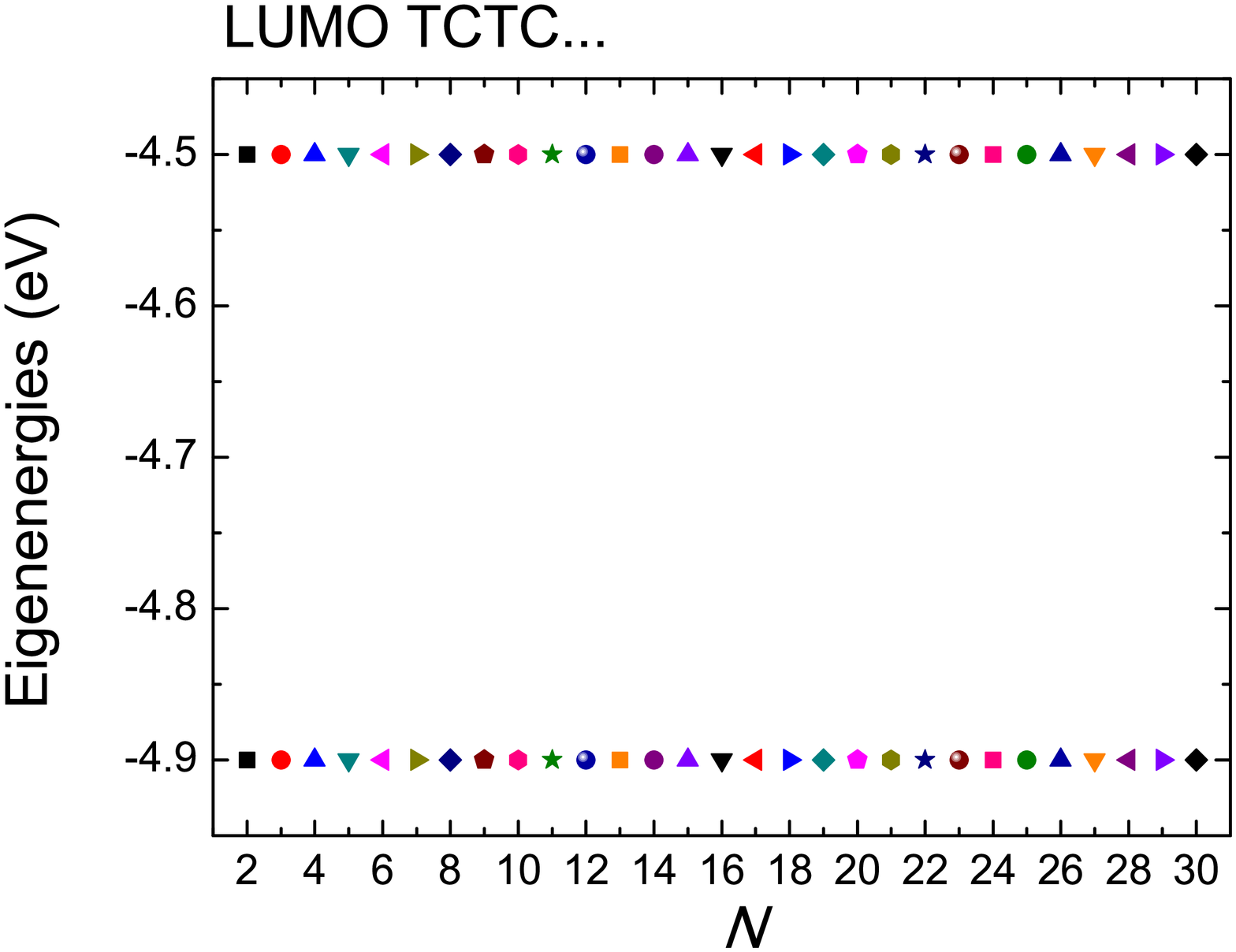}
\includegraphics[width=6cm]{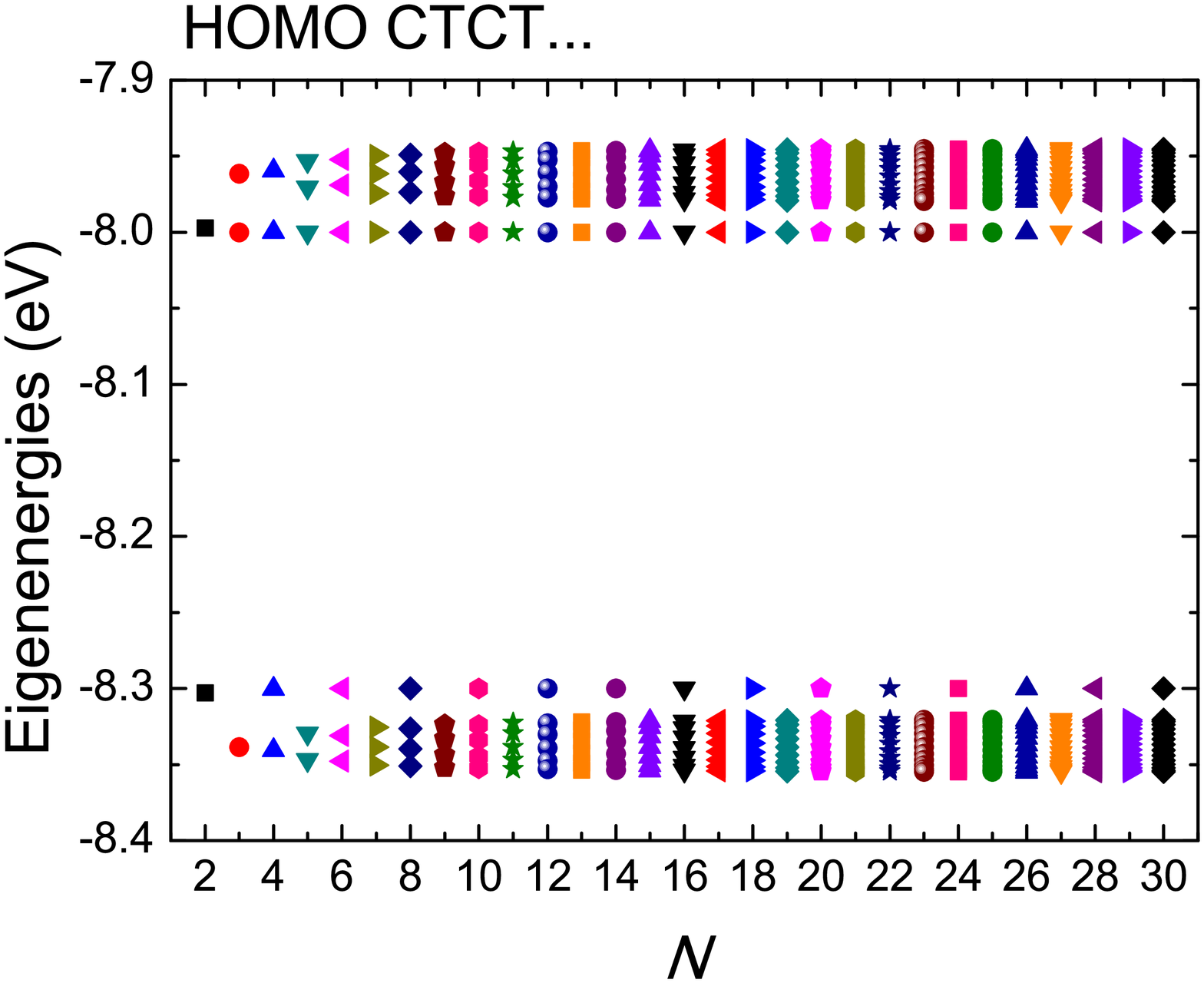}
\includegraphics[width=6cm]{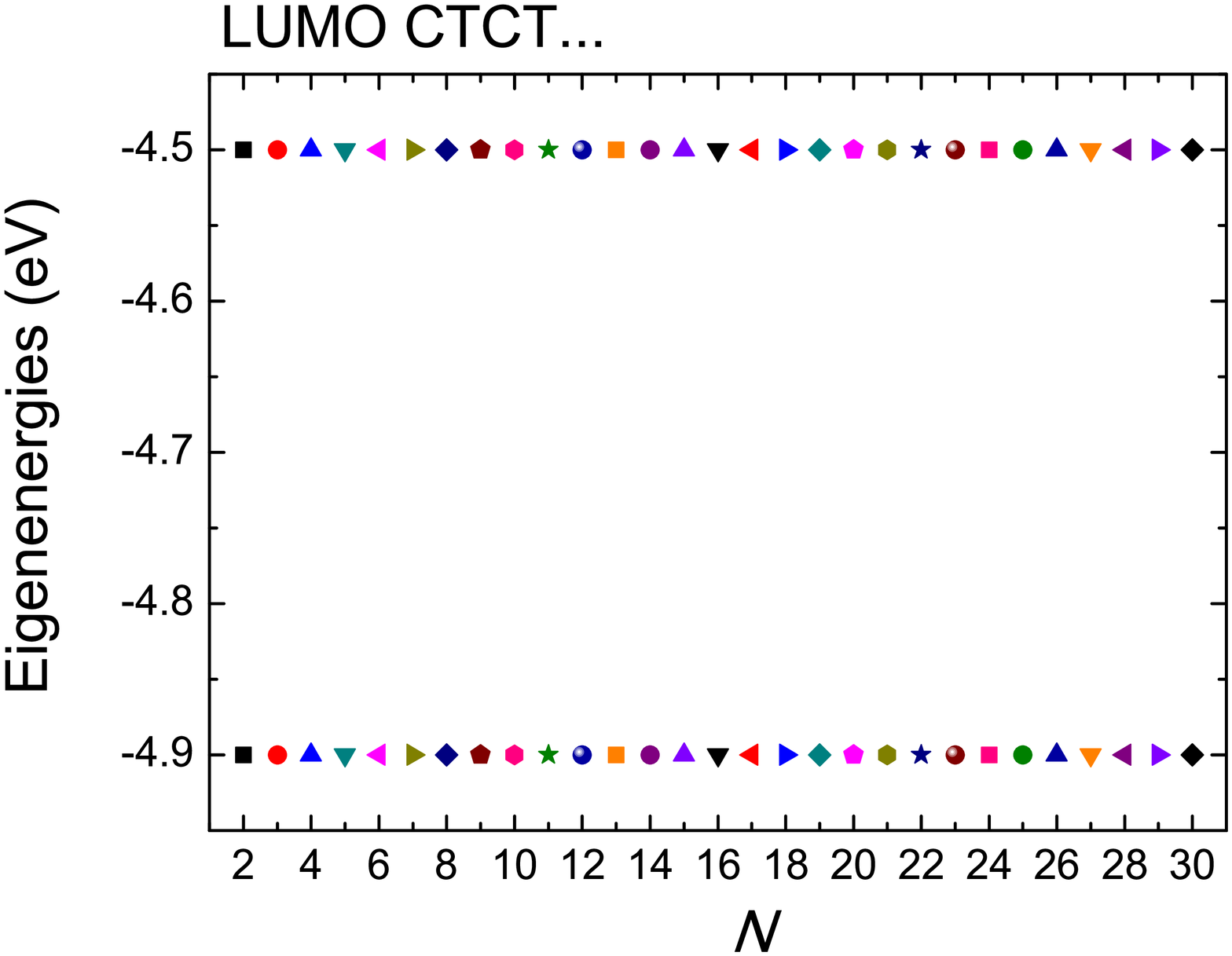}
\includegraphics[width=6cm]{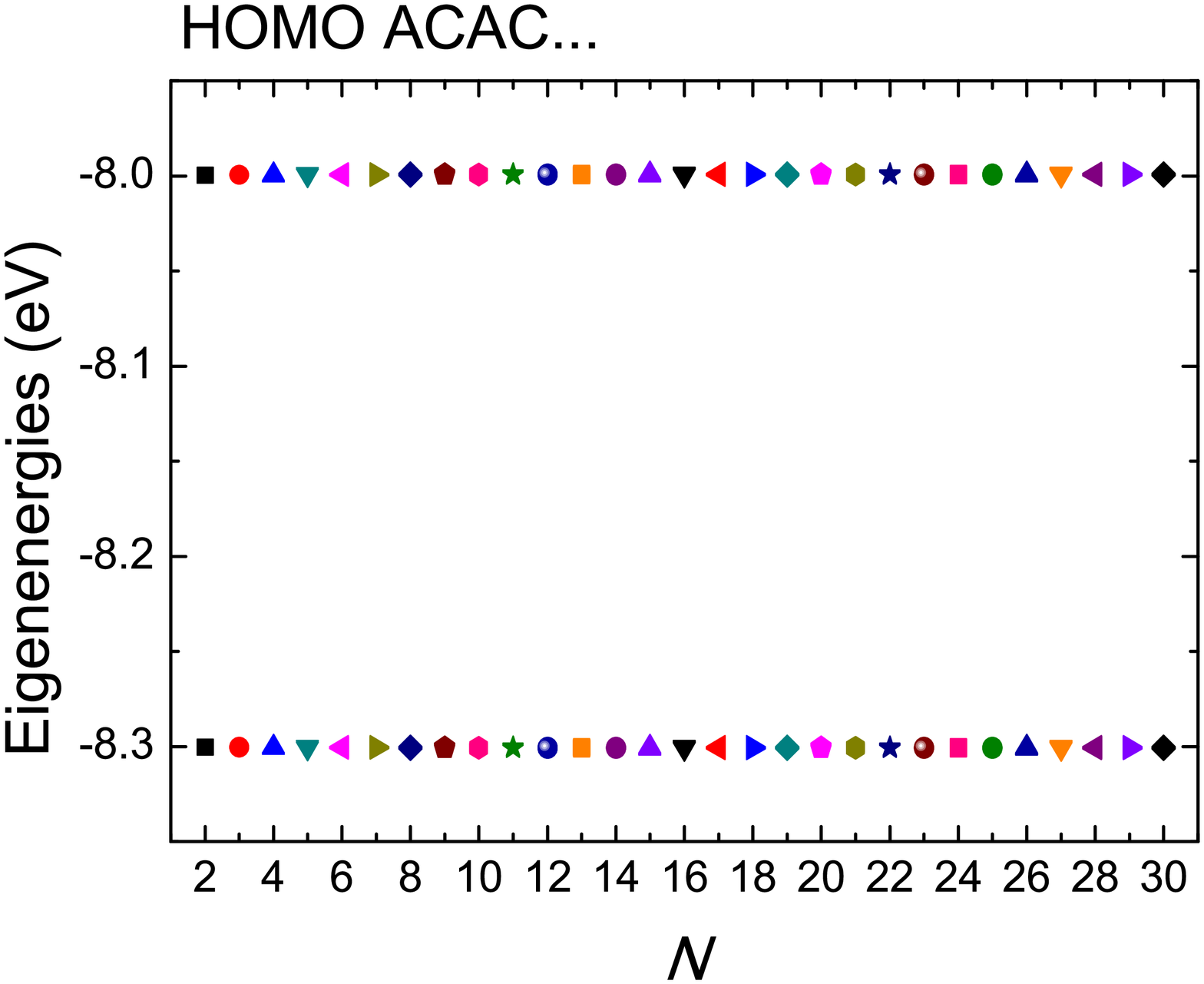}
\includegraphics[width=6cm]{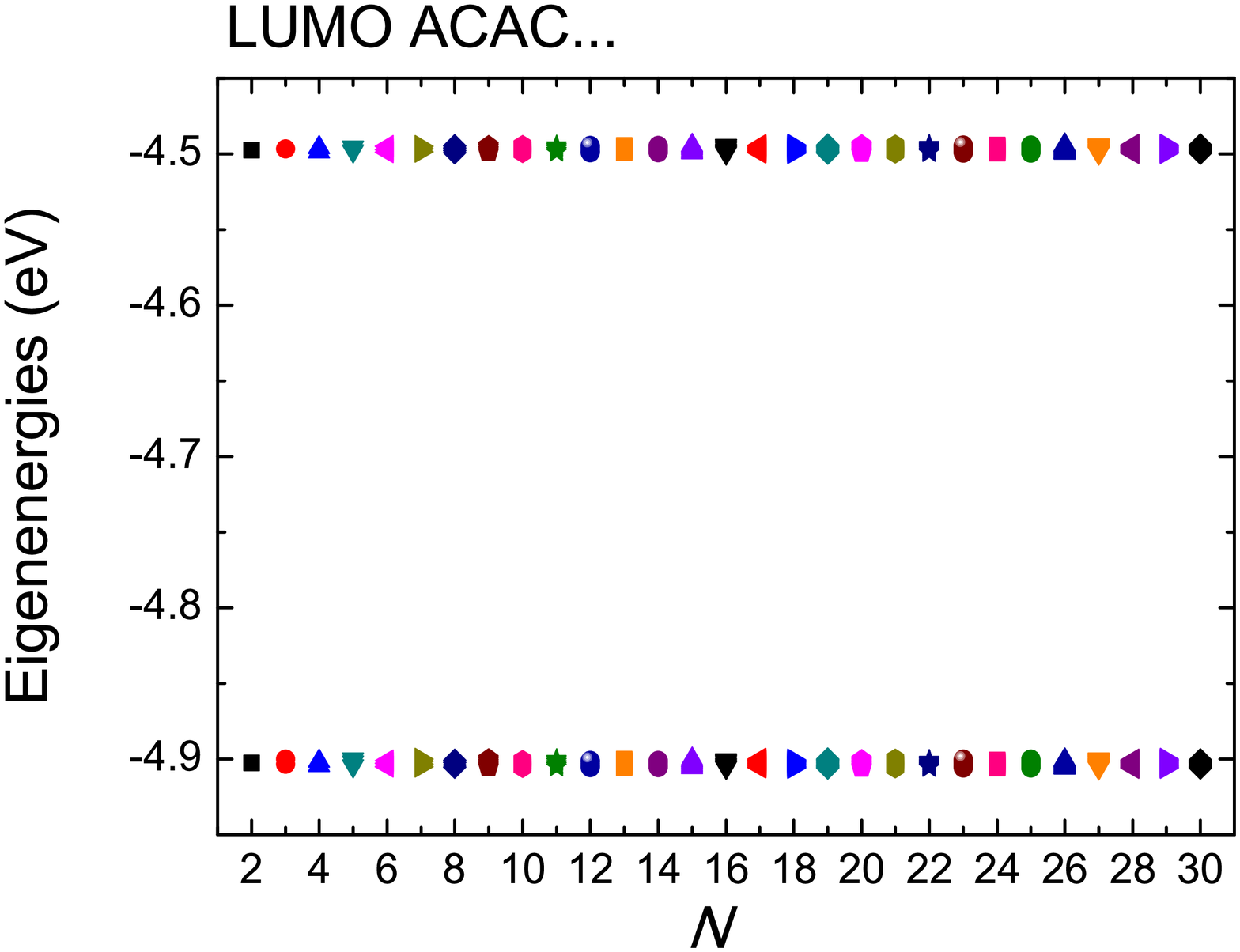}
\includegraphics[width=6cm]{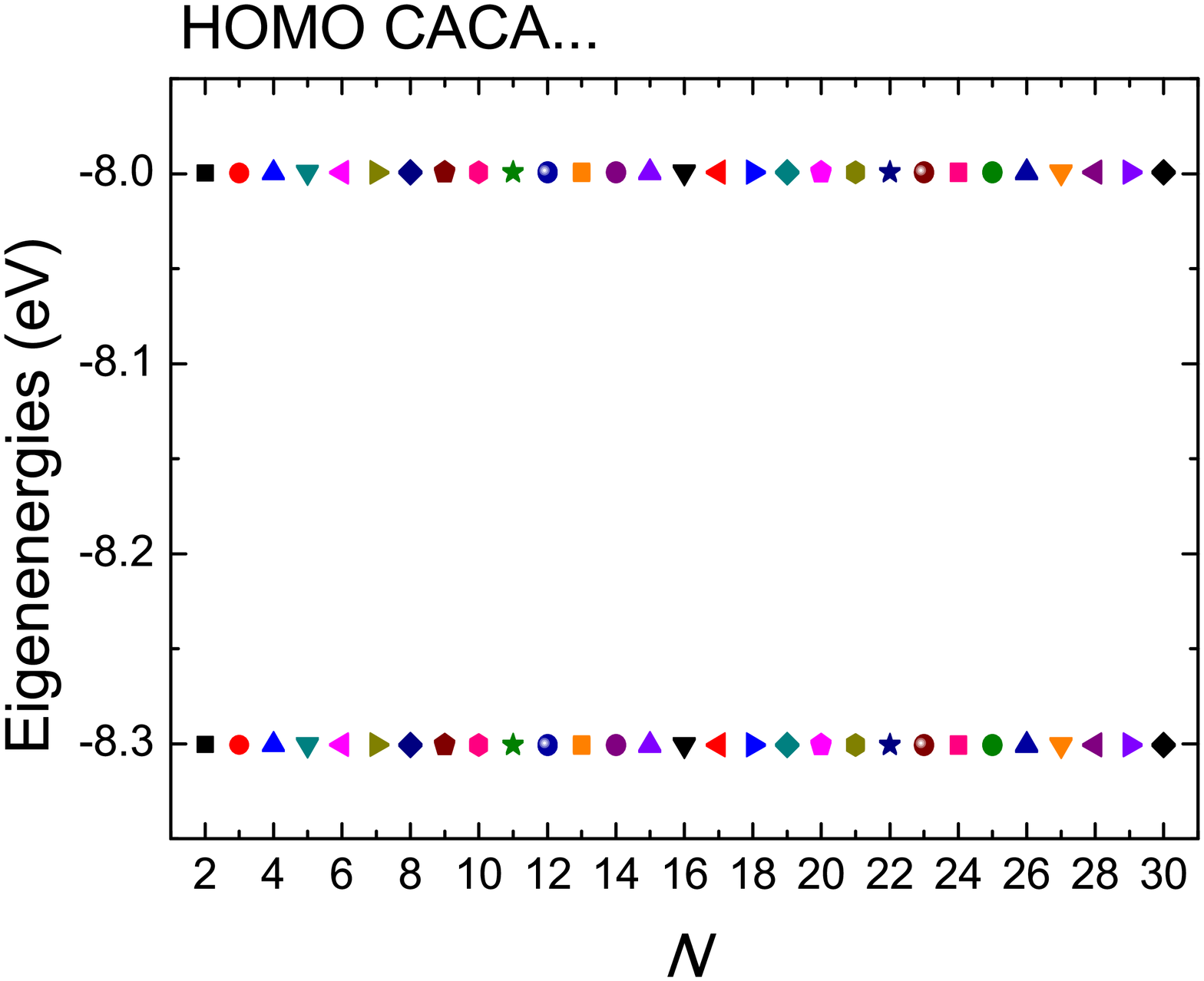}
\includegraphics[width=6cm]{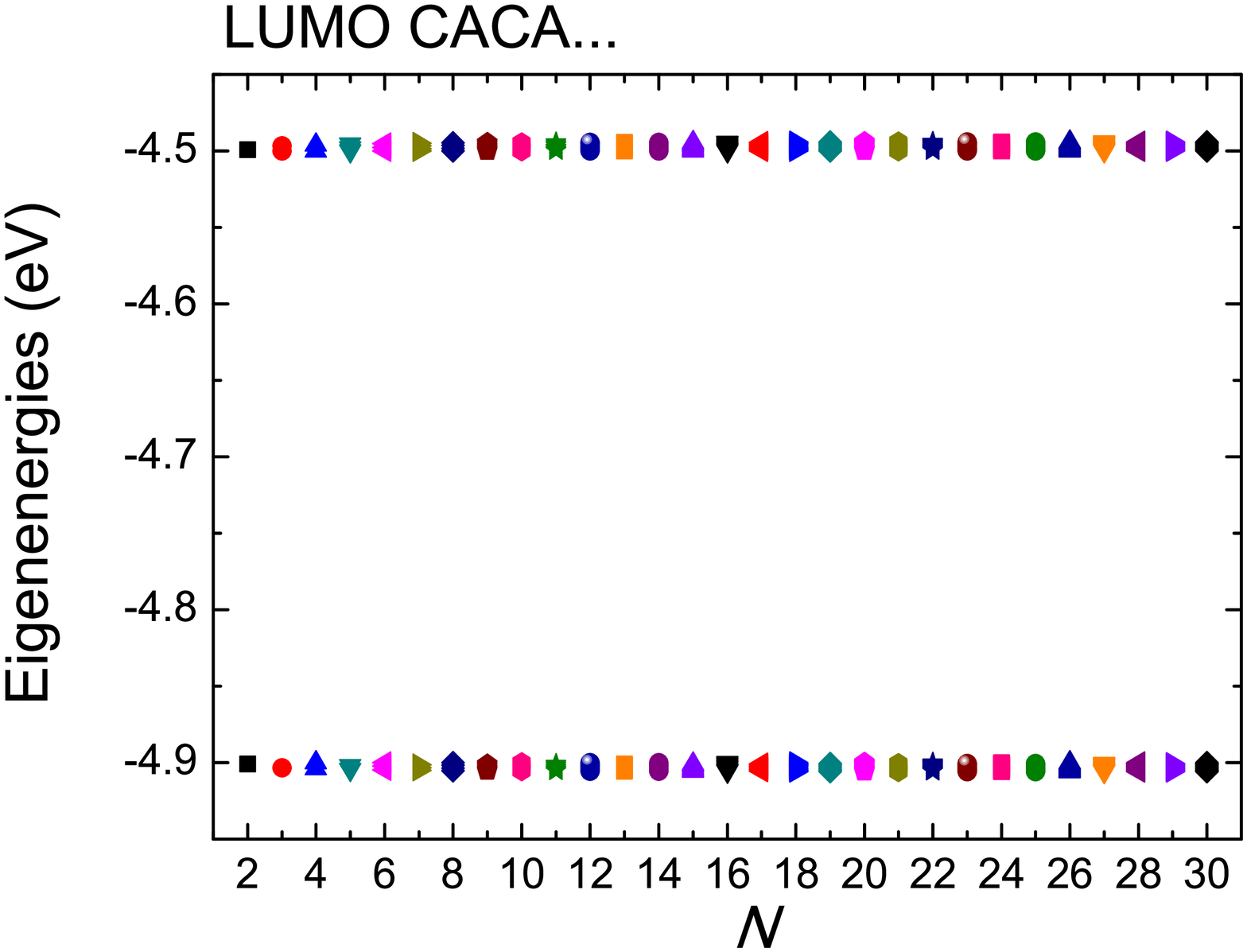}
\caption{\label{fig:eigenspectrum-tc} Eigenspectra of type $\gamma'$ polymers.}
\end{figure*}

For type $\gamma'$ polymers, the matrix $\textrm{A}$ is
\begin{equation}\label{Atypec}
\textrm{A} = \left[
\begin{array}{ccccc}
E^{o} & t     & 0     & 0   & \cdots  \\
t     & E^{e} & t'    & 0   & \cdots  \\
0     & t'    & E^{o} & t   & \cdots  \\
\vdots    & \vdots    & \vdots    & \vdots & \vdots  \end{array} \right]
\end{equation}
For odd $N$, \textrm{A} has the same number of $t$ and $t'$.
For odd $N$, the eigenvalues can be written~\cite{Alvarez:2005} as
\begin{equation}\label{eigenvaluestypec}
\{ \lambda_k \} = \left\{
\begin{array}{l}
E^{\textrm{o}}, \quad \textrm{and} \\
\frac{\Sigma}{2}  \pm \sqrt{\left(\frac{\Delta}{2}\right)^2 + t^2 + {t'}^2 + 2 t t' \cos \left( \frac{r \pi}{m+1} \right)}
\end{array} \right.
\end{equation}
where $m=\frac{N-1}{2}$ and $r = 1, 2, \dots m$. This is in accordance with Ref.~\cite{Gover:1994}.
For odd $N$, analytical expressions for the eigenvectors can be found in Ref.\cite{Alvarez:2005}.
It is worth noting that the eigenvectors $v_{\mu k}$ depend on $E^{o}$, $E^{e}$, $t$ and $t'$, hence, for any $k$,
the probability to find the carrier at a particular monomer $\mu$, $|v_{\mu k}|^2$
also depends on $E^{o}$, $E^{e}$, $t$ and $t'$.
Hence, in contrast to type $\alpha'$ polymers now we have \textbf{eigenspectrum dependence} of the probabilities.
For even $N$, the situation is more complicated~\cite{Gover:1994}.
We have not encountered an analytical solution, in the literature, yet. For even $N$, \textrm{A} does not have the same number of $t$ and $t'$.
The eigenspectra of type $\gamma'$ polymers for odd and even $N$ are shown in Fig.~\ref{fig:eigenspectrum-tc}.

\subsection{\label{subsec:meanprob} Mean --over time-- Probabilities (time-dependent problem)} 
The behavior of the mean --over time-- probabilities to find the carrier at base-pair $\mu$, $\langle |A_{\mu}(t)|^2 \rangle$, is different in types $\alpha'$, $\beta'$, $\gamma'$.

\subsubsection{\label{subsubsec:meanprob-alpha} type $\alpha'$ (and type $\alpha'$ {\it cyclic})} 
For type $\alpha'$ polymers, $\langle |A_{\mu}(t)|^2 \rangle$ are {\it palindromes}
(notice that $v_{\mu k}$ for type $\alpha'$ polymers are palindromes, too) and
they do not depend on the on-site energies and the hopping parameters, but only on $N$.
In other words, \textbf{eigenspectrum independence} and \textbf{palindromicity} of the probabilities are reflected here from the stationary case (cf. subsubsection~\ref{eigen:typea}).

\begin{figure*}[h!]
\includegraphics[width=6cm]{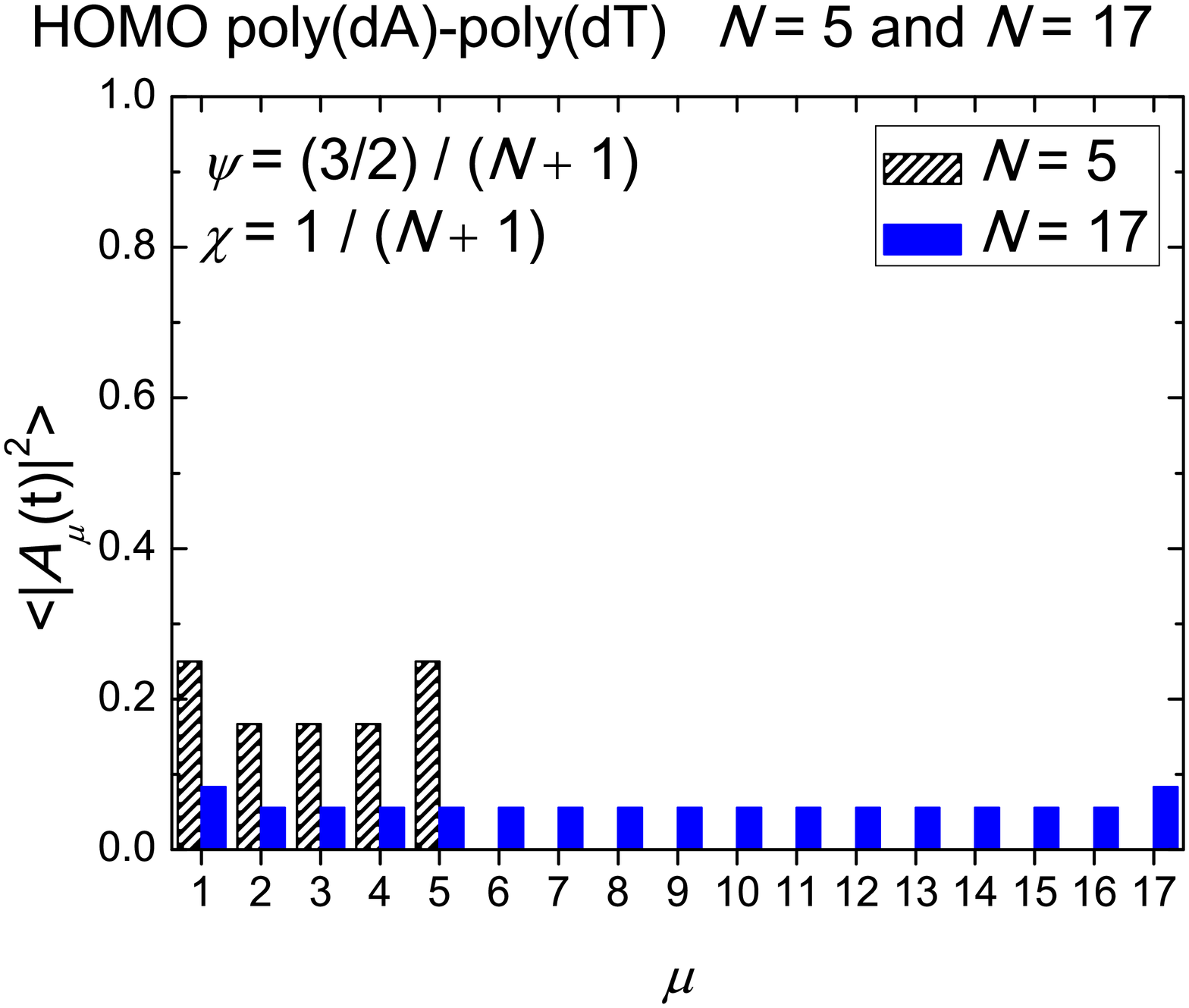}
\includegraphics[width=6cm]{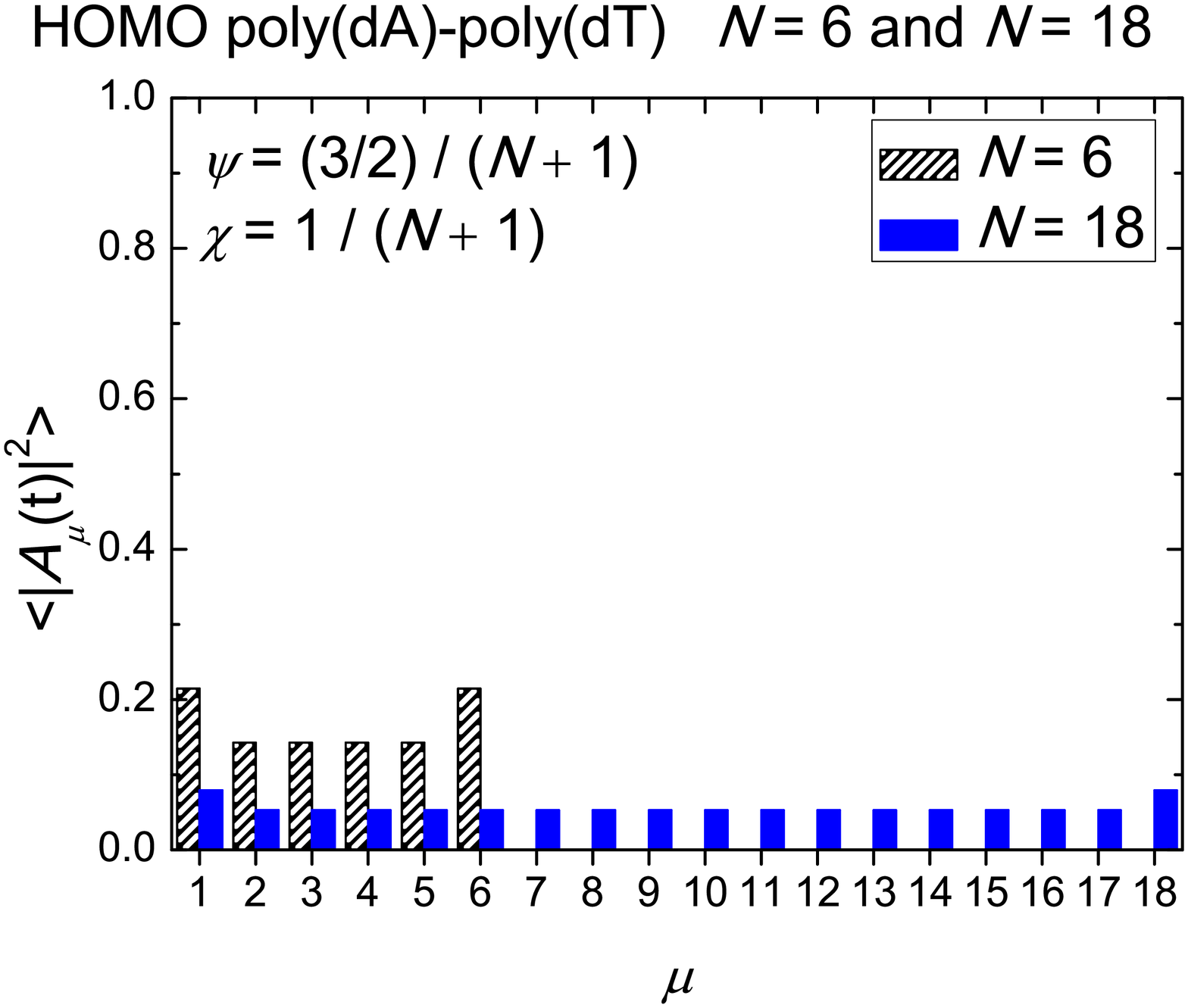}
\includegraphics[width=6cm]{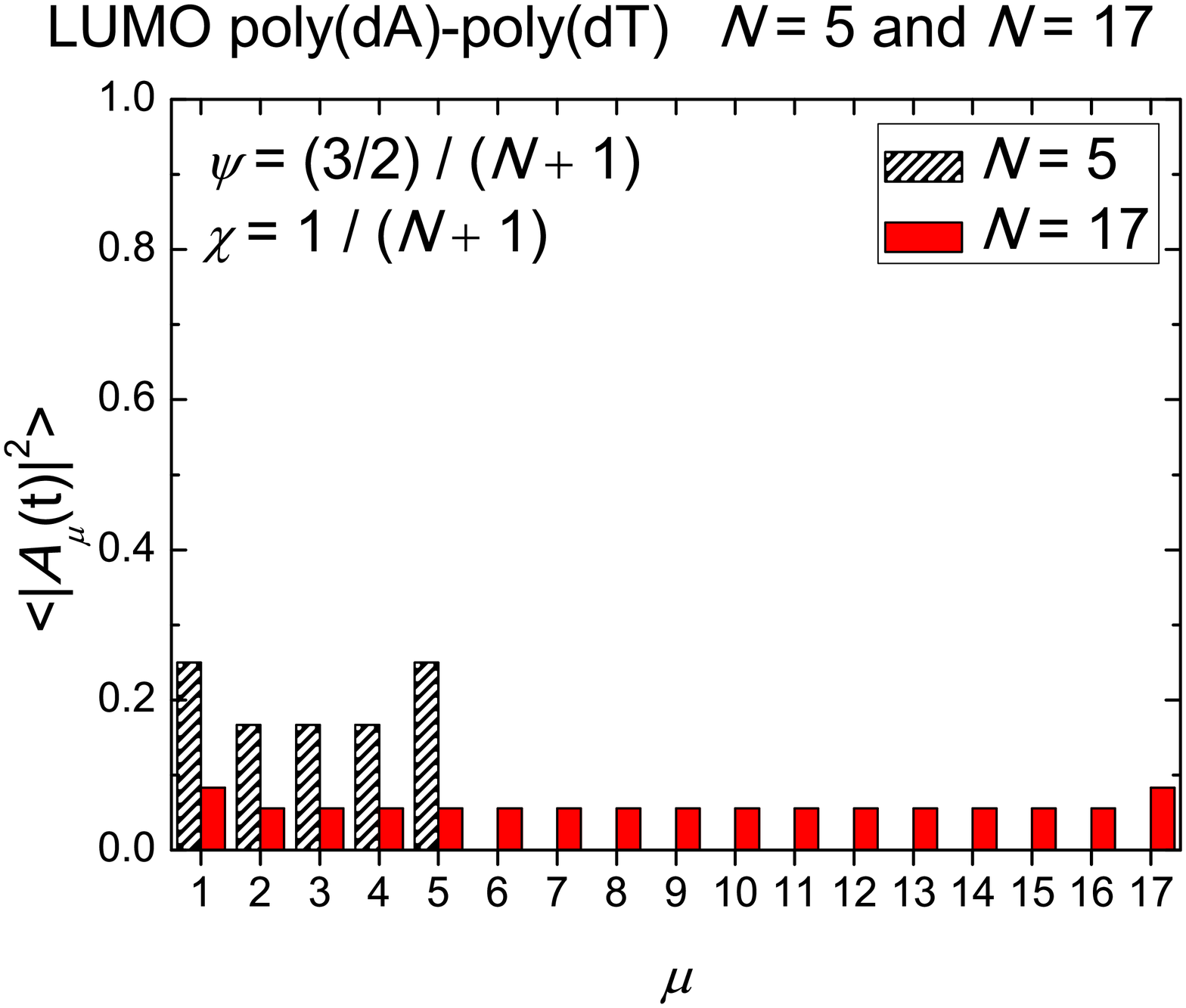}
\includegraphics[width=6cm]{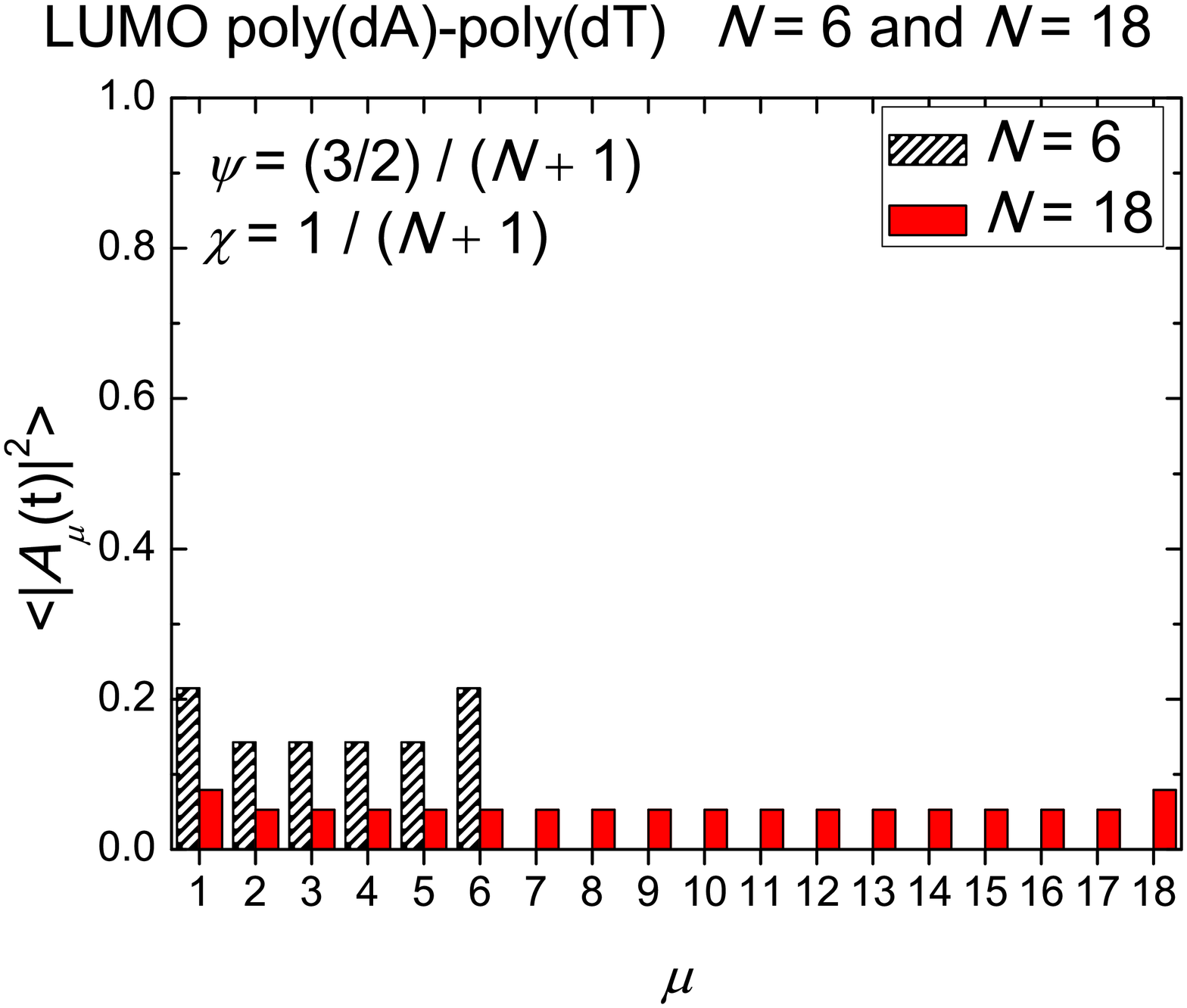}
\caption{\label{fig:N5N6N17N18polydApolydT} Mean --over time-- probabilities $\langle |A_{\mu} (t)|^2 \rangle$ of type $\alpha'$ polymers,
if we initially place the carrier at the 1st monomer.
$\psi$ and $\chi$ denote $\langle |A_{\mu} (t)|^2 \rangle$ at one of the favored monomers and at one of the rest monomers, respectively.
Typically, we illustrate the situation for HOMO and LUMO poly(dA)-poly(dT) but the figures for poly(dG)-poly(dC) are identical.
[Left column]  $N=5$ and $N=17$.
[Right column] $N=6$ and $N=18$.
$\langle |A_{\mu} (t)|^2 \rangle$ follow Eqs.~(\ref{oneedgea})-(\ref{onemiddlea})
i.e. are \textit{palindromes} and depend only on $N$ and not on the on-site energies and the hopping integrals.}
\end{figure*}

If we initially place the carrier at the 1st monomer, then the mean --over time-- probabilities are
\begin{equation}\label{oneedgea}
\langle |A_1 (t)|^2 \rangle = \langle |A_N (t)|^2 \rangle = \frac{3}{2(N+1)}, \forall N \ge 2,
\end{equation}
\begin{equation}\label{onemiddlea}
\langle |A_2 (t)|^2 \rangle = \dots = \langle |A_{N-1} (t)|^2 \rangle = \frac{1}{N+1}, \forall N \ge 3.
\end{equation}
In Fig.~\ref{fig:N5N6N17N18polydApolydT} we illustrate, $\langle |A_{\mu} (t)|^2 \rangle$ for HOMO and LUMO poly(dA)-poly(dT) [the figures are identical for poly(dG)-poly(dC)] if we initially place the carrier at the 1st monomer:
at the left  column for $N=5$ and $N=17$ and at the right column for $N=6$ and $N=18$.
These follow Eqs.~(\ref{oneedgea})-(\ref{onemiddlea}) i.e. depend only on $N$.
In Fig.~\ref{fig:HOMOLUMOpolydApolydT}, in the Appendix~\ref{polydApolydT}, we display other properties of a characteristic polymer of type $\alpha'$ (poly(dA)-poly(dT)) either for hole (left column) or electron (right column) transfer. Again, we initially place the carrier at the 1st monomer.

Generally, for type $\alpha'$ polymers, for initial placement of the carrier at a particular monomer,
we obtain $\frac{1}{2(N+1)}$ additional mean --over time-- probability at the monomer where the initial placement is made {\it and}
at the symmetric relative to the polymer center monomer.
Hence, for $N$ odd, for initial placement at the central monomer, that central monomer obtains $\frac{2}{2(N+1)}$ additional mean --over time-- probability.
In other words, if we call $\psi$ and $\chi$ the mean --over time-- probabilities at the favored and at the rest monomers, respectively, then
$\psi = \chi + \frac{1}{2(N+1)}$ (or $\psi = \chi + \frac{2}{2(N+1)}$ for $N$ odd and initial placement at the central monomer).
Since the sum of all the mean --over time-- probabilities is 1,
we obtain
\begin{equation}\label{psichi-typea-allmostallcase}
\psi=\frac{3}{2(N+1)}, \quad \chi=\frac{1}{N+1},
\end{equation}
except for $N$ odd and initial placement at the central monomer in which case we obtain
$\psi=\frac{2}{N+1}$, $\chi=\frac{1}{N+1}$.

On the contrary, if we imagine to initially distribute the carrier probability equally among monomers, then we obtain the mean --over time-- probabilities (from edge to center monomers) as $\frac{3}{N(N+1)}, \frac{7}{N(N+1)}, \dots$, while for $N$ odd the mean --over time-- probability at the central monomer is $\frac{2N}{N(N+1)}$.

\begin{figure*}[h!]
\includegraphics[width=6cm]{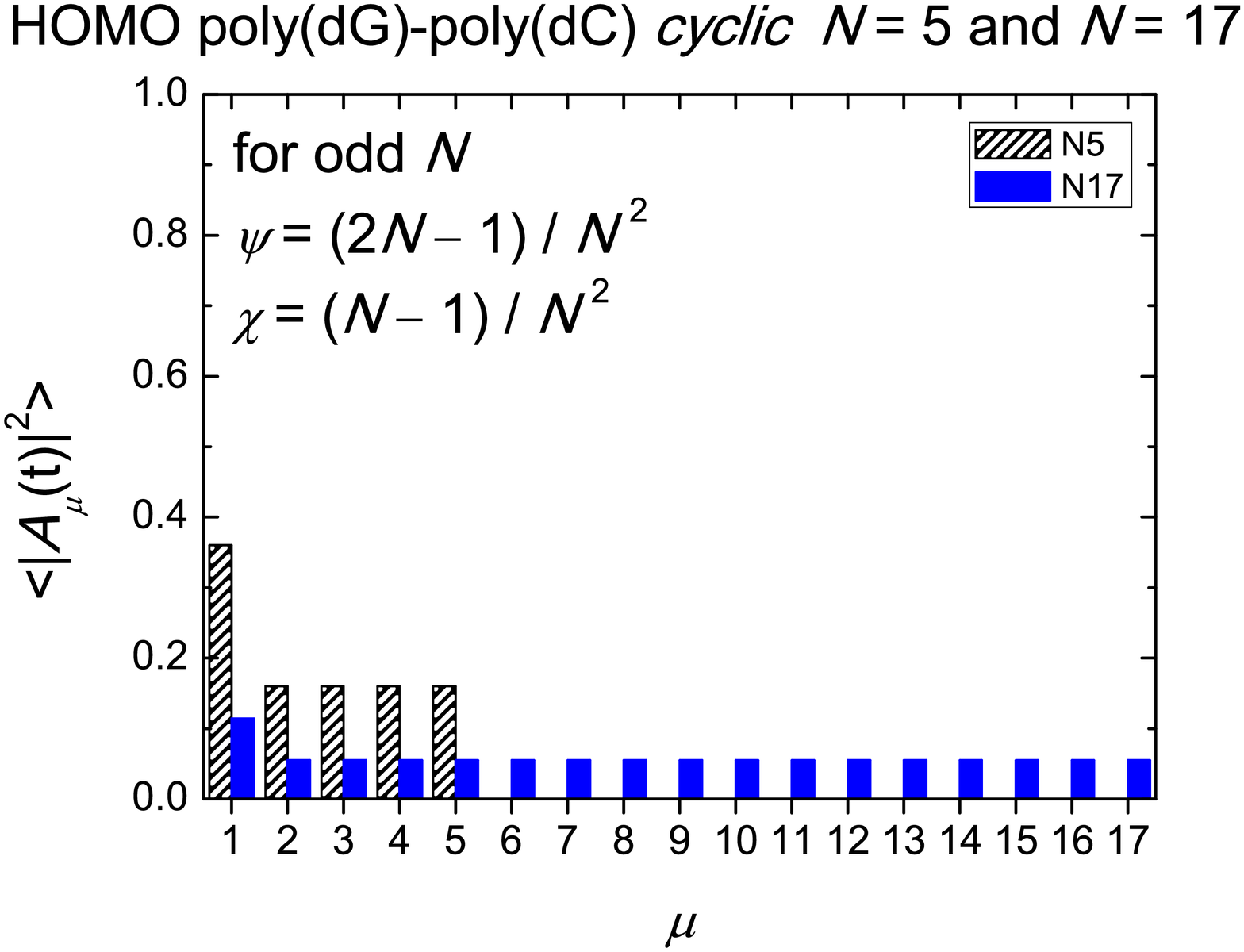}
\includegraphics[width=6cm]{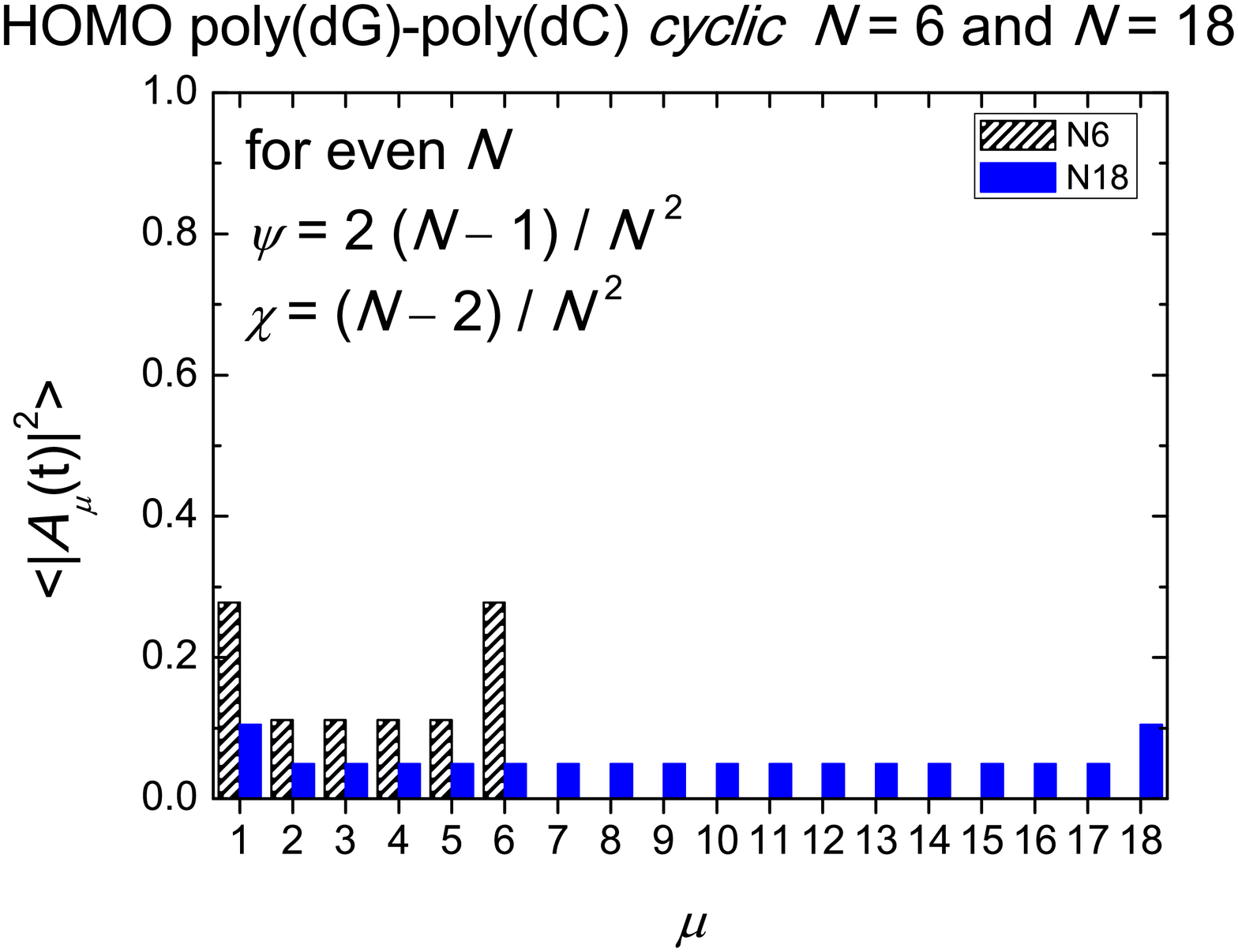}
\includegraphics[width=6cm]{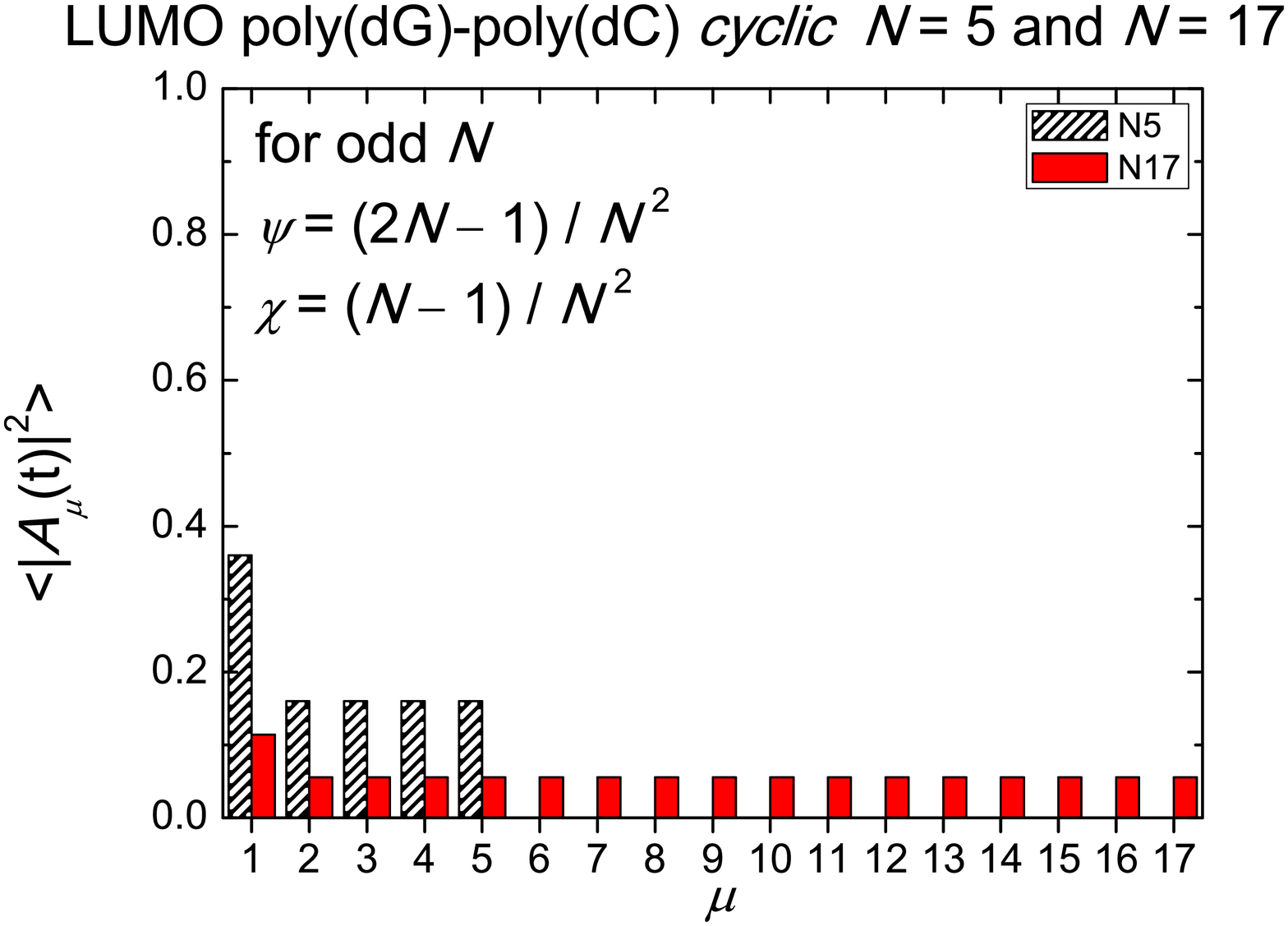}
\includegraphics[width=6cm]{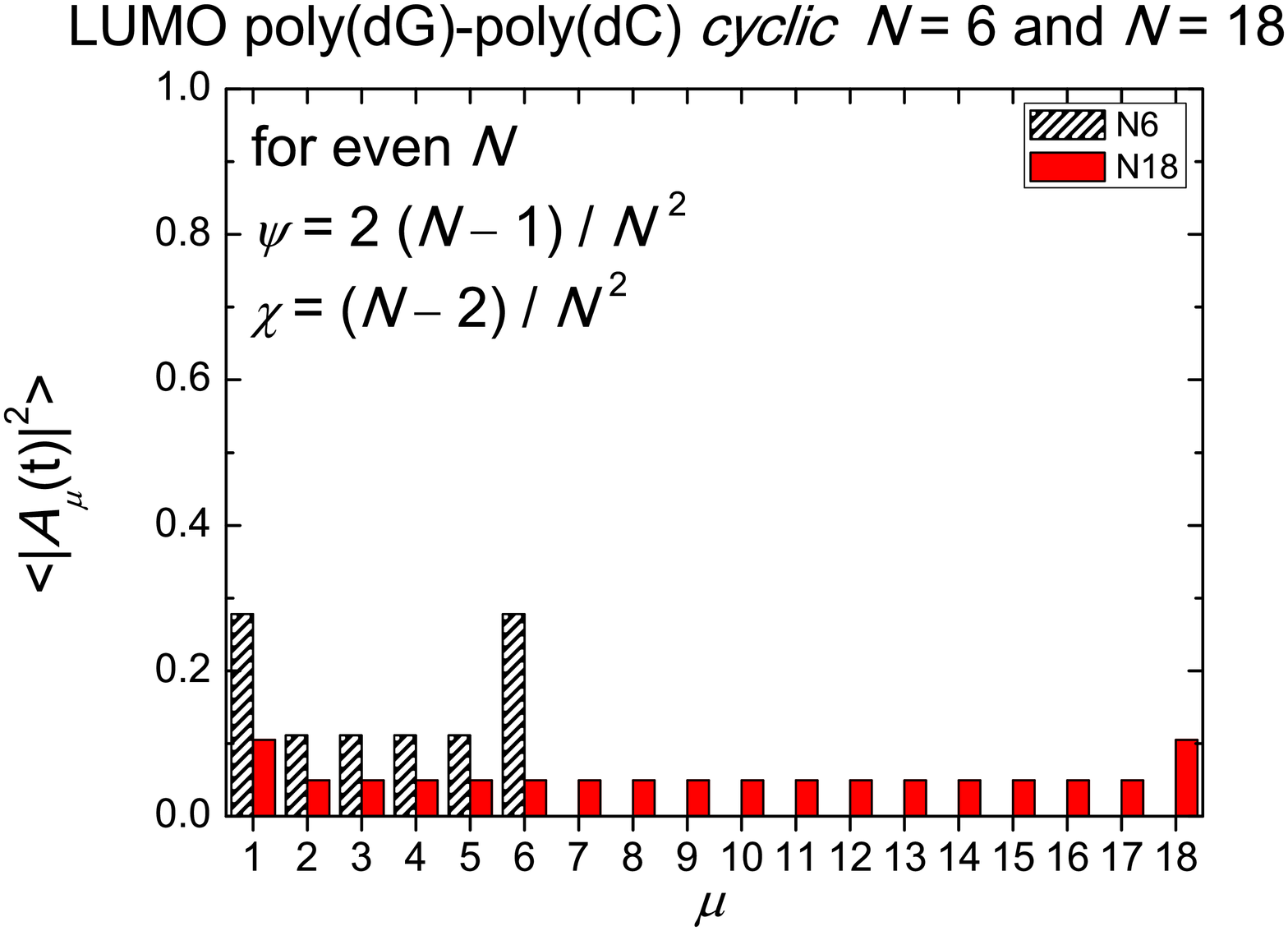}
\caption{\label{fig:typeacyclic}
Mean --over time-- probabilities $\langle |A_{\mu} (t)|^2 \rangle$ of type $\alpha'$ \textit{cyclic} polymers,
if we initially place the carrier at the ``1st monomer''.
$\psi$ and $\chi$ denote $\langle |A_{\mu} (t)|^2 \rangle$ at one of the favored monomers and at one of the rest monomers, respectively.
Typically, we illustrate the situation for HOMO and LUMO poly(dG)-poly(dC) but the figures for poly(dA)-poly(dT) are identical.
[Left column]  $N=5$ and $N=17$.
[Right column] $N=6$ and $N=18$.
$\langle |A_{\mu} (t)|^2 \rangle$ follow Eqs.~\ref{psichi-typeac-even}-\ref{psichi-typeac-odd}, i.e. depend only on $N$ and not on the on-site energies and the hopping integrals.
If --on the contrary-- we imagine to initially distribute the carrier probability equally among monomers,
this initial equidistribution is conserved and no mean carrier movement occurs.}
\end{figure*}

Let us now allow the first monomer to interact with the last monomer with $t^{bp}$, i.e. for {\it cyclic} type $\alpha'$ polymers.
Then, for initial placing of the carrier at a particular monomer, we obtain $\frac{1}{N}$ additional mean --over time-- probability at the monomer where the initial placement is made {\it and} at the diametric monomer if it exists (i.e. for even $N$). In other words, if we call $\psi$ and $\chi$ the mean --over time-- probabilities at the favored and at the rest monomers, respectively, then $\psi = \chi + \frac{1}{N}$. Since the sum of all the mean --over time-- probabilities is 1, we obtain
\begin{eqnarray}
\label{psichi-typeac-even}
\psi=\frac{2(N-1)}{N^2}, \quad \chi=\frac{N-2}{N^2}, \quad \textrm{for even} \; N, \\
\label{psichi-typeac-odd}
\psi=\frac{2N-1}{N^2},   \quad \chi=\frac{N-1}{N^2}, \quad \textrm{for odd}  \; N.
\end{eqnarray}
This is depicted in Fig.~\ref{fig:typeacyclic}.
On the contrary, if we imagine to initially distribute the carrier probability equally among monomers, this initial equidistribution is conserved and no mean carrier movement is observed.

\subsubsection{\label{subsubsec:meanprob-beta} type $\beta'$} 
\begin{figure*}[]
\includegraphics[width=6cm]{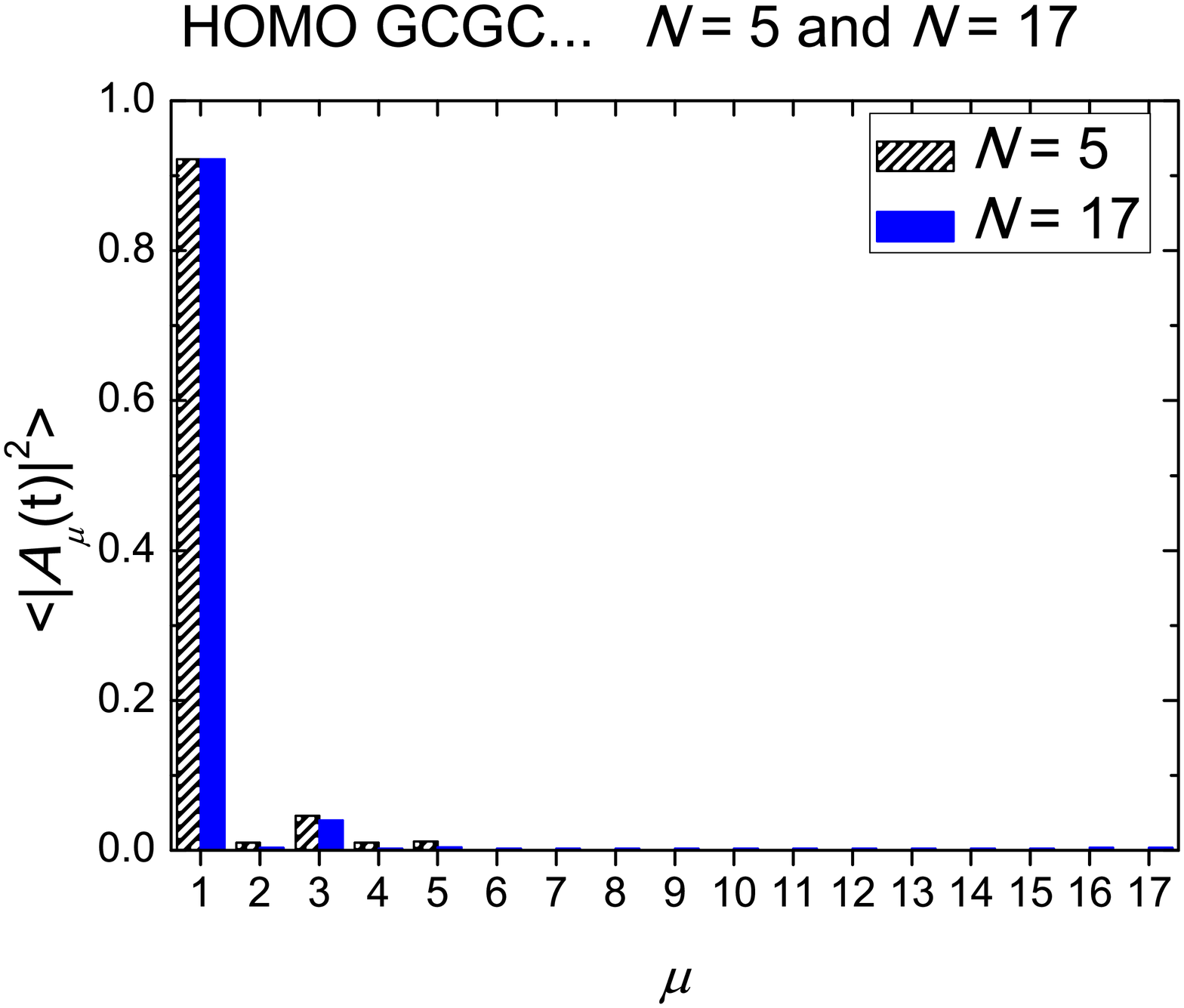}
\includegraphics[width=6cm]{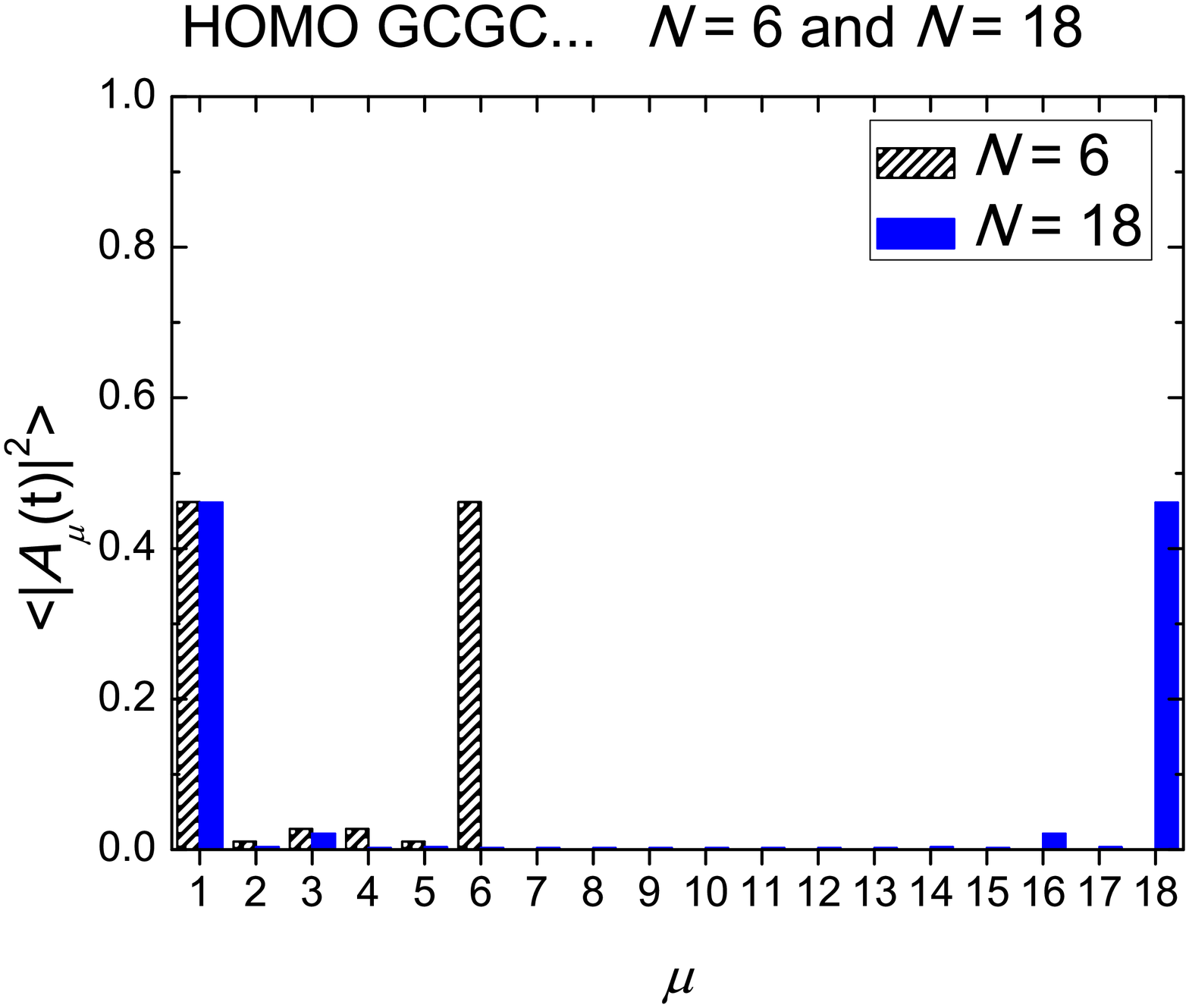}
\includegraphics[width=6cm]{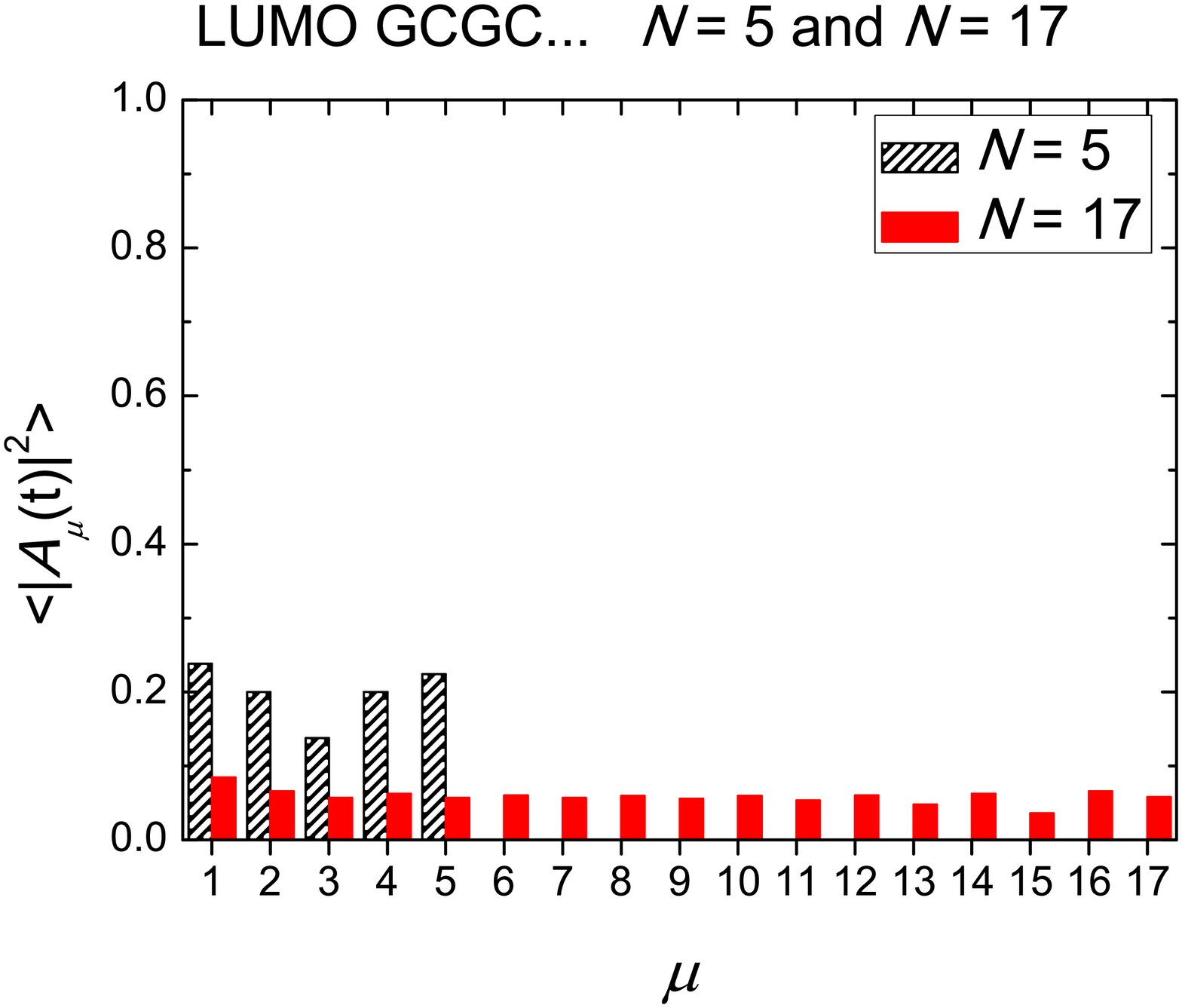}
\includegraphics[width=6cm]{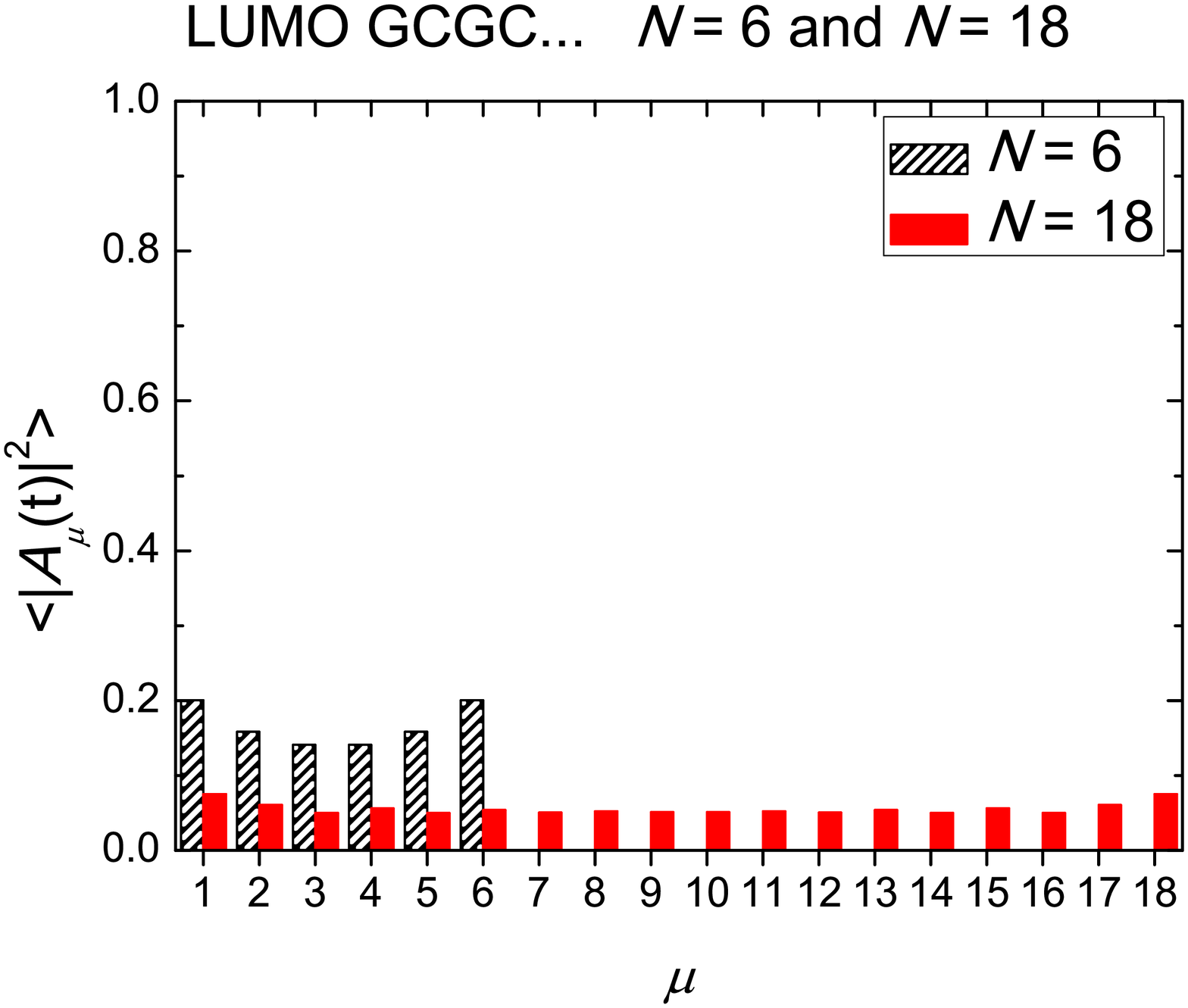}
\caption{\label{fig:N5N6N17N18GCGC} $\langle |A_{\mu} (t)|^2 \rangle$ for HOMO and LUMO GCGC... if we initially place the carrier at the 1st monomer.
[Left column]  $N=5$ and $N=17$.
[Right column] $N=6$ and $N=18$.
For $N$ even $\langle |A_{\mu} (t)|^2 \rangle$ are palindromes; for $N$ odd, this only holds for even $\mu$.
While for holes $|\frac{t^{bp}}{t^{bp'}}| = |\frac{t^{bp}_{GC}}{t^{bp'}_{CG}}|=0.2$,
for electrons $|\frac{t^{bp}}{t^{bp'}}| = |\frac{t^{bp}_{GC}}{t^{bp'}_{CG}}| = 1.25$, i.e.,
the hoping parameters~\cite{Simserides:2014,LKGS:2014}  are quite different in magnitude for holes but rather similar for electrons.
This leads to almost disrupted charge transfer if the number of repetition units is not integer i.e. for odd $N$.
}
\end{figure*}

Let us put the carrier initially at the first monomer.
For type $\beta'$ polymers, $\langle |A_{\mu} (t)|^2 \rangle$ do not depend only on $N$ in contrast to type $\alpha'$ polymers, i.e.,
for type $\beta'$ polymers \textbf{eigenspectrum independence of the probabilities does not hold}.
However, interestingly, for $N$ even, $\langle |A_{\mu}(t)|^2 \rangle$ are palindromes, while
for $N$ odd, this only holds for even $\mu$.
In other words, we have~\textbf{palindromicity for $N$ even, but only partial palindromicity for $N$ odd}.
These symmetry properties, can be summarized as
\begin{eqnarray}\label{edgesmiddlesb}
\langle |A_{1+i} (t)|^2 \rangle & = & \langle |A_{N-i} (t)|^2 \rangle, \\ \nonumber
N &=& \textrm{even}, \; i = 0, 1, \dots N-1, \quad  \textrm{or} \\ \nonumber
N &=& \textrm{odd}, \;\; i = 1, 3, \dots N-2.
\end{eqnarray}
In Fig.~\ref{fig:N5N6N17N18GCGC} we depict $\langle |A_{\mu} (t)|^2 \rangle$ for HOMO and LUMO GCGC...;
at the left  column for $N=5$ and $N=17$ and
at the right column for $N=6$ and $N=18$.
For GCGC..., the hoping parameters \cite{Simserides:2014,LKGS:2014} are quite different in magnitude for holes but rather similar for electrons, i.e.
while for holes $|\frac{t^{bp}}{t^{bp'}}| = |\frac{t^{bp}_{GC}}{t^{bp'}_{CG}}|=0.2$,
for electrons $|\frac{t^{bp}}{t^{bp'}}| = |\frac{t^{bp}_{GC}}{t^{bp'}_{CG}}| = 1.25$.
This leads to almost disrupted hole transfer when the number of repetition units is not integer i.e. for odd $N$.

\subsubsection{\label{subsubsec:meanprob-gamma} type $\gamma'$} 
\begin{figure*}[]
\includegraphics[width=6cm]{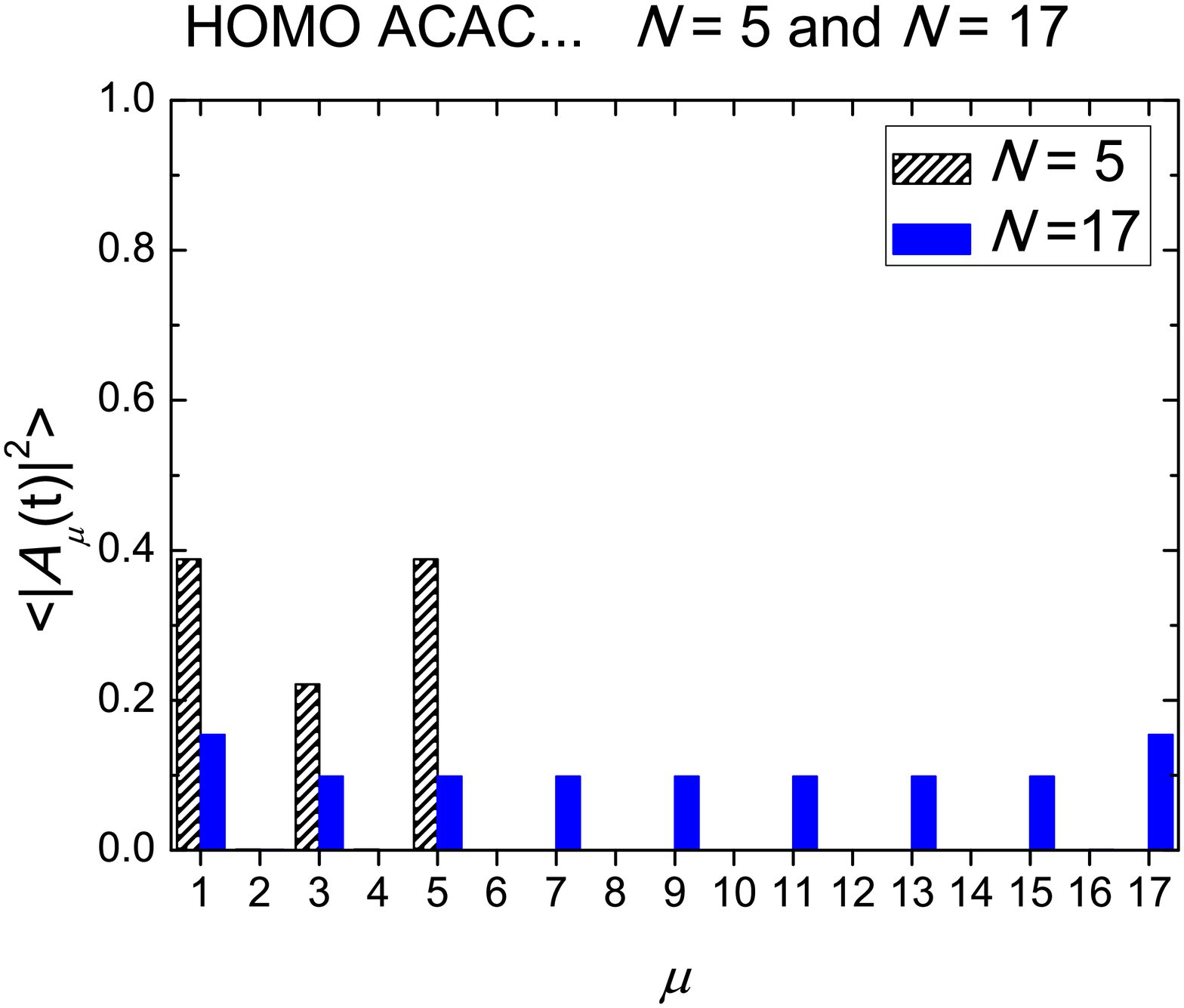}
\includegraphics[width=6cm]{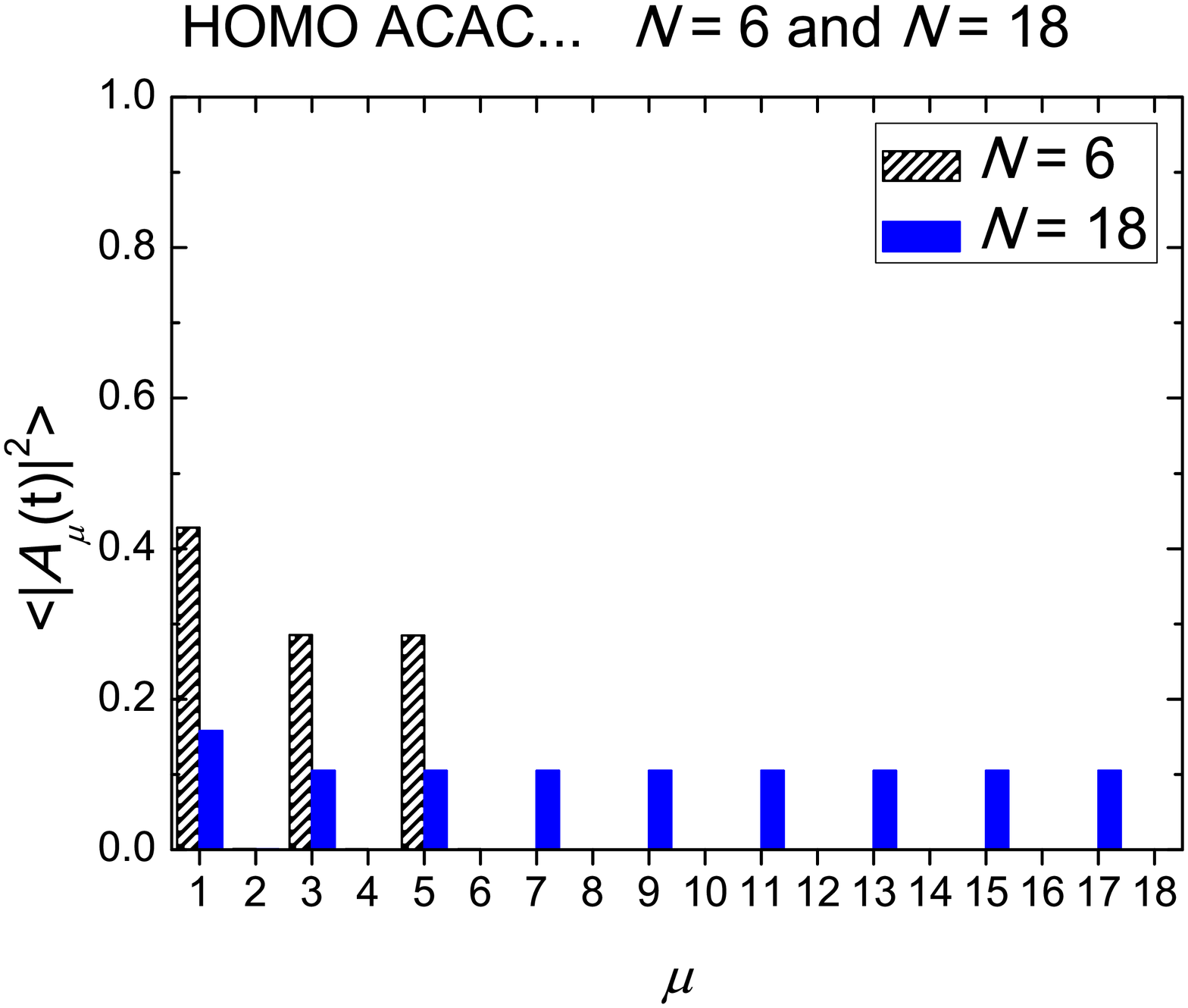}
\includegraphics[width=6cm]{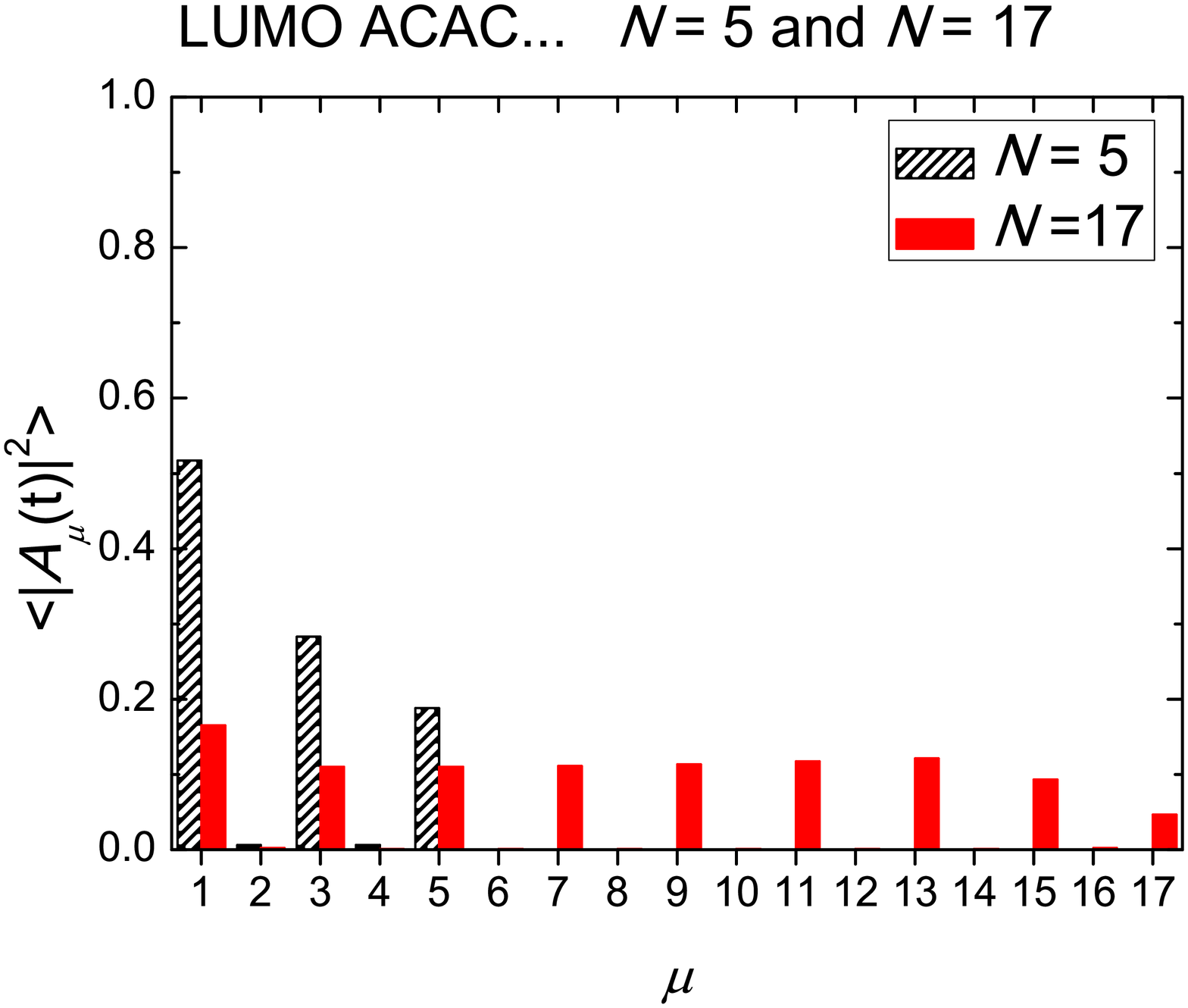}
\includegraphics[width=6cm]{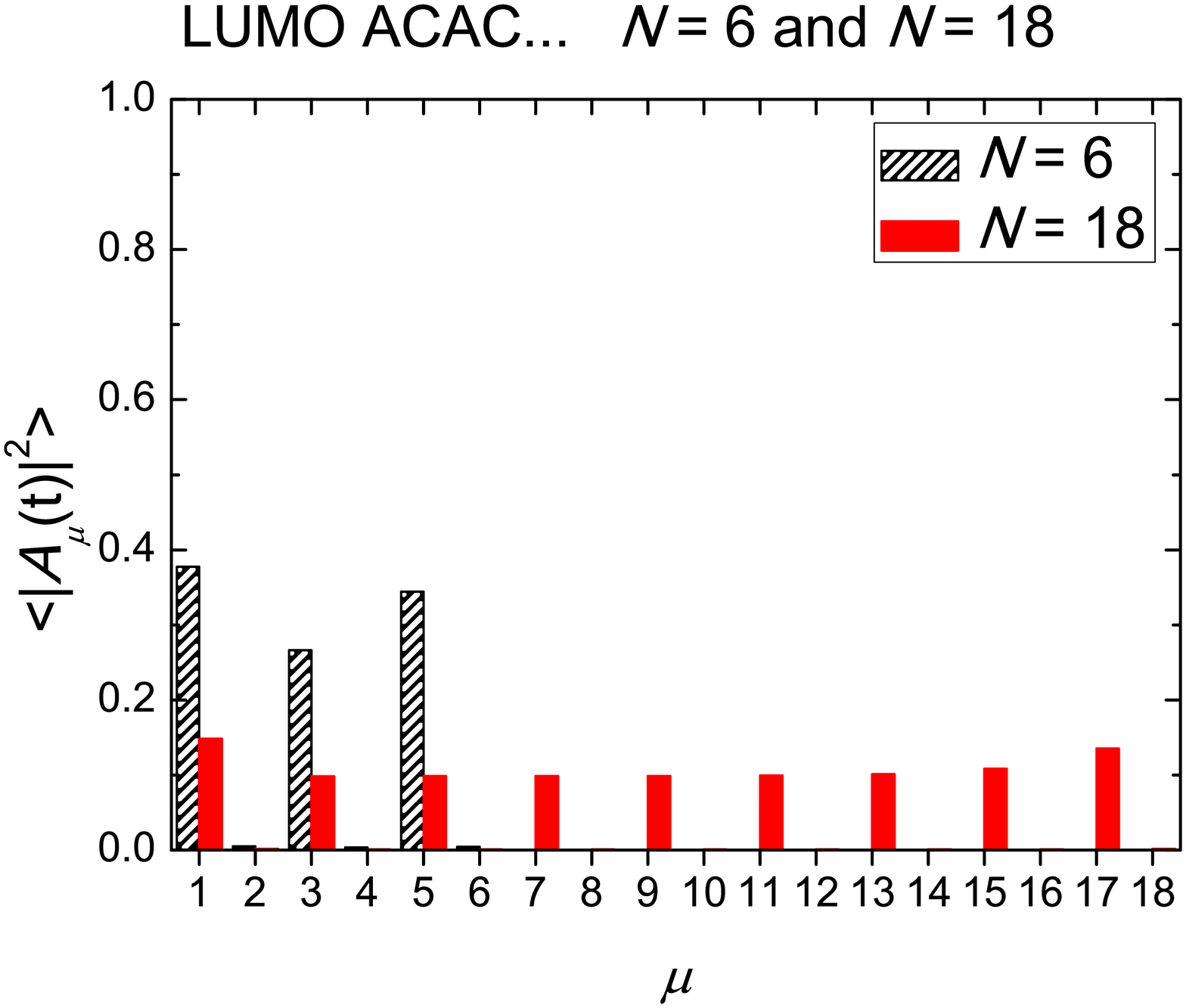}
\caption{\label{fig:N5N6N17N18ACAC} $\langle |A_{\mu} (t)|^2 \rangle$ for HOMO and LUMO ACAC... if we initially place the carrier at the 1st monomer.
[Left column]  $N=5$ and $N=17$.
[Right column] $N=6$ and $N=18$.
We notice that for HOMO ACAC... where $|t^{bp}|=|t^{bp'}|$, for $N$ odd, $\langle |A_{\mu} (t)|^2 \rangle$ are incidentally palindromes.
This is a property mirrored from the stationary case since at this specific case $|v_{\mu k}|^2$ are palindromes.}
\end{figure*}

Let us put the carrier initially at the first monomer.
For type $\gamma'$ polymers, $\langle |A_{\mu} (t)|^2 \rangle$ do not depend only on $N$ in contrast to type $\alpha'$ polymers, i.e.,
for type $\gamma'$ polymers \textbf{eigenspectrum independence of the probabilities does not hold}.
In Fig.~\ref{fig:N5N6N17N18ACAC} we depict $\langle |A_{\mu} (t)|^2 \rangle$ for HOMO and LUMO ACAC...;
at the left  column for $N=5$ and $N=17$ and
at the right column for $N=6$ and $N=18$.
For HOMO ACAC... where incidentally $|t^{bp}|=|t^{bp'}|$, for $N$ odd, $\langle |A_{\mu}(t)|^2 \rangle$ are palindromes.

\subsection{\label{subsec:expvspow} Pure mean transfer rate fits} 
\begin{figure*}[h!]
\includegraphics[width=8.5cm]{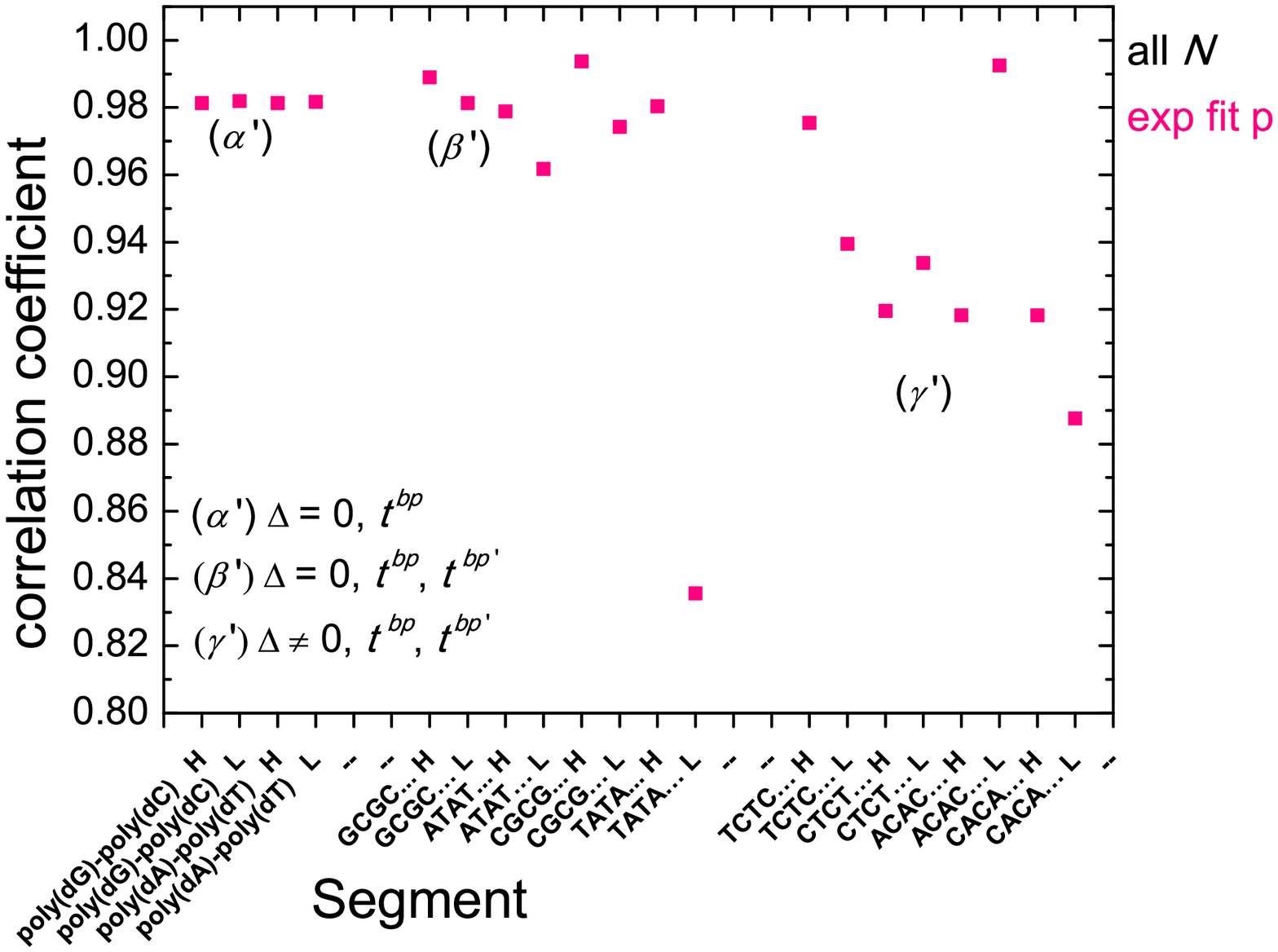}
\includegraphics[width=8.5cm]{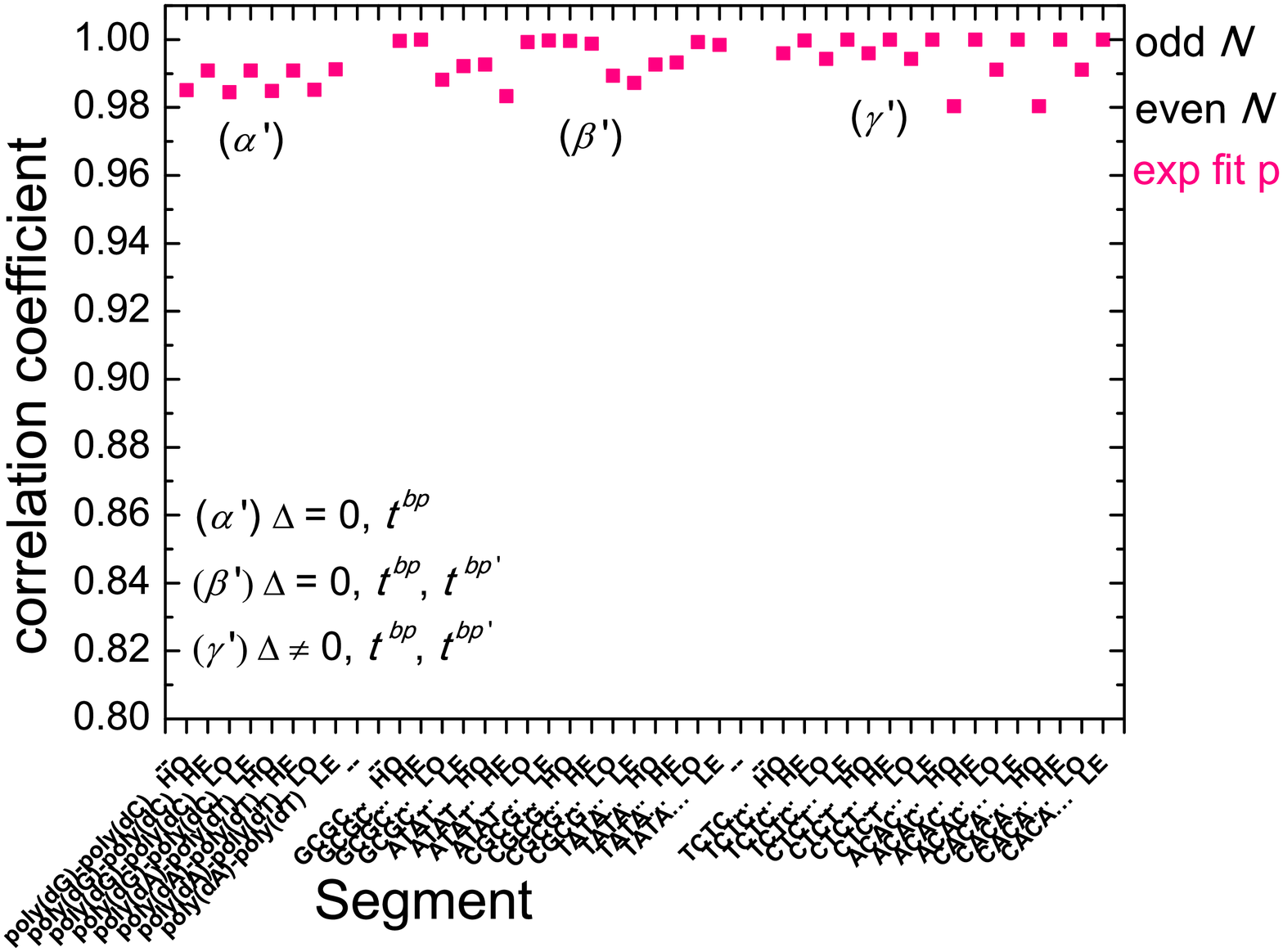}
\includegraphics[width=8.5cm]{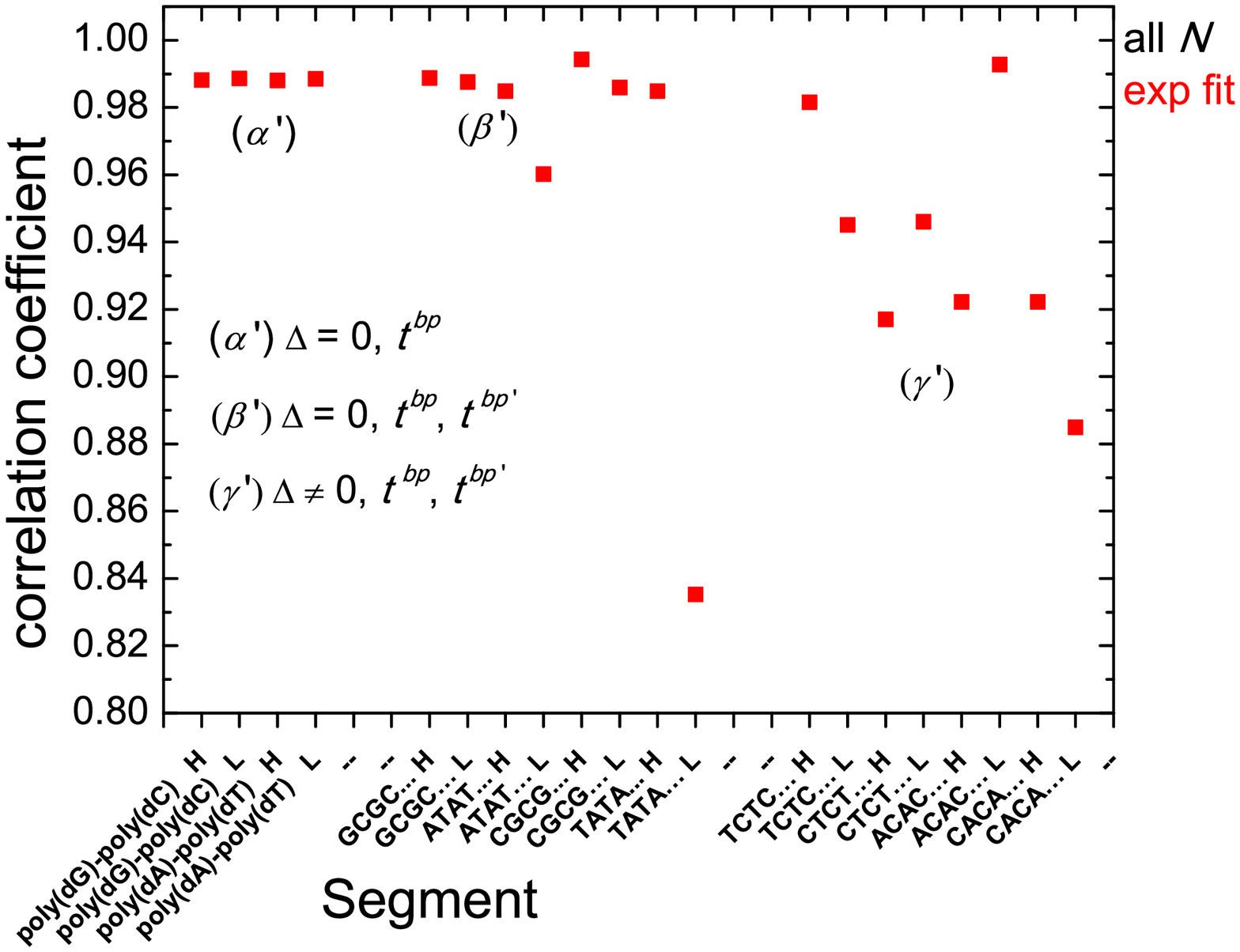}
\includegraphics[width=8.5cm]{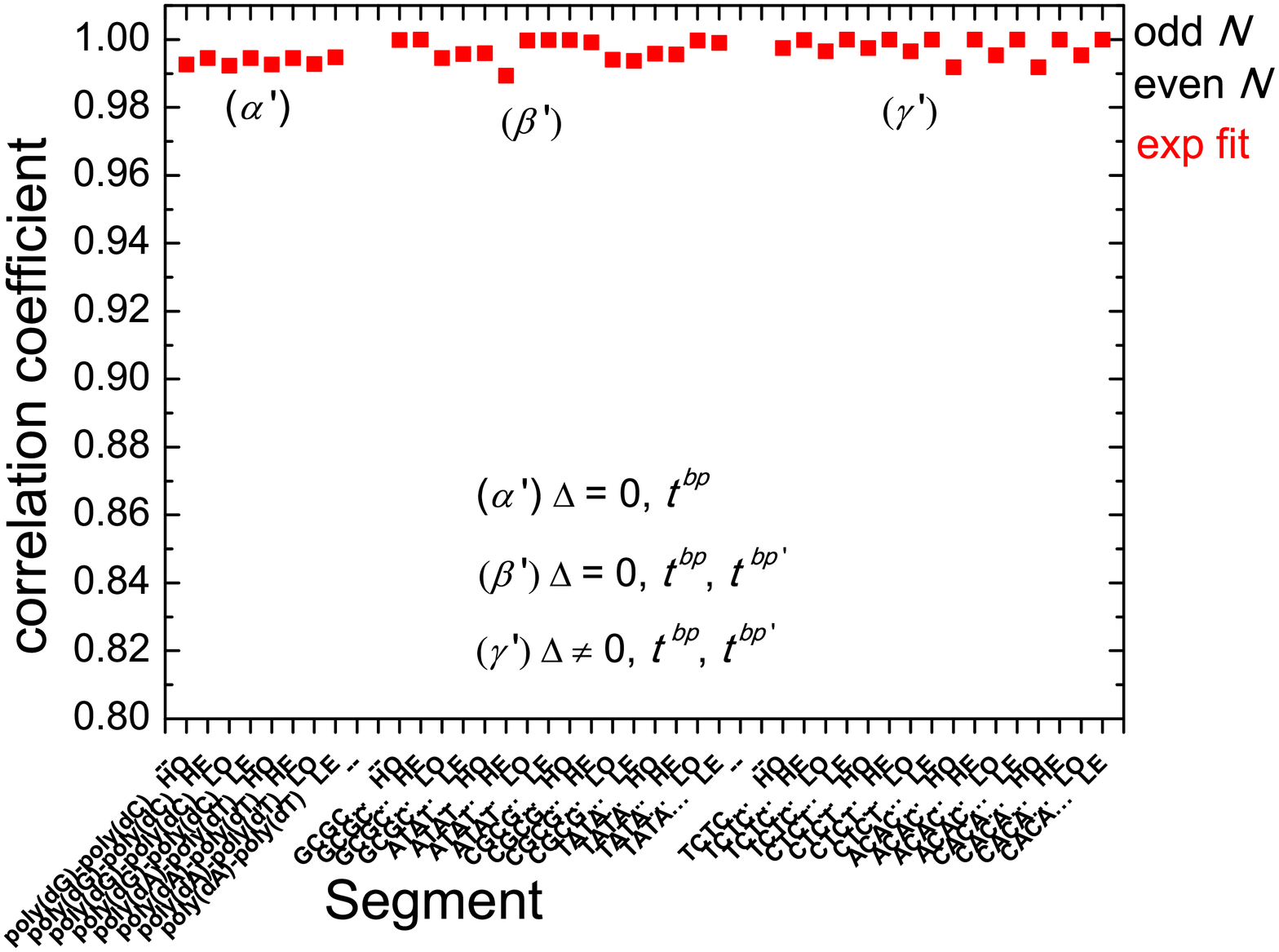}
\includegraphics[width=8.5cm]{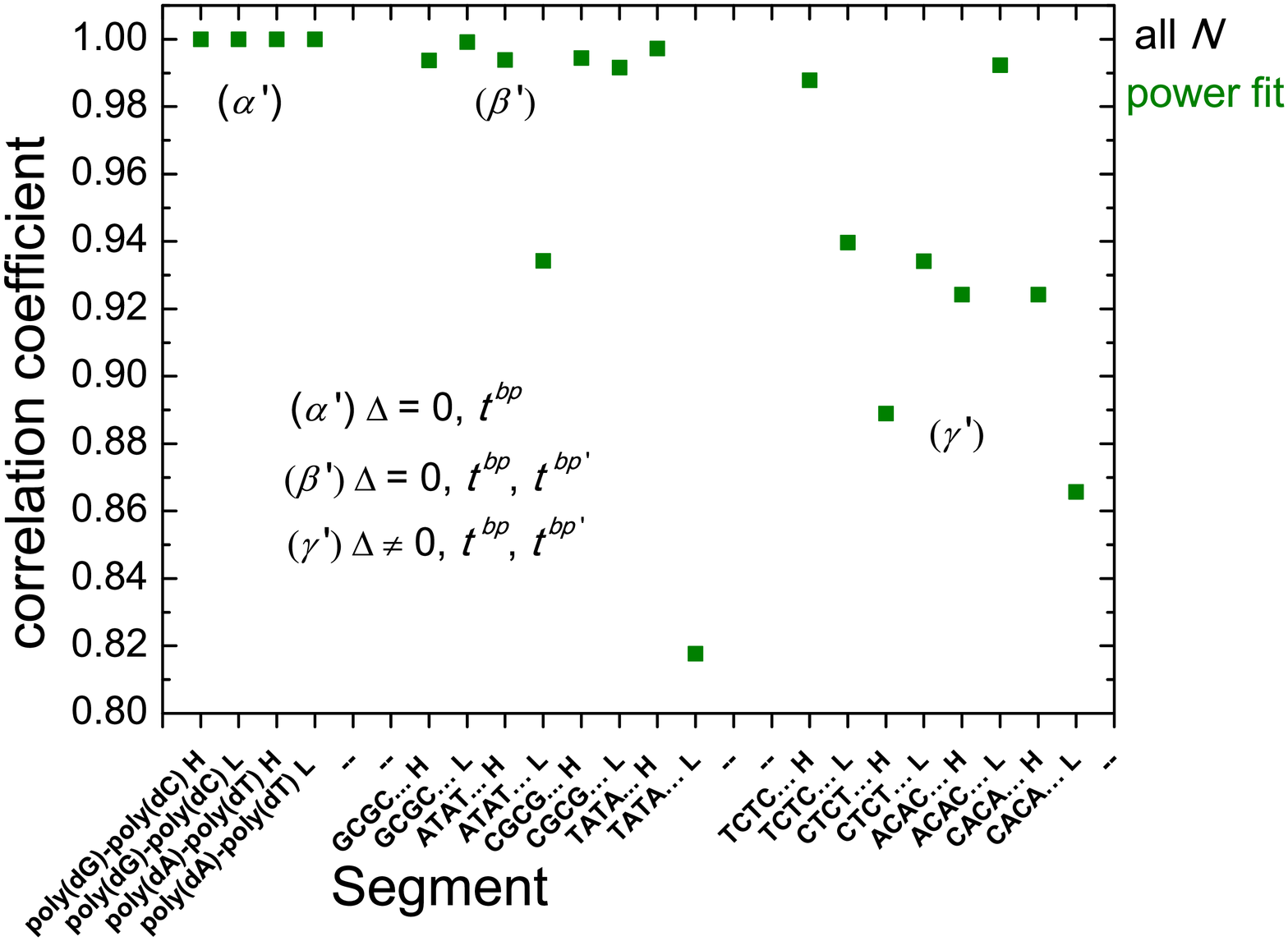}
\includegraphics[width=8.5cm]{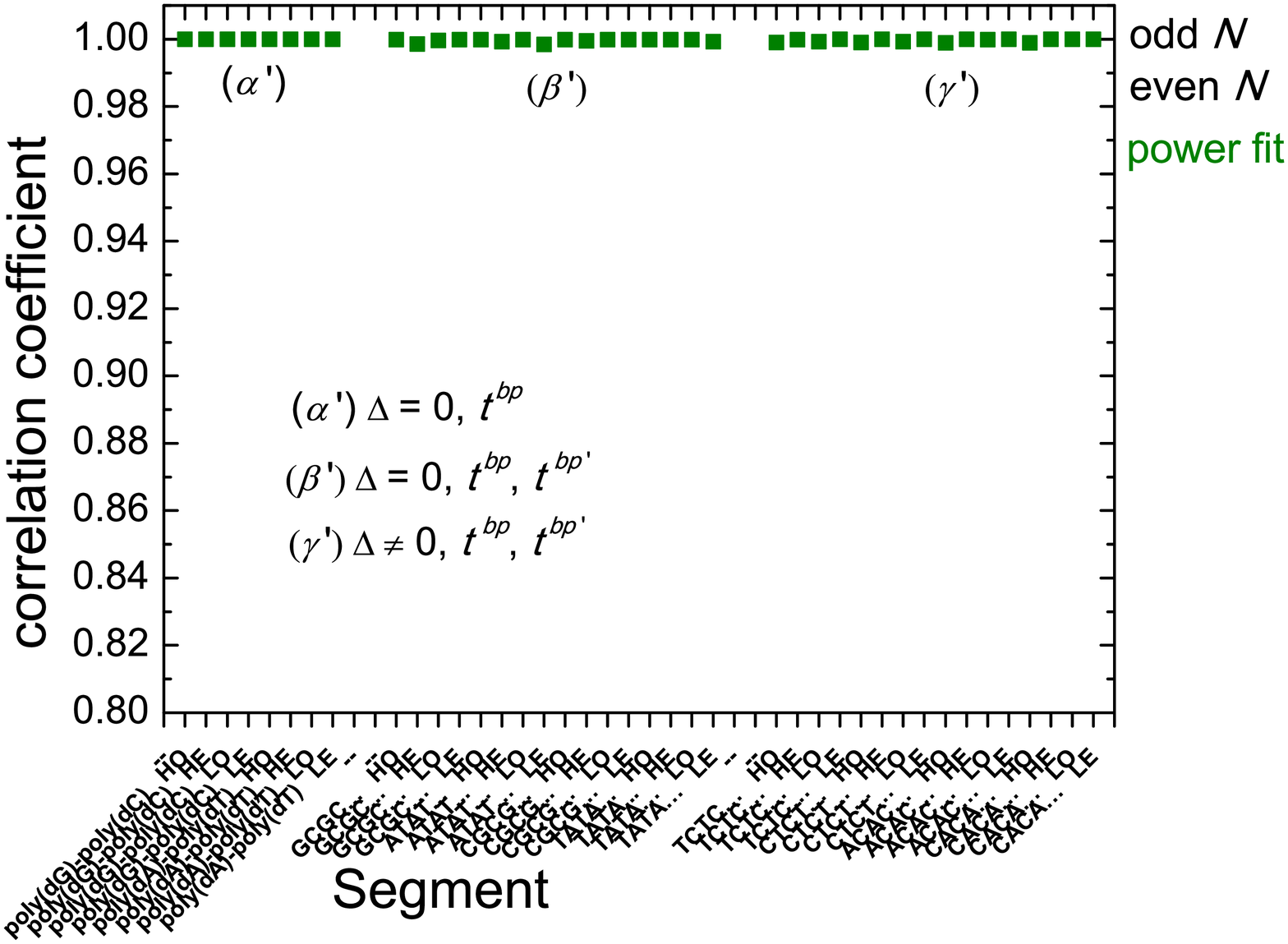}
\caption{\label{fig:expvspowcc} We compare the correlation coefficients of type $\alpha'$, $\beta'$, and $\gamma'$ polymers for:
the exponential fit $k =     k_0 e^{- \beta d}$ (1st row),
the exponential fit $k = A + k_0 e^{- \beta d}$ (2nd row), and
the power law fit $k = k_0' N^{- \eta}$ (3rd row).
In the 1st column the fits include {\it all} $N$ while in the 2nd column we fit separately {\it even} and {\it odd} $N$.
The power law fits are significantly better.
For type $\alpha'$, a fitting $k(N)$ does not really depend on which $N$ -- even or odd -- we include.
This does not hold for types $\beta'$ and $\gamma'$; since the repetition unit is a dimer, we need to separate fits for odd and even $N$.}
\end{figure*}

\begin{figure*}[h!]
\includegraphics[width=8.5cm]{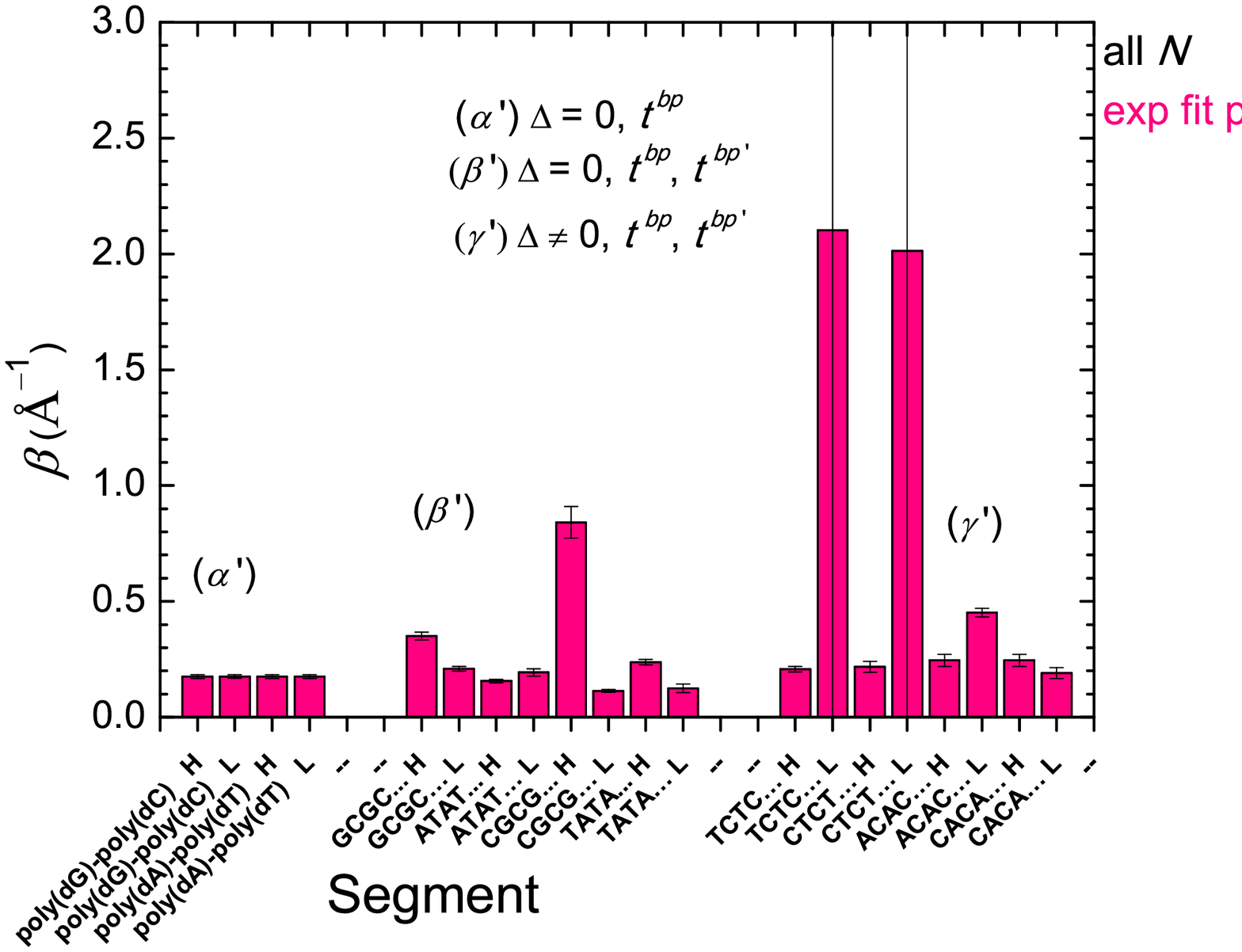}
\includegraphics[width=8.5cm]{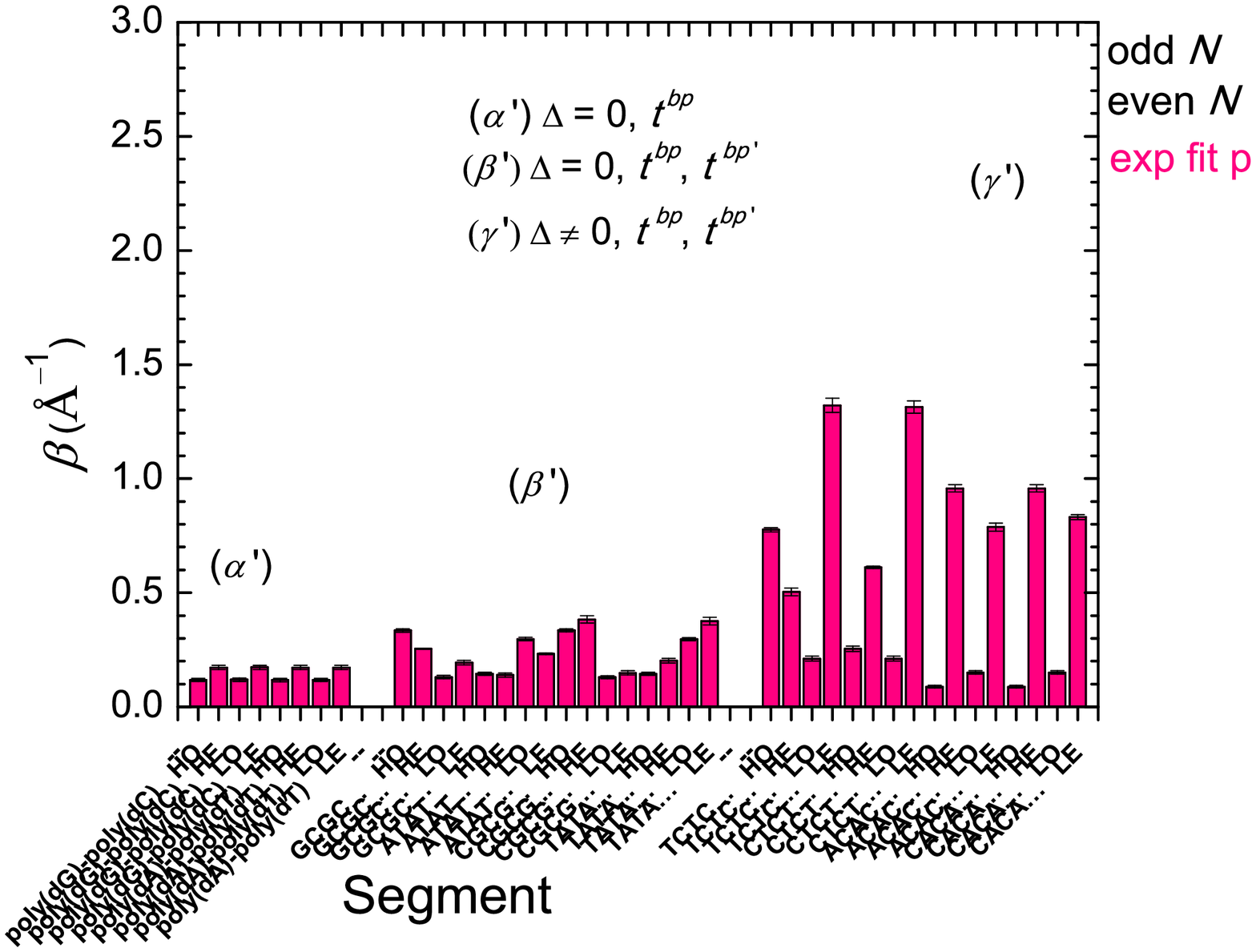}
\includegraphics[width=8.5cm]{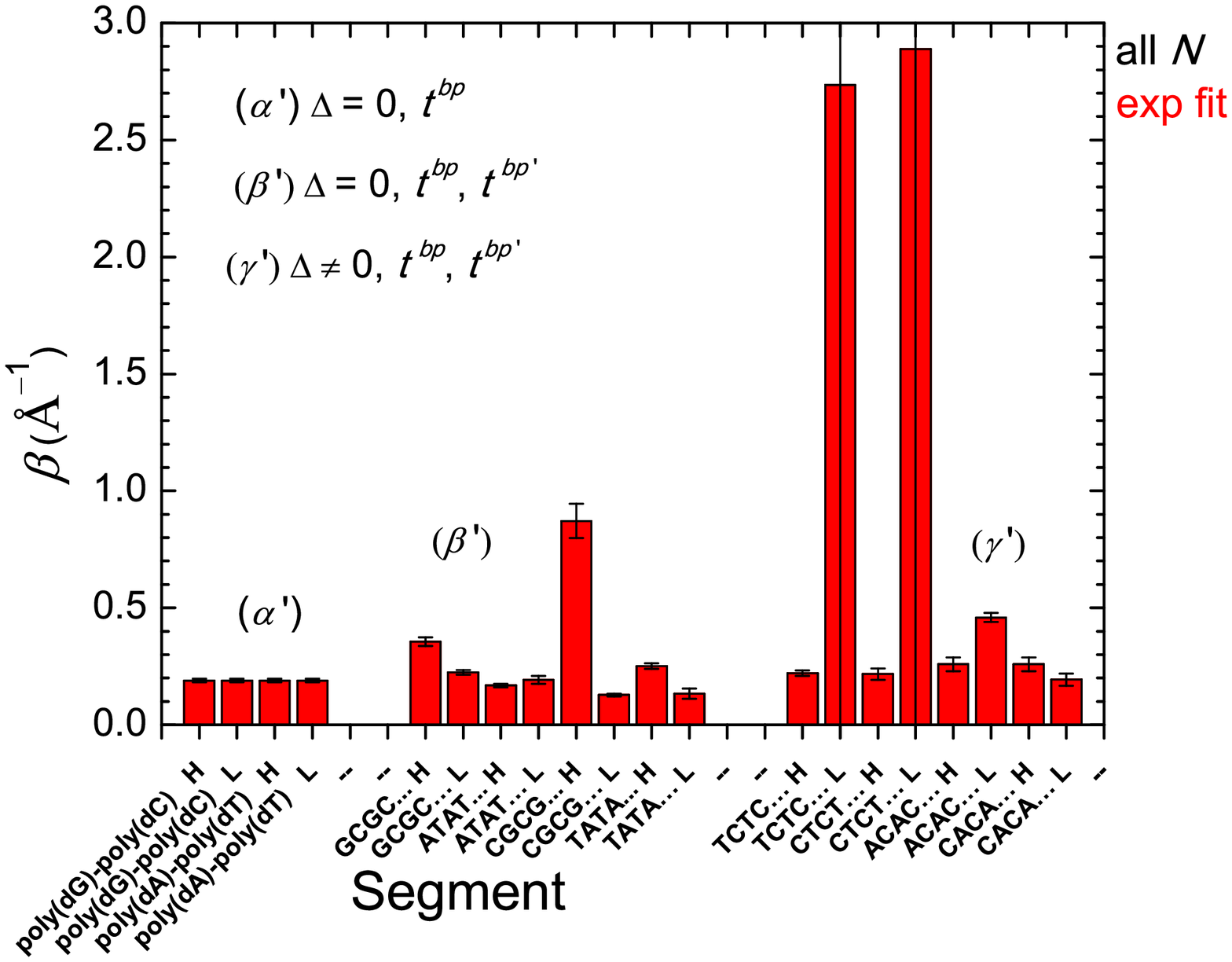}
\includegraphics[width=8.5cm]{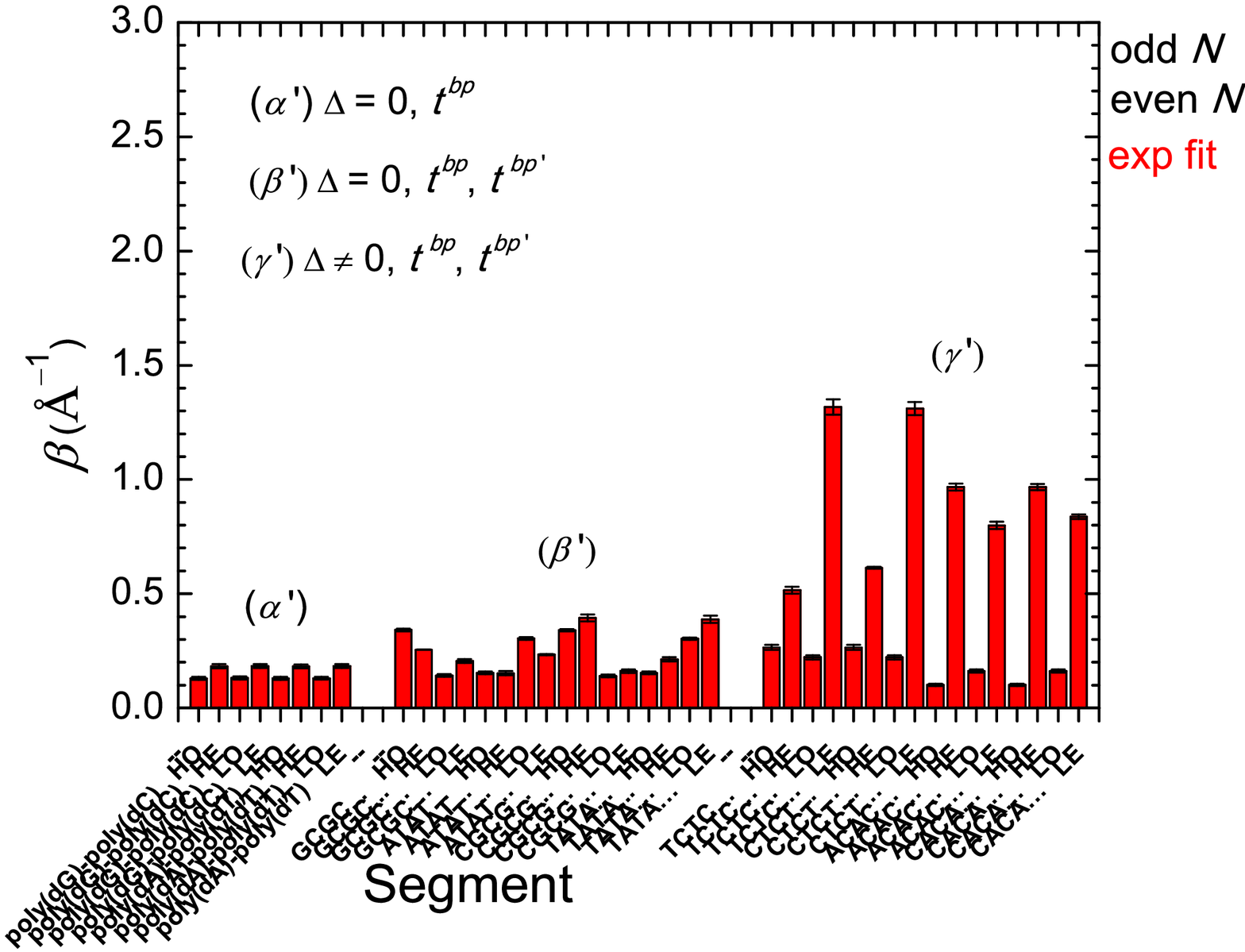}
\includegraphics[width=8.5cm]{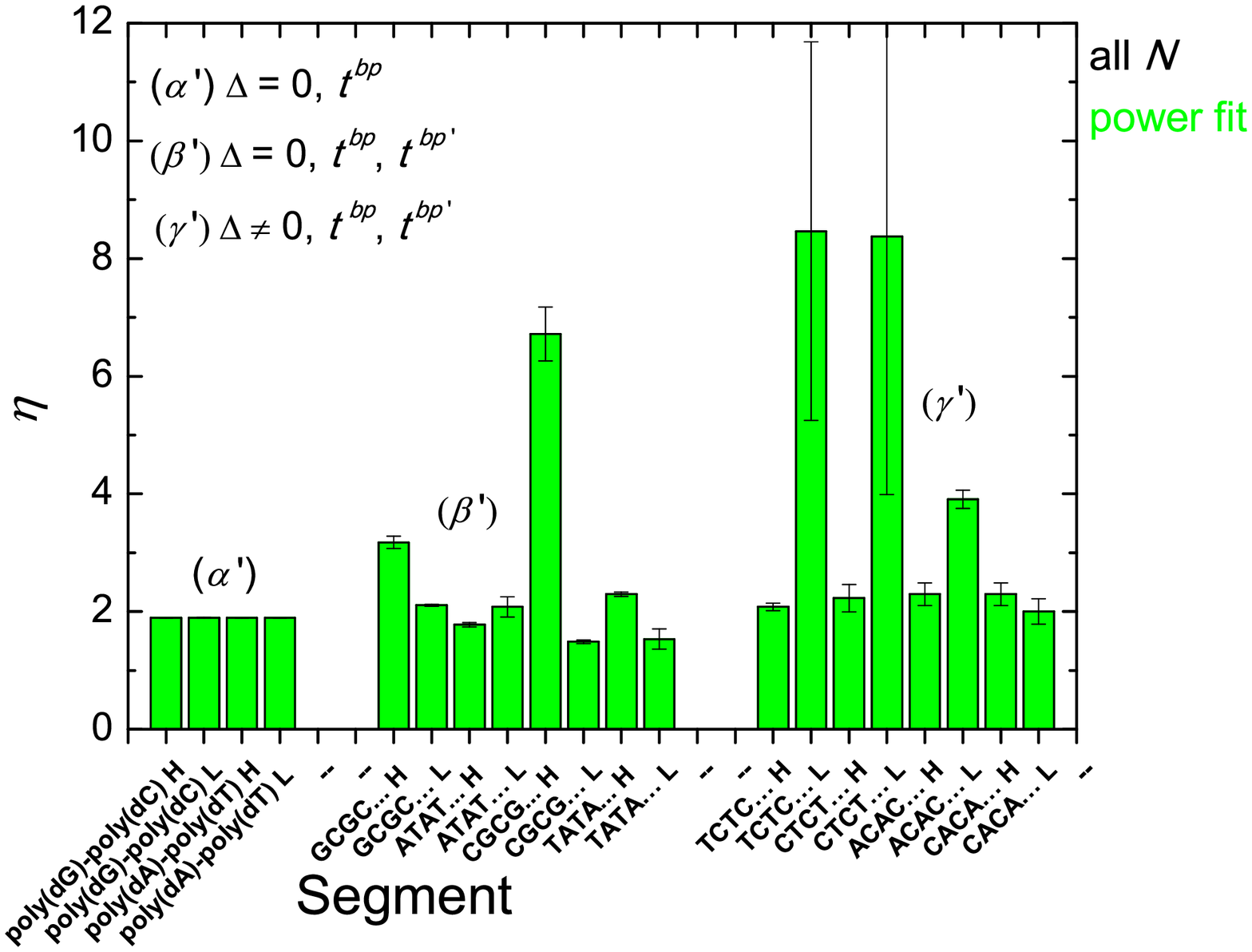}
\includegraphics[width=8.5cm]{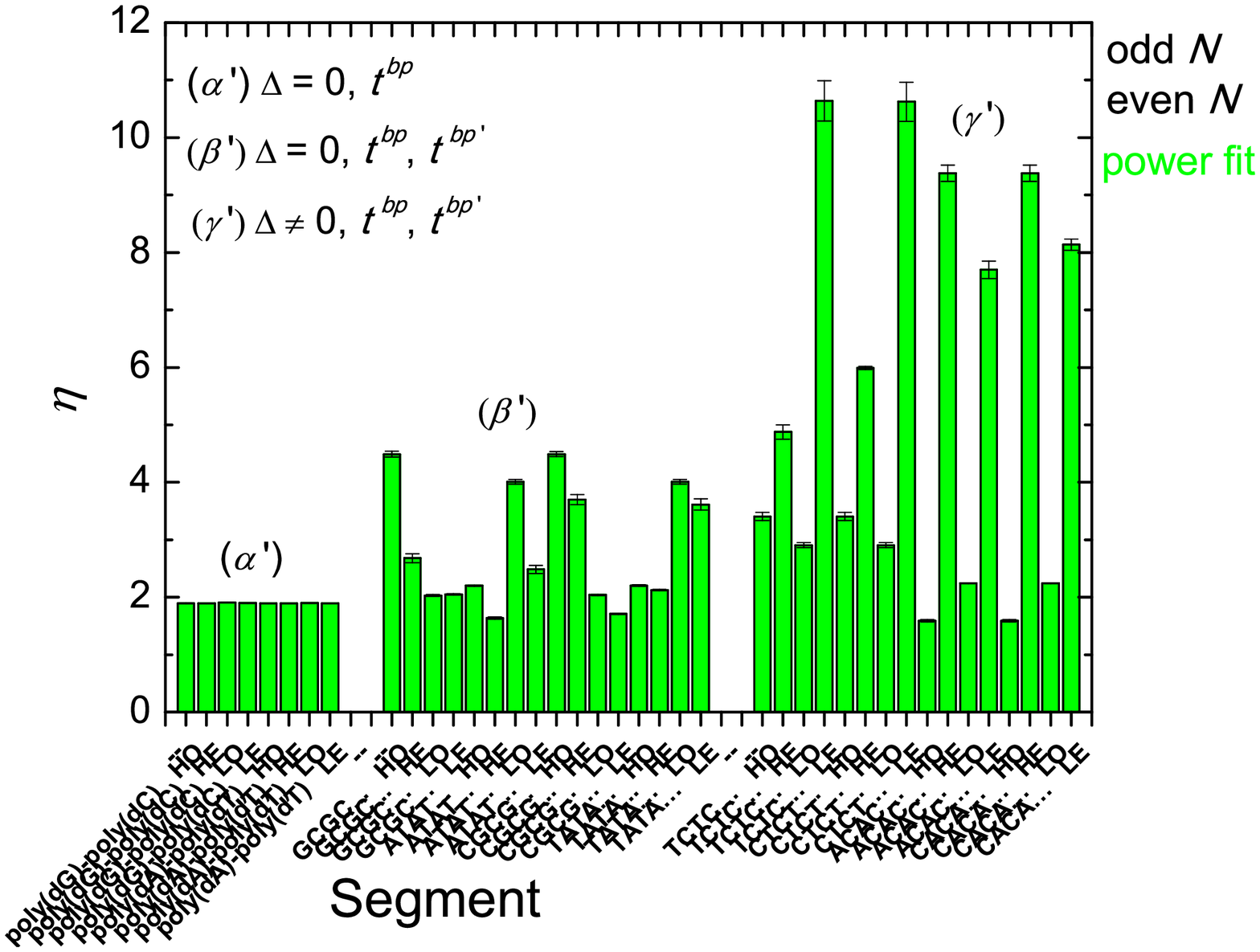}
\caption{\label{fig:expvspowbetaeta} We compare $\beta$ and $\eta$ of type $\alpha'$, $\beta'$, and $\gamma'$ polymers for:
the exponential fit $k =     k_0 e^{- \beta d}$ (1st row),
the exponential fit $k = A + k_0 e^{- \beta d}$ (2nd row), and
the power law fit $k = k_0' N^{- \eta}$ (3rd row).
In the 1st column the fits include {\it all} $N$ while in the 2nd column we fit separately {\it even} and {\it odd} $N$.
The power law fits are significantly better.
For type $\alpha'$, a fitting $k(N)$ does not really depend on which $N$ -- even or odd -- we include.
This does not hold for types $\beta'$ and $\gamma'$; since the repetition unit is a dimer, we need to separate fits for odd and even $N$.
By the way, including {\it all} $N$, in two cases we obtain very large error bars.}
\end{figure*}

Next, we initially place the carrier at the 1st monomer and examine the pure mean transfer rates. Specifically:
In
Fig.~\ref{fig:expvspowcc} (correlation coefficients) and
Fig.~\ref{fig:expvspowbetaeta} ($\beta$ and $\eta$),
we compare
the exponential fit $k =     k_0 e^{- \beta d}$ (1st row),
the exponential fit $k = A + k_0 e^{- \beta d}$ (2nd row, $A$ typically results tiny), and
the power law fit $k = k_0' N^{-\eta}$ (3rd row)
for type $\alpha'$, $\beta'$ and $\gamma'$ polymers.
These fits are carried out up to $N = $ 60, i.e., $d = $ 200.6 {\AA},
since it is known~\cite{Meggers:1998,Henderson:1999,KawaiMajima:2013} that a carrier can migrate along DNA over 200 {\AA}.
The 1st column refers to fits including {\it all} $N$ while in the 2nd column we fit separately {\it even} and {\it odd} $N$.
We observe that, generally, the fits improve when we separate {\it even} and {\it odd} $N$.

Furthermore, it is evident that the power law fits are significantly better.
This agrees with the assertion~\cite{Giese:1999,Giese:2002} that
when every single hopping step occurs over the same distance, the hopping mechanism is described better by a power-law fit. Since here we study periodic polymers, it seems that the above scenario holds. 
For type $\alpha'$, since $\langle |A_{\mu} (t)|^2 \rangle$, both for even and odd $N$, follow the same Eqs.~(\ref{oneedgea})-(\ref{onemiddlea}),
a fitting $k(N)$ does not really depend on which $N$ -- even or odd -- we include.
This is obvious in terms of correlation coefficients or $\eta$ at the last row of
Figs.~\ref{fig:expvspowcc},~\ref{fig:expvspowbetaeta}.
This does not hold for types $\beta'$ and $\gamma'$ where --since the repetition unit is a dimer-- we need to separate fits for odd and even $N$.

Finally, it is evident that, as a general trend, the fall of $k$ as a function of $d$ or $N$ becomes steeper when the intricacy of the energy structure is increased i.e. from type $\alpha'$ to type $\beta'$ and further to type $\gamma'$.

\section{\label{sec:conclusion} Conclusion} 
We have systematically studied electron or hole oscillations in B-DNA monomer-polymers and dimer-polymers, i.e., periodic sequences with repetition unit made of one or two monomers, where monomer is a base-pair.
We used a tight-binding approach at the base-pair level to determine the temporal and spatial evolution of a single extra carrier along the $N$ base-pair DNA polymer.
We studied the HOMO and LUMO eigenspectra as well as the mean over time probabilities to find the carrier at a particular monomer.

Furthermore, we used the pure mean transfer rate $k$ to evaluate the easiness of charge transfer and estimated the inverse decay length $\beta$ for exponential fits $k(d)$, where $d = (N-1) \times $ 3.4 {\AA} is the charge transfer distance, and the exponent $\eta$ for power law fits $k(N)$.
It seems that the power law fits are significantly better. We also illustrated that increasing the energy structure intricacy i.e. the number of different parameters involved in the TB description, the fall of $k$ as a function of $d$ or $N$ becomes steeper, and
we showed the range covered by $\beta$ and $\eta$.

Finally, we combined analytical and numerical solutions for the time-independent and the time-dependent problem and
analyzed palindromicity and degree of eigenspectrum independence of the probabilities to find the carrier at a particular monomer.
{\it Eigenspectrum independence} means that the probability to find the carrier at a particular monomer does not depend on the on-site energies and the hopping integrals.
{\it Palindromicity} means that the occupation probability for the $\mu$-th monomer is equal to the occupation probability of the $(N-\mu+1)$-th monomer.
Type $\alpha'$ polymers display both
palindromicity and eigenspectrum independence of the probabilities.
Type $\beta'$ polymers display partial eigenspectrum dependence, and palindromicity for $N$ even but only partial palindromicity for $N$ odd.
Generally, type $\gamma'$ polymers do not have either eigenspectrum independence or palindromicity of the probabilities.

\begin{acknowledgments}
A. Morphis wishes to thank the State Scholarships Foundation-IKY, for the scholarship he has been offered for conducting Ph.D research in Greece through the ``IKY Fellowships of Excellence for Postgraduate Studies in Greece-Siemens Program'' in the framework of Hellenic Republic-Siemens Settlement Agreement.
\end{acknowledgments}

\clearpage

\begin{widetext}

\appendix

\section{\label{apA} matrix A} 
\begin{equation}\label{A}
\textrm{A} = \left[
\begin{array}{ccccccc}
  E^{bp(1)}_{H/L}&t^{bp(1;2)}_{H/L}& 0                &\cdots&     0 &      0 & 0    \\
t^{bp(2;1)}_{H/L}&  E^{bp(2)}_{H/L}& t^{bp(2;3)}_{H/L}&\cdots&     0 &      0 & 0    \\
\vdots           & \vdots          & \vdots           &\vdots&\vdots & \vdots &\vdots\\
0      & 0     & 0      &\cdots&t^{bp(N-1;N-2)}_{H/L}&  E^{bp(N-1)}_{H/L} & t^{bp(N-1;N)}_{H/L}    \\
0      & 0     &  0     &\cdots&                   0 &t^{bp(N;N-1)}_{H/L} & E^{bp(N)}_{H/L} \end{array} \right].
\end{equation}

\section{\label{polydApolydT} $\textrm{poly(dA)-poly(dT)}$ as an example of type $\alpha'$ polymers} 
In Fig.~\ref{fig:HOMOLUMOpolydApolydT} we show some HOMO and LUMO properties of poly(dA)-poly(dT).
We call \textit{Edge Group} the first and the last monomer, and \textit{Middle Group} the rest of the monomers.
The total probability at the {\it Edge Group}, $e(N) = \frac{3}{N+1}$, and at the {\it Middle Group}, $m(N)=\frac{N-2}{N+1}$
[cf. Eqs.~(\ref{oneedgea})-(\ref{onemiddlea})]. It seems that a power-law describes better the situation. This agrees with the claim that when every single hopping step occurs over the same distance, the hopping mechanism is described better by a power-law fit~\cite{Giese:1999,Giese:2002}. Since here we have the simplest periodic type one could imagine, the above claim evidently holds.
In the last row we show $ u = k d$ versus $d$, which is less than 10$^4$ m/s even for small oligomers.
\begin{figure*}[h!]
\includegraphics[width=6cm]{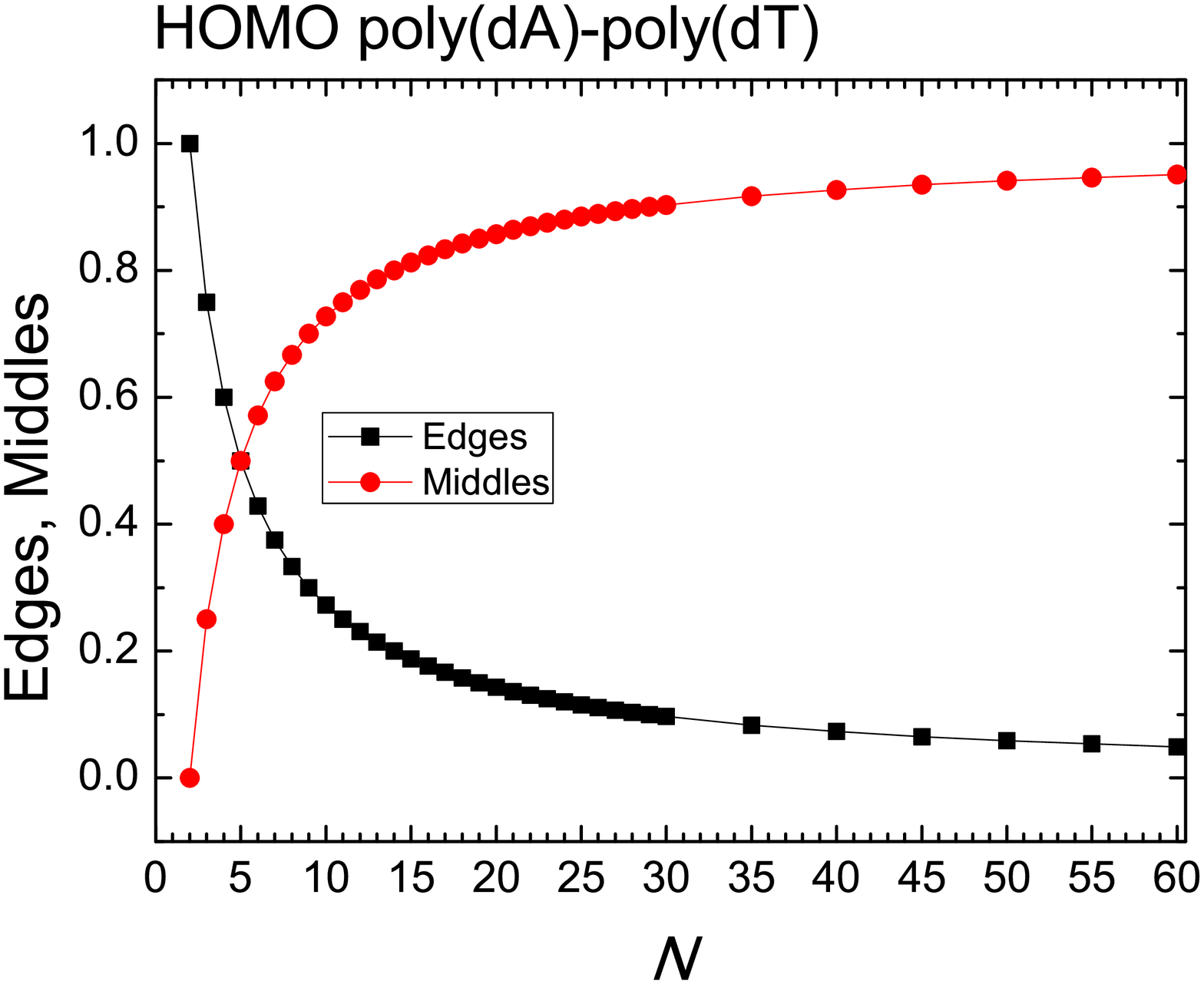}
\includegraphics[width=6cm]{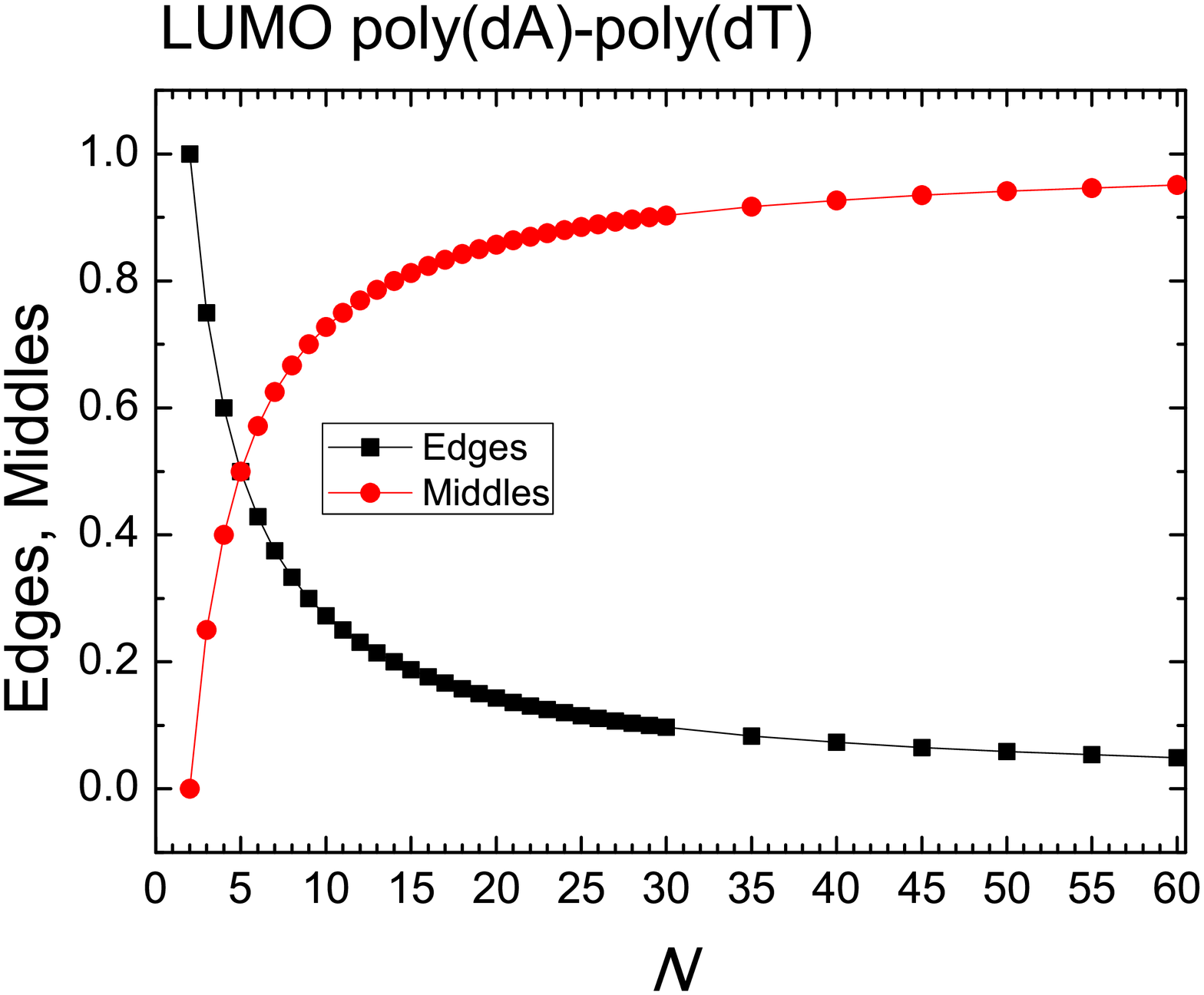}
\includegraphics[width=6cm]{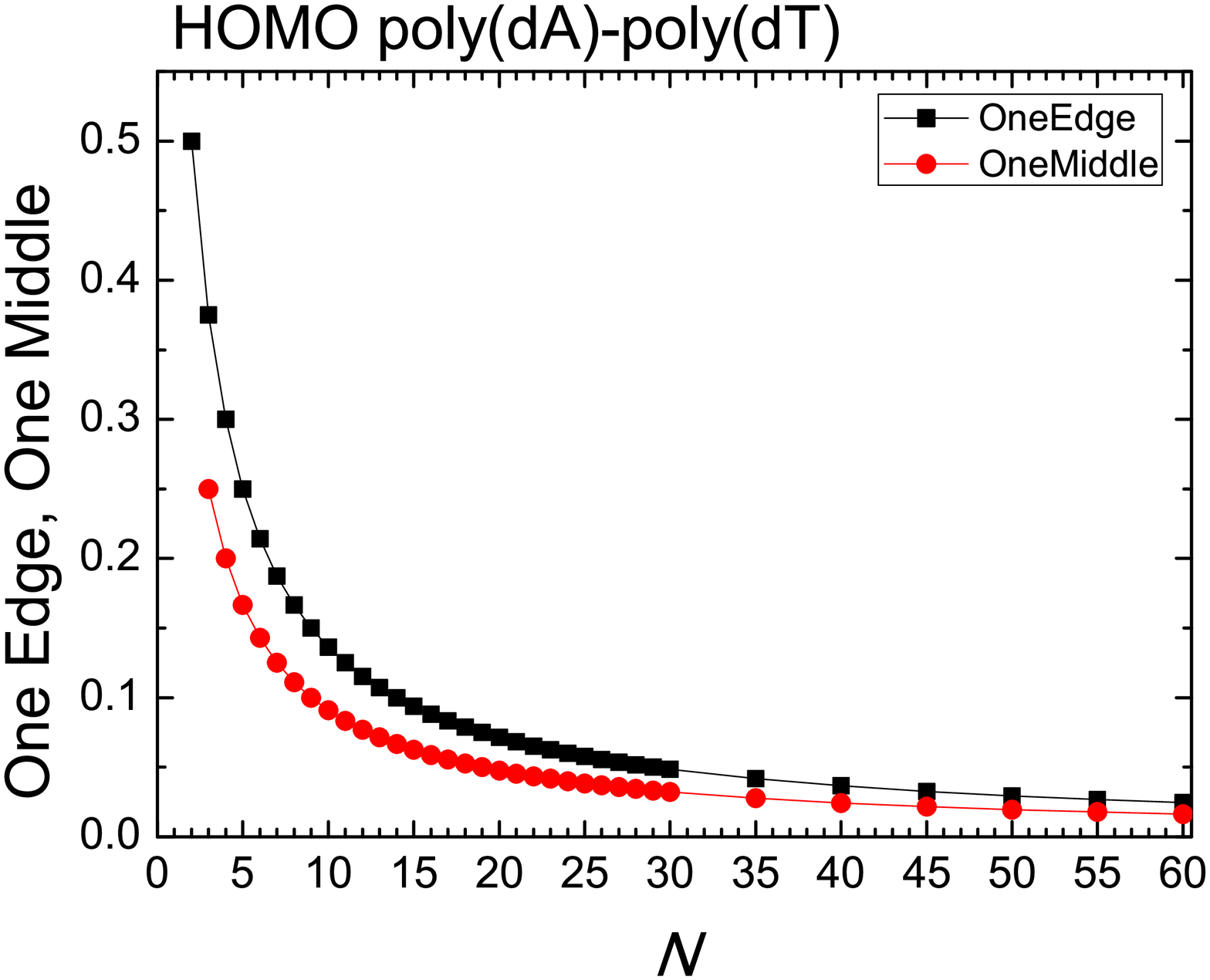}
\includegraphics[width=6cm]{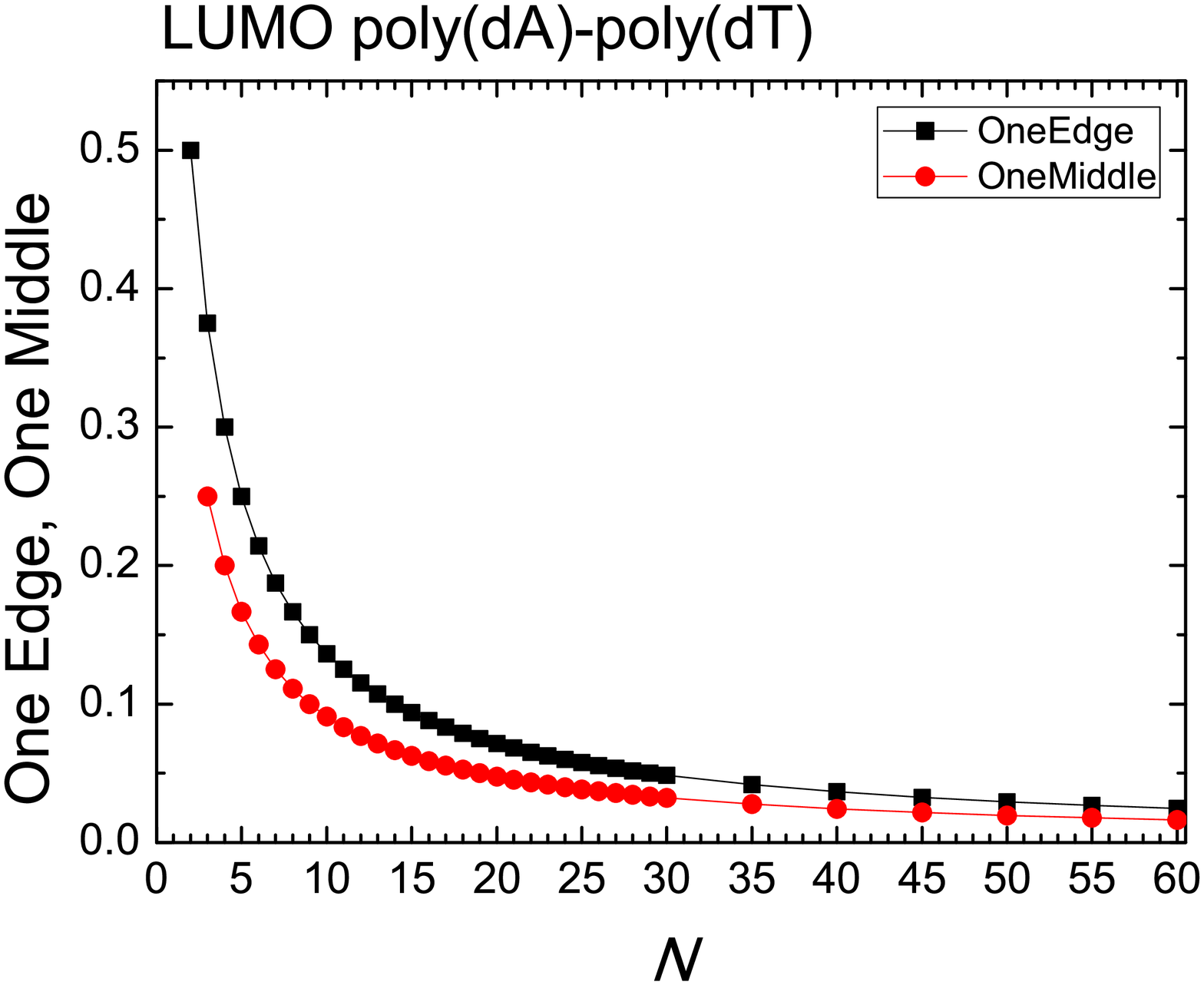}
\includegraphics[width=6cm]{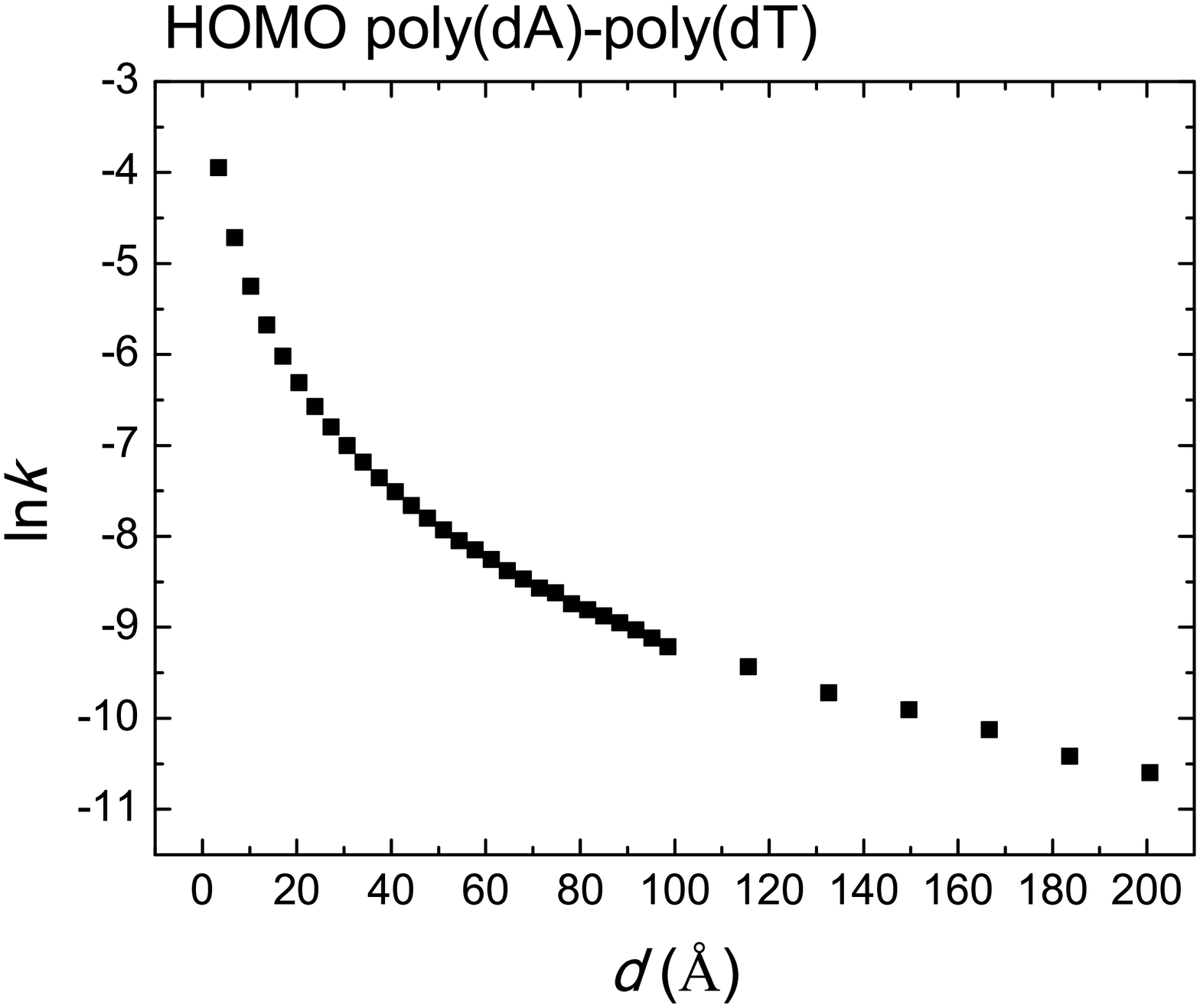}
\includegraphics[width=6cm]{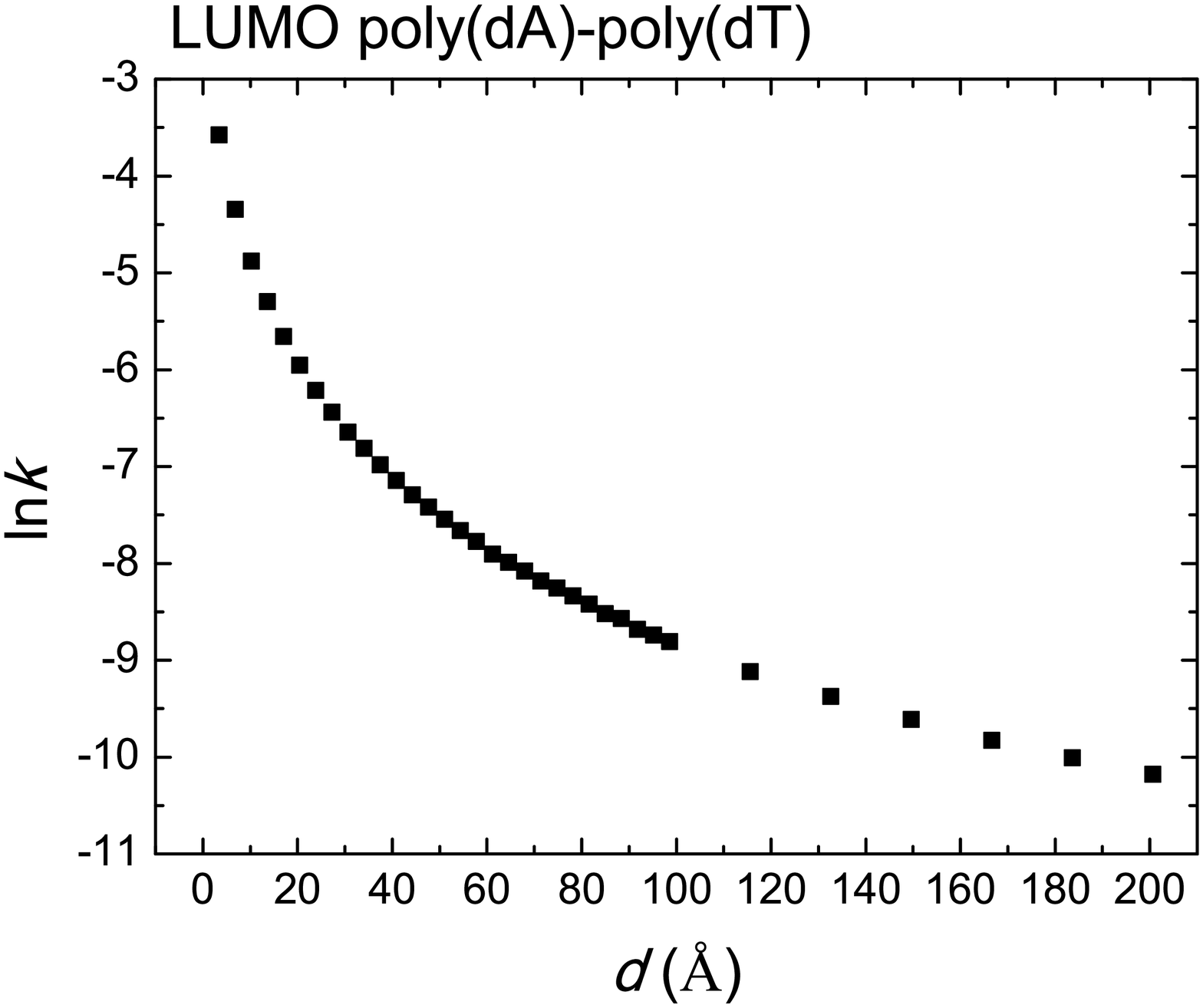}
\includegraphics[width=6cm]{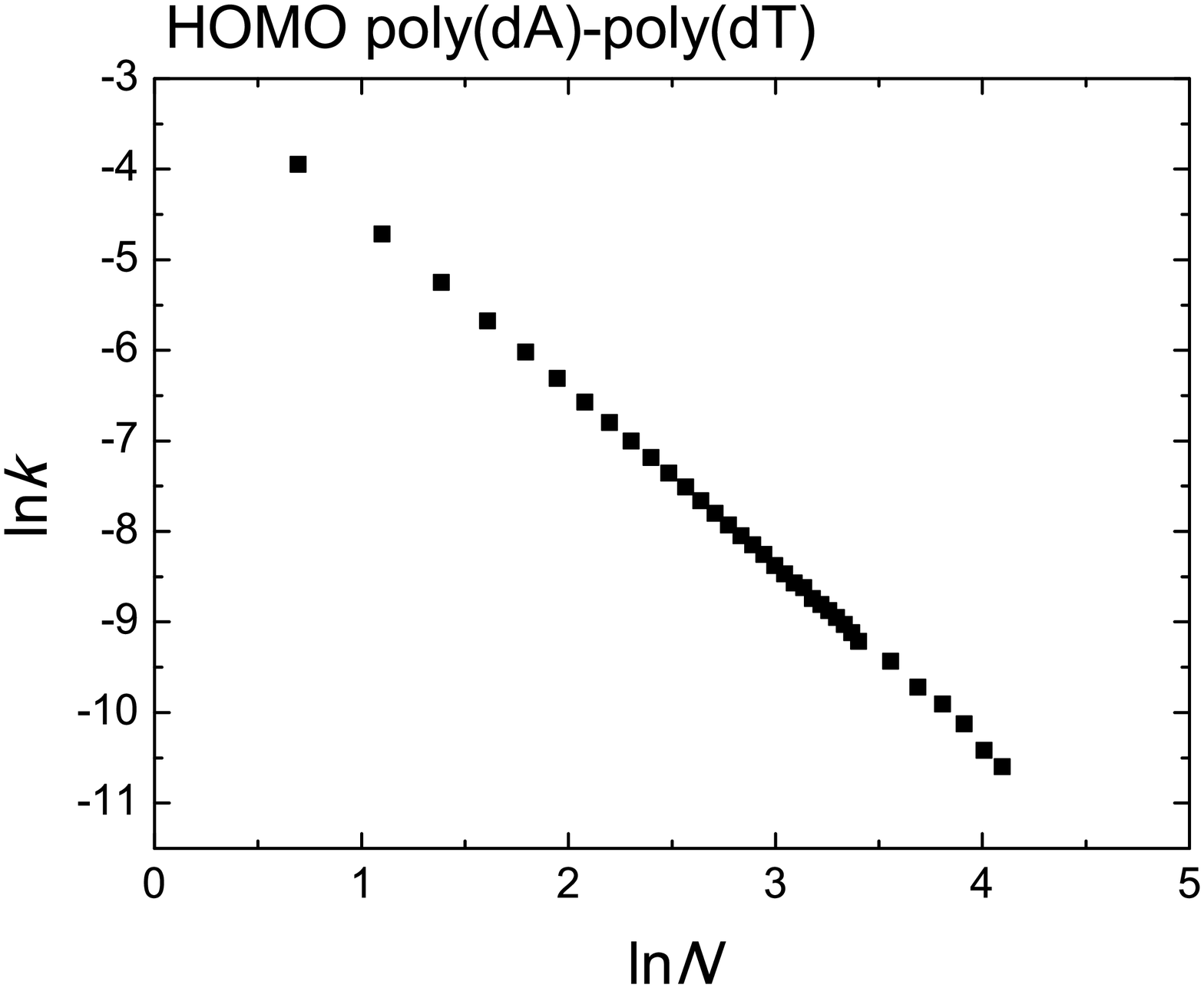}
\includegraphics[width=6cm]{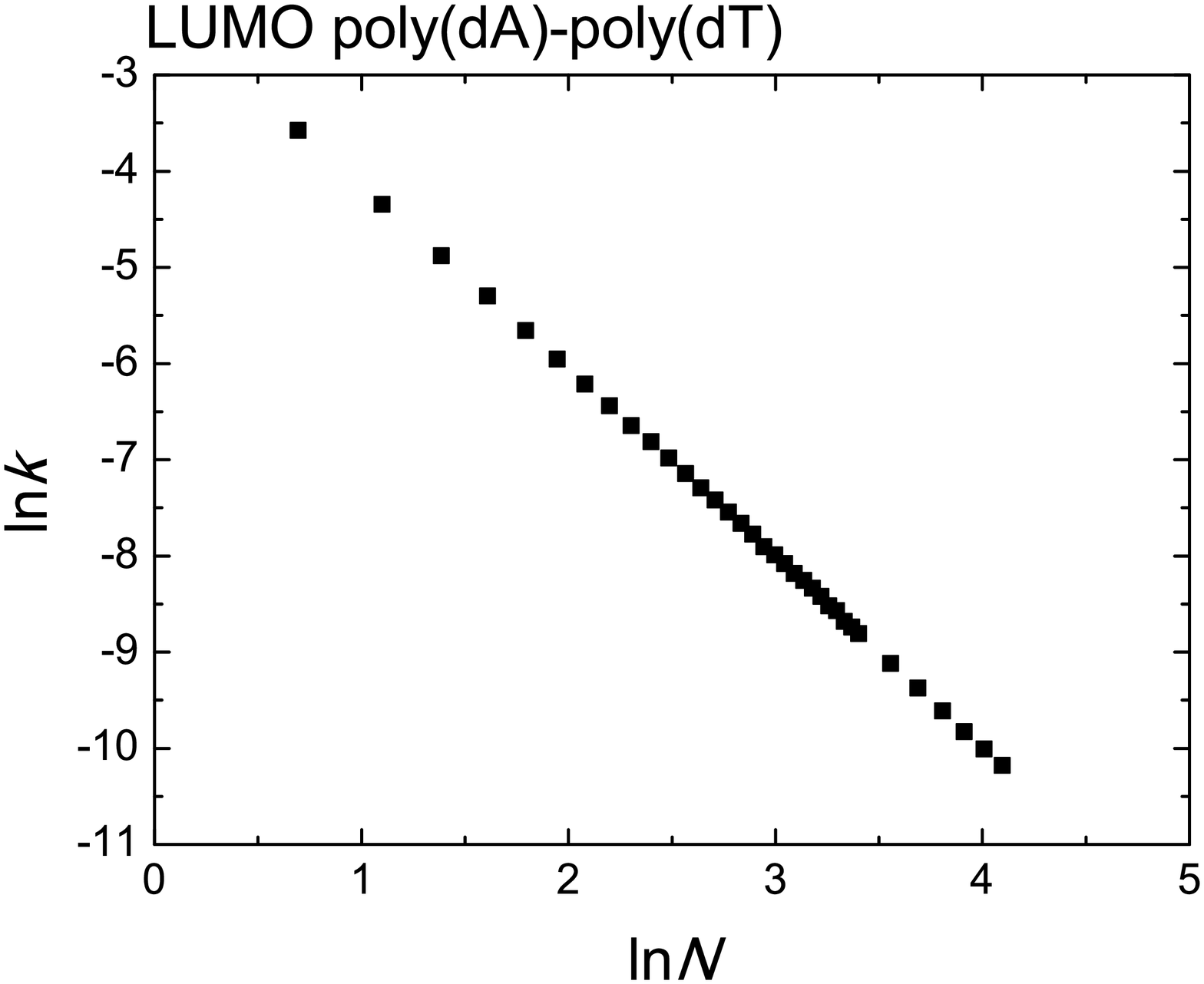}
\includegraphics[width=6cm]{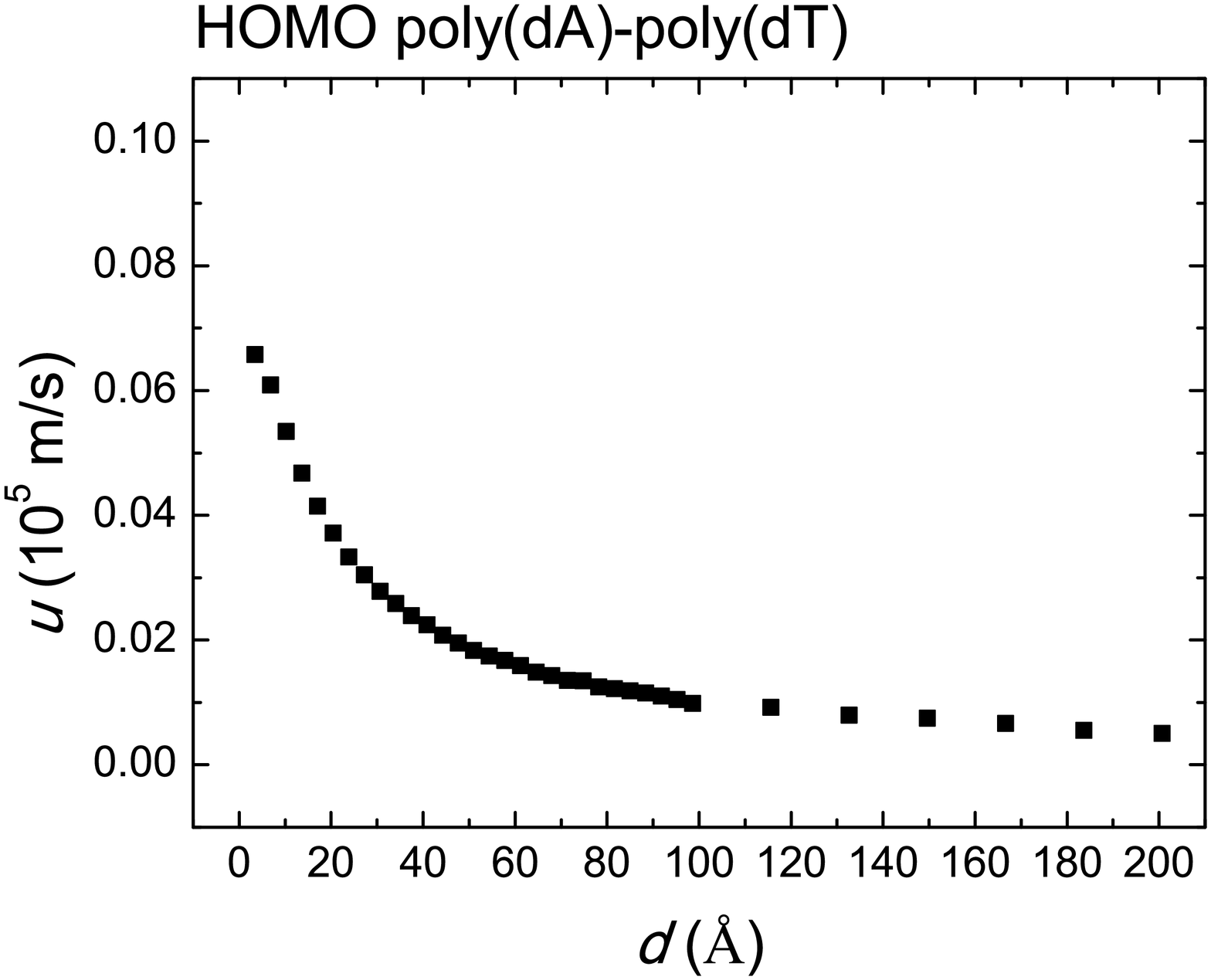}
\includegraphics[width=6cm]{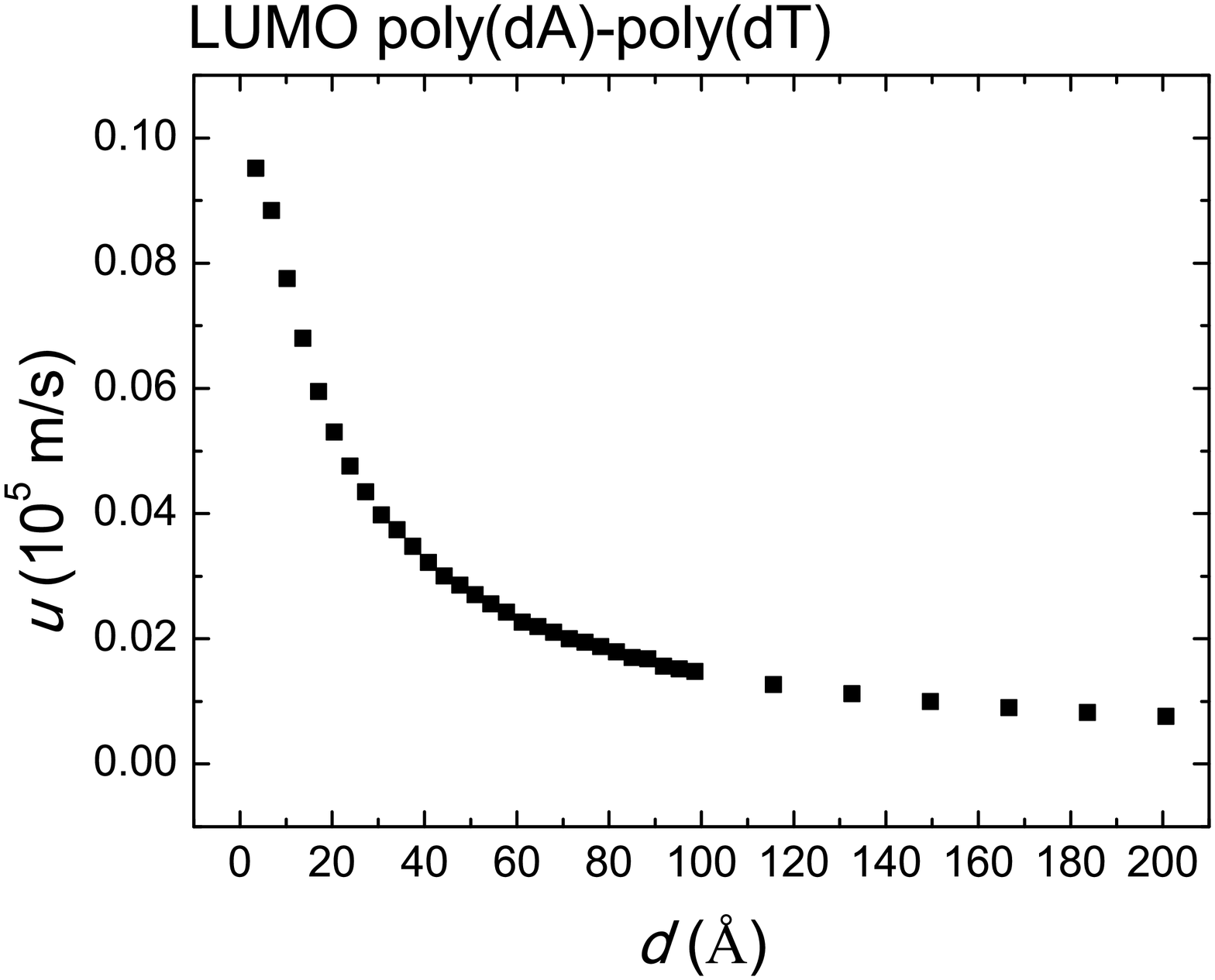}
\caption{\label{fig:HOMOLUMOpolydApolydT}
Hole (left column) and electron (right column) transfer in poly(dA)-poly(dT).
[1st row] The total probability at the {\it Edge Group}, $e(N) = \frac{3}{N+1}$, and at the {\it Middle Group}, $m(N)=\frac{N-2}{N+1}$.
[2nd row] For type $\alpha'$ polymers these probabilities are equally distributed among the monomers of these Groups.
The probability at each of the members of the Groups is shown [cf. Eqs.~(\ref{oneedgea})-(\ref{onemiddlea})].
[3rd row] The logarithm of the \textit{pure} mean transfer rate $k$ as a function of the distance from the first to the last monomer i.e. the charge transfer distance $d = (N-1) \times$ 3.4 {\AA}.
[4th row] The logarithm of $k$ as a function of the logarithm of the number of monomers $N$.
[5th row] The speed of charge transfer $ u = k d$ versus $d$.}
\end{figure*}
\end{widetext}

\end{document}